\documentclass[english,11pt]{article}

\usepackage{geometry}
\usepackage[sort,comma]{natbib}
\usepackage[pagebackref=false,colorlinks]{hyperref}
\hypersetup{pdffitwindow=true,linkcolor=blue,citecolor=blue,urlcolor=blue}
\usepackage{amsmath}
\usepackage{bbm}
\usepackage{multirow}
\usepackage{graphicx}
\DeclareGraphicsExtensions{.pdf,.jpg,.png}
\usepackage{subfigure}

\usepackage{amsfonts}
\usepackage{mathdots}
\usepackage{amsthm}
\usepackage{newtxmath}

\usepackage{multirow}
\usepackage{centernot}
\usepackage{todonotes}

\usepackage[font={footnotesize,it}, sc, bf]{caption}
\renewenvironment{abstract}{%
  \noindent\textbf\abstractname .\hspace{1pt}
}{
  \endlist \par\bigskip\bigskip
}

\usepackage{lastpage}
\RequirePackage[hyperpageref]{backref}
\renewcommand*{\backref}[1]{} 
\renewcommand*{\backrefalt}[4]{
    \ifcase #1
       No referred.
    \or
       \emph{Referred to on page #2.}
    \else
       \emph{Referred to on pages #2.}
    \fi}

\usepackage[shortlabels]{enumitem}

\usepackage[titletoc,title]{appendix}

\binoppenalty=\maxdimen
\relpenalty=\maxdimen

\begin{document}

\begin{center}
{\LARGE\bf Shrinkage estimation of large covariance matrices\\[4pt] using multiple shrinkage targets}
\end{center}
\medskip
\begin{center}
{\large Harry Gray$^{1}$, Gwena\"el G.R. Leday$^{1,*}$, \\Catalina A. Vallejos$^{2,3,*}$ and Sylvia Richardson$^{1,*}$ \\[15pt]
\emph{$^{1}$MRC Biostatistics Unit, University of Cambridge, Cambridge, United Kingdom}\\
\emph{$^{2}$MRC Human Genetics Unit, University of Edinburgh, Edinburgh, United Kingdom}\\
\emph{$^{3}$The Alan Turing Institute, London, United Kingdom}\\
\emph{$^*$ Corresponding author: gwenael.leday@mrc-bsu.cam.ac.uk (GGRL); catalina.vallejos@igmm.ed.ac.uk (CAV); sylvia.richardson@mrc-bsu.cam.ac.uk (SR).}\\
}
\end{center}

\bigskip

\begin{center}
Draft, \today
\end{center}
\bigskip\bigskip

\begin{abstract}
Linear shrinkage estimators of a covariance matrix --- defined by a weighted average of the sample covariance matrix and a pre-specified shrinkage \emph{target} matrix --- are popular when analysing high-throughput molecular data. However, their performance strongly relies on an appropriate choice of target matrix. This paper introduces a more flexible class of linear shrinkage estimators that can accommodate multiple shrinkage target matrices, directly accounting for the uncertainty regarding the target choice. This is done within a conjugate Bayesian framework, which is computationally efficient. Using both simulated and real data, we show that the proposed estimator is less sensitive to target misspecification and can outperform state-of-the-art (nonparametric) single-target linear shrinkage estimators. Using protein expression data from The Cancer Proteome Atlas we illustrate how multiple sources of prior information (obtained from more than 30 different cancer types) can be incorporated into the proposed multi-target linear shrinkage estimator. In particular, it is shown that the target-specific weights can provide insights into the differences and similarities between cancer types. Software for the method is freely available as an R-package at \href{http://github.com/HGray384/TAS}{http://github.com/HGray384/TAS}.
\end{abstract}
%\null\bigskip
%\tableofcontents

%%%%%%%%%%%%%%%%%%%%%%%%%%%%%%%%%%%%%%%%%%%%%%%%
%%%%%%%%%%%%%%%%%%%%%%%%%%%%%%%%%%%%%%%%%%%%%%%%

\section{Introduction}

Covariance matrix estimation plays a central role in statistical analyses. In molecular biology, for instance, covariance estimation facilitates the identification of dependence structures between molecular variables that shed light on the underlying molecular or cellular processes \citep{gaiteri2014, schafer2005}. %It also allows the prediction of phenotypes and classification of patients. 
Because high-throughput omics experiments typically measure a large number of molecular variables (e.g.~gene expression) on relatively few samples, the sample covariance is generally singular or ill-conditioned. This means that the sample covariance matrix suffers from high estimation error that can affect subsequent numerical tasks, such as computing its useful matrix inverse (precision matrix). This problem has been well studied \citep{daniels2001, pourahmadi2013, fan2016, engel2017} and many solutions have been proposed over the last decades. These usually modify the sample covariance so as to stabilise estimation. Some solutions adopt sparse, lasso-type, regularisation that enforces most entries of the estimated covariance matrix to be equal to zero \citep{bickel2008, cai2011, bien2011}, whereas other solutions adopt non-sparse, ridge-type, regularization that does not yield zero entries \citep{ledoit2004, warton2008, won2013, vanWieringen2016}. The choice of a particular form of regularization typically depends on the statistical goals and computational constraints \citep{bickel2006}.

Single-target linear shrinkage (STS) estimators are ridge-type estimators, which are defined as a convex combination between the sample covariance matrix and a pre-specified positive definite \emph{target} matrix. These estimators are very popular in practice due to their simplicity, ease of interpretation and computational efficiency \citep{schafer2005}. For these reasons, they have also been theoretically well studied \citep{ledoit2004, touloumis2015, fisher2011, ikeda2015, chen2010}. The performance of STS estimators, however, is highly dependent on the choice of an appropriate target matrix (see Section~\ref{Sec:modelsim}). Different target matrices have been proposed in the literature, but the choice is ultimately guided by the application and the presumed structure of the unknown covariance matrix \citep{engel2017}. 

Despite a large literature, surprisingly little has been done to extend STS estimators to allow shrinkage towards multiple shrinkage targets. To the best of our knowledge, only \cite{bartz2014} and \cite{lancewicki2014} have proposed multi-target linear shrinkage estimators. These estimators represent optimal convex combinations, in the mean square sense, between the sample covariance matrix and multiple shrinkage targets. However, for these methods, analytical derivations of the shrinkage weights are tied to a particular shrinkage target set and there is no software available.

In this article, we introduce a linear shrinkage estimator that can accommodate multiple general shrinkage target matrices, and thereby incorporate uncertainty about the target choice. The proposed estimator is obtained within a conjugate Bayesian framework which is computationally efficient, even when the number of samples, variables or shrinkage targets is relatively large. Using both simulated and real data, we show that the multi-target estimator is less sensitive to the misspecification of some of its targets and can outperform state-of-the-art (nonparametric) STS estimators. Moreover, we show that the target-specific weights can be usefully interpreted. We apply our approach to a pan-cancer proteomic data set where we illustrate how multiple sources of external information, obtained from different cancer types, can be incorporated within the target set. In particular, it is shown that target-specific shrinkage weights can provide insights into the differences and similarities between cancer types. The method proposed in this paper is implemented as an R package and freely available at \href{http://github.com/HGray384/TAS} {http://github.com/HGray384/TAS}. 

This article is organised as follows. In Section \ref{Sec:methods}, we describe STS estimators and introduce a Bayesian counterpart that we generalise to allow multiple shrinkage targets. Section~\ref{Sec:modelsim} and~\ref{Sec:datasim} compare the performance of the proposed estimator to state-of-the-art STS estimators using simulated  and real data, respectively. We apply our approach in Section~\ref{Sec:app} to a pan-cancer proteomic data set from The Cancer Proteome Atlas. Last, Section \ref{Sec:disc} discusses linear shrinkage estimation by means of multiple targets and concludes on future directions. All code used to produce the results shown in this manuscript is available at \href{http://github.com/HGray384/TAS-paper-code}{http://github.com/HGray384/TAS-paper-code}.

%%%%%%%%%%%%%%%%%%%%%%%%%%%%%%%%%%%%%%%%%%%%%%%%
%%%%%%%%%%%%%%%%%%%%%%%%%%%%%%%%%%%%%%%%%%%%%%%%

\section{Methods}
\label{Sec:methods}

Let $\boldsymbol{X}=(\boldsymbol{x}_1,\ldots ,\boldsymbol{x}_n)$ be a matrix containing $n$ independent observations drawn from a $p$-variate Normal distribution with zero mean vector and positive definite covariance matrix $\boldsymbol{\Sigma}$ (hereby denoted $\boldsymbol{\Sigma}\succ 0$). The maximum likelihood estimator (MLE) of $\boldsymbol{\Sigma}$ is $\boldsymbol{S}=\boldsymbol{X}\boldsymbol{X}^{\top} /n$, which is ill-conditioned or singular whenever $n$ is small relative to $p$ (see Supplementary Material~\ref{supp:MLE}). This section describes the class of single-target linear shrinkage estimators as a solution to this problem, as well as a Bayesian counterpart which we generalise to accommodate multiple shrinkage target matrices. The latter provides a more flexible framework while retaining computational efficiency.

\subsection{Single-target linear shrinkage covariance estimation}

An STS estimator is defined as a weighted average between the MLE and a single pre-specified matrix $\boldsymbol{\Delta}$, often referred to as the \textit{shrinkage target}, i.e.:
\begin{equation}
\hat{\boldsymbol{\Sigma}} = \alpha \boldsymbol{\Delta}+(1-\alpha)\boldsymbol{S}, \quad \text{with } \alpha\in (0,1) \text{ and } \boldsymbol{\Delta}\succ 0.
\label{Eq:linearshrinkage}
\end{equation}
This estimator can be thought of in terms of a bias-variance trade-off \citep{ledoit2004}, which is calibrated through the shrinkage intensity or weight $\alpha$. Values of $\alpha$ close to one define a low-variance but high-bias estimator ($\hat{\boldsymbol{\Sigma}} \approx \boldsymbol{\Delta}$), whilst values of $\alpha$ closer to zero define a low-bias but high-variance estimator ($\hat{\boldsymbol{\Sigma}} \approx \boldsymbol{S}$). The optimal balance for this trade-off often lies away from these limiting cases and analytical solutions have been proposed under different assumptions \citep{schafer2005, chen2010, fisher2011, touloumis2015}. The estimator in \eqref{Eq:linearshrinkage} can also be viewed as a penalized MLE under a specific ridge-type penalty, where the choice of $\alpha$ relates to a regularisation parameter \citep{vanWieringen2016}.

\subsection{Conjugate Bayesian framework}\label{Sec:gcshrink}

In a Bayesian framework, an STS estimator of the covariance  matrix can be obtained in closed-form by placing an inverse-Wishart prior on $\boldsymbol{\Sigma}$ \citep{chen1979, hannart2014}. Adopting the parametrisation of \citet{hannart2014} (Supplementary Material~\ref{supp:reparametrisation}), we denote $\boldsymbol{\Sigma} | \alpha, \boldsymbol{\Delta} \sim \text{Inv-Wishart}(\alpha, \boldsymbol{\Delta})$ with $\alpha\in (0,1)$ and $\boldsymbol{\Delta}\succ 0$. Under this parametrisation it follows that $\mathbb{E}(\boldsymbol{\Sigma} | \alpha, \boldsymbol{\Delta}) = \boldsymbol{\Delta}$ and
\begin{equation} \label{Eq:expectation}
\mathbb{E}(\boldsymbol{\Sigma} | \boldsymbol{X}, \alpha, \boldsymbol{\Delta}) = \alpha \boldsymbol{\Delta}+(1-\alpha)\boldsymbol{S},
\end{equation} thereby making explicit that the marginal posterior expectation $\mathbb{E}(\boldsymbol{\Sigma} | \boldsymbol{X}, \alpha, \boldsymbol{\Delta})$ of $\boldsymbol{\Sigma}$ is an STS estimator with shrinkage target equal to the prior expectation of $\boldsymbol{\Sigma}$.

In recent work, \cite{hannart2014} introduced a general framework for empirical Bayes estimation (through marginal likelihood maximisation) of $\alpha$ and $\boldsymbol{\Delta}(\boldsymbol{\theta})$ when the shrinkage target is parametrised in terms of a low-dimensional vector $\boldsymbol{\theta}$. In the particular case where the shrinkage target is fully specified a priori, the problem of estimating $\alpha$ reduces to the optimisation of a univariate concave objective function. \cite{hannart2014} observed that the empirical Bayes estimate of $\alpha$ is often close to the value that minimises the mean square error. However, the uncertainty regarding this estimate can be large in some cases (see Supplementary Material~\ref{supp:eb}).

\subsection{Incorporating uncertainty about $\alpha$ and $\boldsymbol{\Delta}$} \label{Sec:alpha}

In this section, we hierarchically extend the conjugate model introduced in Section \ref{Sec:gcshrink} by placing independent hyper-prior distributions on $\alpha$ and $\boldsymbol{\Delta}$, such that the posterior expectation of $\boldsymbol{\Sigma}$ remains available in closed-form. We place a uniform discrete prior on $\alpha$ over the support $\mathcal{A}=\{a_1, \ldots, a_{K}\}$, where $0 < a_1 < \cdots < a_K < 1$ and $\text{p}(\alpha=a_k)=1/K$ for $k\in\{1,\ldots,K \}$. Similarly, we place a uniform discrete prior on $\boldsymbol{\Delta}$ over the support $\mathcal{D}=\{\boldsymbol{D}_1, \ldots, \boldsymbol{D}_L\}$, hereafter referred to as the \emph{target set}. We assume that $\boldsymbol{D}_l\succ 0$ and $\text{p}(\boldsymbol{\Delta}=\boldsymbol{D}_l)=1/L$ for $l\in\{1,\ldots,L\}$. Under these priors, the marginal posterior expectation of $\boldsymbol{\Sigma}$ is given by
\begin{equation}
	\mathbb{E}[\boldsymbol{\Sigma} | \boldsymbol{X}]=\sum_{l=1}^L \sum_{k=1}^K \mathbb{E}[\boldsymbol{\Sigma}|\boldsymbol{X}, \alpha = a_k, \boldsymbol{\Delta}=\boldsymbol{D}_l]\text{p}(\alpha = a_k, \boldsymbol{\Delta}=\boldsymbol{D}_l | \boldsymbol{X}),
	\label{Eq:margexp}
\end{equation}
where
\begin{equation}
	\text{p}(\alpha = a_k, \boldsymbol{\Delta} = \boldsymbol{D}_l | \boldsymbol{X}) = \frac{\text{p}(\boldsymbol{X}|\alpha = a_k, \boldsymbol{\Delta}=\boldsymbol{D}_l) \text{p}(\alpha = a_k) \text{p}(\boldsymbol{\Delta}=\boldsymbol{D}_l)}{\sum_{q=1}^L \sum_{k=1}^K \text{p}(\boldsymbol{X}|\alpha  = a_k, \boldsymbol{\Delta}=\boldsymbol{D}_q) \text{p}(\alpha = a_k) \text{p}(\boldsymbol{\Delta}=\boldsymbol{D}_q)}.
	\label{Eq:margweights}
\end{equation}

Note that \eqref{Eq:margexp} is akin to a model average estimator \citep{hoeting1999}, combining individual STS estimators obtained from the statistical models indexed by the support of $(\alpha, \boldsymbol{\Delta})$. The estimator in \eqref{Eq:margexp} can also be re-formulated as
\begin{equation} 
	\mathbb{E}[\boldsymbol{\Sigma} | \boldsymbol{X}]=\sum_{l=1}^L w_l \boldsymbol{D}_l + \Bigg(1-\sum_{l=1}^L w_l\Bigg)\boldsymbol{S},
	\label{Eq:TAS}
\end{equation}
where
\begin{equation} 
w_l= \sum_{k=1}^K  a_k \text{p}(\alpha = a_k, \boldsymbol{\Delta}=\boldsymbol{D}_l | \boldsymbol{X})
\label{Eq:TAS_weights}
\end{equation}
is a target-specific posterior weight synthesizing the contribution of the target $\boldsymbol{D}_l$ relative to the target set $\mathcal{D}$. This reformulation shows that $\mathbb{E}[\boldsymbol{\Sigma} | \boldsymbol{X}]$ lies within the family of multi-target linear shrinkage estimators: it is a convex combination between the MLE and the target matrices $\boldsymbol{D}_1, \ldots, \boldsymbol{D}_L$. We refer to the estimator in \eqref{Eq:TAS} as the Target-Averaged linear Shrinkage (TAS) estimator, hereafter denoted by $\boldsymbol{\hat{\Sigma}}_{\text{TAS}}$.

The proposed estimator has several desirable properties. First, it provides a generic framework where any positive definite target matrix can be incorporated in the target set $\mathcal{D}$. Second, it is computationally attractive since the computation of \eqref{Eq:TAS} only requires $K\times L$ evaluations of the marginal likelihood of a Gaussian conjugate model, which is available in closed-form (see Supplementary Material~\ref{supp:ml}). Also, when an additional target matrix $\boldsymbol{D}_{L+1}$ is added to the set $\mathcal{D}$, updating \eqref{Eq:TAS} only requires $K$ new marginal likelihood evaluations and subsequently re-distributing the weights. Third, the target-specific weights $w_l$ may provide valuable insights (see Section \ref{Sec:modelsim}, \ref{Sec:datasim}, \ref{Sec:app}).

\subsection{Choice of shrinkage target matrices}
\label{Sec:targetset}

The performance of the TAS estimator depends on the choice of the set of target matrices $\mathcal{D}$, much alike the performance of STS estimators depends on the choice of the target matrix $\boldsymbol{\Delta}$. Here, we discuss the choice of $\mathcal{D}$.

In the absence of prior information, the set $\mathcal{D}$ may include, for example, the nine target matrices described in Table \ref{Tab:targets}. Such choice may be seen as a sensible starting point due to the popularity of these nine targets in the literature. Note, however, that some of the targets can be nearly identical in some cases (e.g. $\boldsymbol{T}_2$ and $\boldsymbol{T}_5$ when $\bar{r}\approx 0$), so the posterior weights in~\eqref{Eq:TAS_weights} must be interpreted with care.
It is also possible to further enrich this set with any covariance structures not listed in Table \ref{Tab:targets}. Examples include Toeplitz, higher-order autoregressive, or latent factor structures \citep[e.g.][]{chen1979, ledoit2003}.

The set $\mathcal{D}$ may also be used to incorporate external information about $\boldsymbol{\Sigma}$, provided this can be translated into a positive definite covariance matrix. The availability of such information may arise in situations where the same set of molecular variables has been measured on an independent sample that is thought to be biologically related (e.g. similar disease). In this case, a target matrix may be constructed using the sample covariance matrix of the auxiliary data, or regularised versions thereof. This is illustrated in Section \ref{Sec:app} using data from The Cancer Proteome Atlas.\\

\begin{table}[ht]\small
\centering
\begin{tabular}{rccc}
	\hline\hline\\[-8pt]
    % & & & Correlation & \\
    & zero correlation & constant correlation & decaying correlations \\
    & ($r_{ij} = 0$) & ($r_{ij} = \bar{r}$) & ($r_{ij} = \bar{r}^{|i-j|}$)\\[2pt] \hline\\[-6pt]
    unit variance ($v_{i} = 1$)  & $\boldsymbol{T}_1$    & $\boldsymbol{T}_4$ &   $\boldsymbol{T}_7$\\
    common variance ($v_{i} = \bar{s}$) & $\boldsymbol{T}_2$  & $\boldsymbol{T}_5$   & $\boldsymbol{T}_8$  \\
    unequal variances ($v_{i} = s_{ii}$) & $\boldsymbol{T}_3$ & $\boldsymbol{T}_6$ &  $\boldsymbol{T}_9$  \\[2pt]  \hline\hline
\end{tabular}
\caption{Popular choices of shrinkage target matrices for STS estimators. A shrinkage target $\boldsymbol{T}=\boldsymbol{V}^{1/2} \boldsymbol{R} \boldsymbol{V}^{1/2}$, with $\boldsymbol{V} = \text{diag}\{v_1, \ldots, v_p\}$ a diagonal variance matrix and $\boldsymbol{R} = (r_{ij})_{1\leq i < j \leq p}$ a correlation matrix. Here, $s_{ij}$ denotes the $(i,j)^{\text{th}}$ element of the sample covariance matrix $\boldsymbol{S}$; $\bar{s}$ and $\bar{r}$ are the averages of the empirical variances and correlations, respectively.}
\label{Tab:targets}
\end{table}

\subsection{Implementation}

The proposed method is freely available as an R package at \href{http://github.com/HGray384/TAS} {http://github.com/HGray384/TAS}. As default, $\mathcal{D}$ comprises the nine shrinkage targets defined in Table \ref{Tab:targets} and the support $\mathcal{A}$ is set as $\{a_1=0.01, a_2=0.02, \ldots, a_{99}=0.99\}$ (note that increasing the granularity of this grid does not affect results; see Supplementary Material~\ref{supp:card}). However, these choices can easily be modified when using the software. We remark that the $K \times L$ marginal likelihood evaluations that are required to compute \eqref{Eq:TAS} can easily be parallelised to further reduce computational time. We observe, however, that this is not critical in practice (see Table \ref{Tab:time}).

\begin{table}[ht]\small
\centering
\begin{tabular}{cccc}
	\hline\hline\\[-8pt]
    & $p=100$ & $p=500$ & $p=1000$\\ \hline
    $n=100$  & 0.08 & 4.46 & 33.61 \\
    $n=250$ & 0.09 & 4.68 & 33.21  \\
    $n=500$ & 0.10 & 5.14 & 34.14 \\[3pt]  \hline\hline
\end{tabular}
\caption{Average time in seconds (over 100 repetitions) to compute the TAS estimate (using the nine targets in Table \ref{Tab:targets}) as a function of the number $n$ of samples and $p$ of variables. Timings were measured on a Dell OptiPlex7040 with Intel Core i7-6700CPU.}
\label{Tab:time}
\end{table}

%%%%%%%%%%%%%%%%%%%%%%%%%%%%%%%%%%%%%%%%%%%%%%%%
%%%%%%%%%%%%%%%%%%%%%%%%%%%%%%%%%%%%%%%%%%%%%%%%
%\newpage
\section{Model-based simulation study}
\label{Sec:modelsim}

In this section, we study the performance of the proposed estimator using simulated data. We generate $M=100$ data sets of size $n\in \{25, 50, 75\}$ from a $p$-variate Gaussian distribution with zero mean vector and covariance matrix $\Sigma$, where $p=100$. Four distinct covariance structures are considered, yielding the following four simulation scenarios:
\begin{itemize}
	\item \textbf{Scenario 1: common variance, zero correlation.} $\boldsymbol{\Sigma}_1 = 5 \times \boldsymbol{I}_{p\times p}$,
	\item \textbf{Scenario 2: unit variance, constant correlation.} $\boldsymbol{\Sigma}_2 = \boldsymbol{I}_{p\times p} + 0.3 \times ( \mathbf{1}_{p\times p} - \boldsymbol{I}_{p\times p})$, where $\mathbf{1}_{q\times r}$ is the $q\times r$ unit matrix with elements all equal to one.
 \item \textbf{Scenario 3: unequal variances, decaying correlations.} $\boldsymbol{\Sigma}_3 = \boldsymbol{D}^{1/2}\boldsymbol{C}\boldsymbol{D}^{1/2}$, where the $(i,j)^{\text{th}}$ entry of $\boldsymbol{C}$ equals $(-0.7)^{\vert i-j \vert}$ and $\boldsymbol{D}=\text{diag}(d_1, \ldots, d_p)$ with $d_i\sim \mathcal{U}(1,5)$.
	\item \textbf{Scenario 4: unit variance, block-diagonal correlation.} $\boldsymbol{\Sigma}_4 \sim \text{Inv-Wishart}$, such that $\mathbb{E}[\boldsymbol{\Sigma}_4] \propto \boldsymbol{B}$, where $\boldsymbol{B}$ is a block-diagonal matrix with two identical $p/2 \times p/2$ blocks, each with the same constant correlation structure that was used in scenario 2. 
\end{itemize}

These scenarios have been chosen to capture distinct covariance structures that are represented in the default target set $\mathcal{D}=\{\boldsymbol{T}_1, \ldots, \boldsymbol{T}_9\}$ (i.e.~$\boldsymbol{T}_2$, $\boldsymbol{T}_4$ and $\boldsymbol{T}_9$ for scenarios 1, 2 and 3 respectively), as well as to include a case (scenario 4) that is not captured by the target set $\mathcal{D}$. Using data simulated under these scenarios, we compare the performance of the multi-target shrinkage estimator $\boldsymbol{\hat{\Sigma}}_{\text{TAS}}$, with target set $\mathcal{D}$ and the nine STS estimators obtained when using each of the shrinkage targets in $\mathcal{D}$ separately. These are denoted by $\boldsymbol{\hat{\Sigma}}_{\text{ST1}}$, ..., $\boldsymbol{\hat{\Sigma}}_{\text{ST9}}$. We also consider the estimators of \citet{schafer2005}  and \citet{touloumis2015}, respectively implemented in the R packages \emph{corpcor} and \emph{ShrinkCovMat}. The estimator of \citet{schafer2005} is an STS estimator obtained via a two-step approach in which the sample variances are shrunk towards their median and the sample correlations shrunk towards zero. We denote this estimator by $\boldsymbol{\hat{\Sigma}}_{\text{cpc}}$. The estimators proposed by \citet{touloumis2015} are three non-parametric STS estimators (i.e.~they do not rely on distributional assumptions) with shrinkage targets $\boldsymbol{T}_1$, $\boldsymbol{T}_2$, and $\boldsymbol{T}_3$. We denote these by $\boldsymbol{\hat{\Sigma}}_{\text{AT1}}$, $\boldsymbol{\hat{\Sigma}}_{\text{AT2}}$, and $\boldsymbol{\hat{\Sigma}}_{\text{AT3}}$, respectively. We remark that the estimators of \citet{touloumis2015} were reported to outperform those of \citet{chen2010} and \citet{fisher2011}, while being comparable to that of \citet{ikeda2015}.

To assess the performance of these 14 estimators, we report the Percentage Relative Improvement in Average Loss (PRIAL) \citep{touloumis2015, ikeda2015}:
\begin{equation}
\label{prial} 
	\frac{\sum_{m=1}^M\|\boldsymbol{\Sigma}-\boldsymbol{S^{(m)}}\|_{F}^2-\sum_{m=1}^M\|\boldsymbol{\Sigma}-\boldsymbol{\hat{\Sigma}^{(m)}}\|_{F}^2}{\sum_{m=1}^M\|\boldsymbol{\Sigma}-\boldsymbol{S^{(m)}}\|_{F}^2}*100,
\end{equation}
where $\| \cdot \|_{F}$ denotes the Frobenius norm. The PRIAL measures the relative improvement of an estimator $\boldsymbol{\hat{\Sigma}}$ over the sample covariance matrix $\boldsymbol{S}$, across the $M$ simulated data sets. A negative value indicates that the estimator $\boldsymbol{\hat{\Sigma}}$ does not improve upon $\boldsymbol{S}$, whereas a positive value indicates an improvement. The improvement is relatively small when the PRIAL value is close to 0\% (in which case $\boldsymbol{\hat{\Sigma}}$ is relatively closer to $\boldsymbol{S}$) and relatively large when the PRIAL value is close to 100\% (in which case $\boldsymbol{\hat{\Sigma}}$ is relatively closer to $\boldsymbol{\Sigma}$). The PRIAL can also be interpreted as the improvement of performing shrinkage versus no shrinkage.

\begin{figure}[ht]
	\begin{minipage}[c]{0.5\linewidth}
    \null\hfill
  		\subfigure[][Scenario 1: PRIAL]{
  			\includegraphics[width=0.8\textwidth]{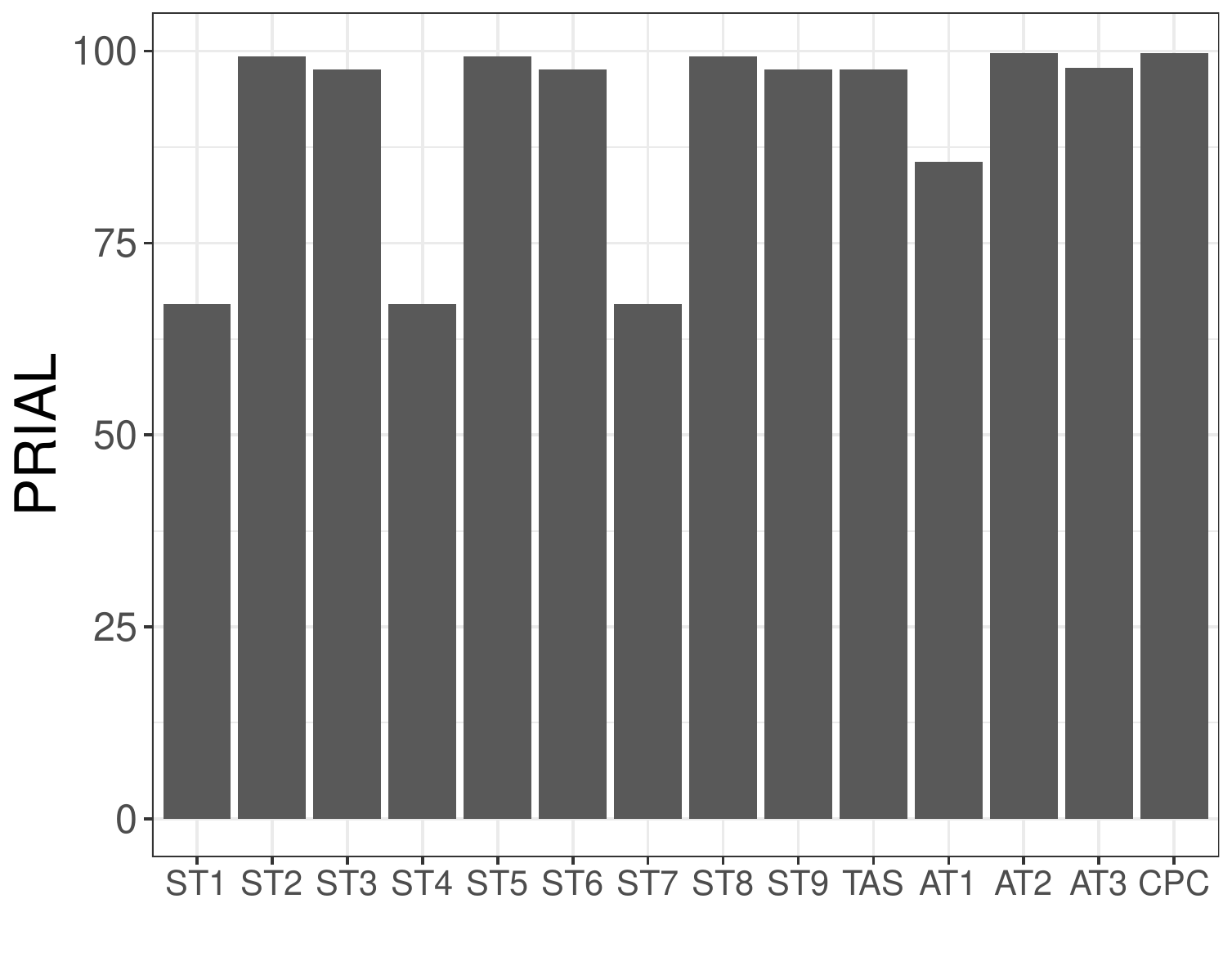}
  			\label{Fig:modelSim1:sub1}
  		}
  \end{minipage}\hfill
  	\begin{minipage}[c]{0.5\linewidth} 
    		\subfigure[][Scenario 1: target-specific posterior weights]{
  			\includegraphics[width=0.8\textwidth]{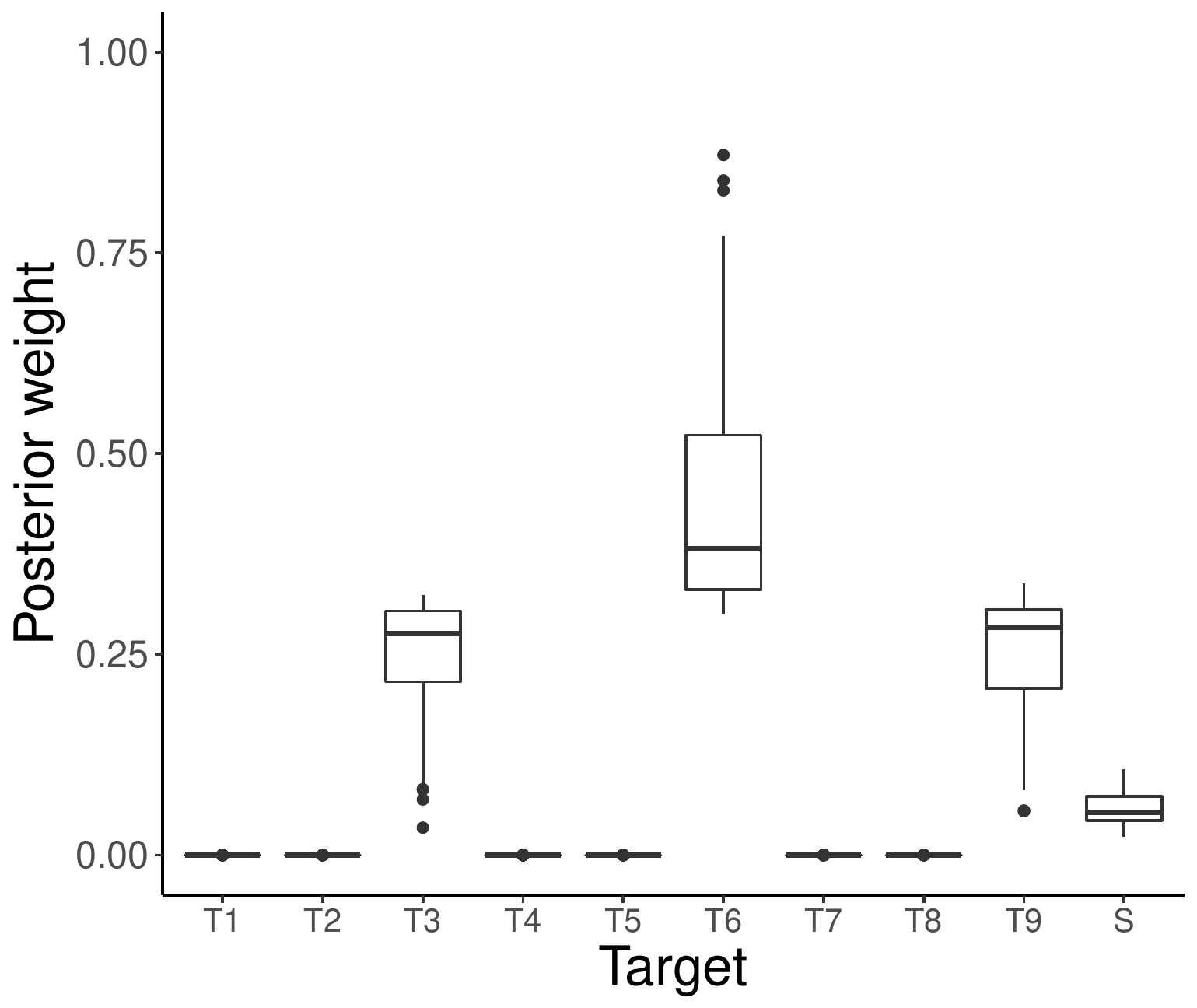}
  			\label{Fig:modelSim1:sub1w}
  		}
  \hfill\null
  \end{minipage}\hfill
	\begin{minipage}[c]{0.5\linewidth}
    \null\hfill
  		\subfigure[][Scenario 2: PRIAL]{
  			\includegraphics[width=0.8\textwidth]{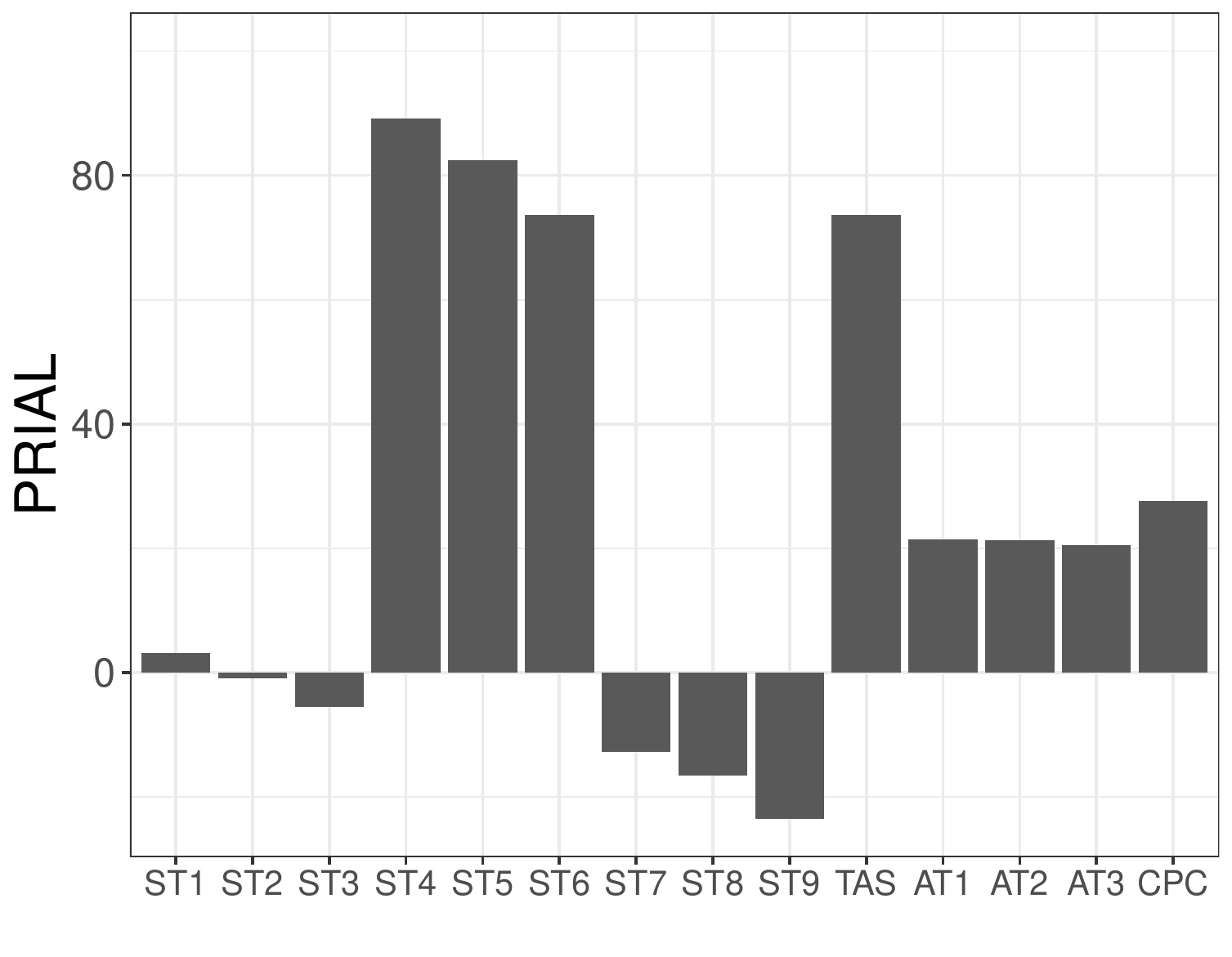}
  			\label{Fig:modelSim1:sub2}
  		}
  \end{minipage}\hfill
    	\begin{minipage}[c]{0.5\linewidth} 
    		\subfigure[][Scenario 2: target-specific posterior weights]{
  			\includegraphics[width=0.8\textwidth]{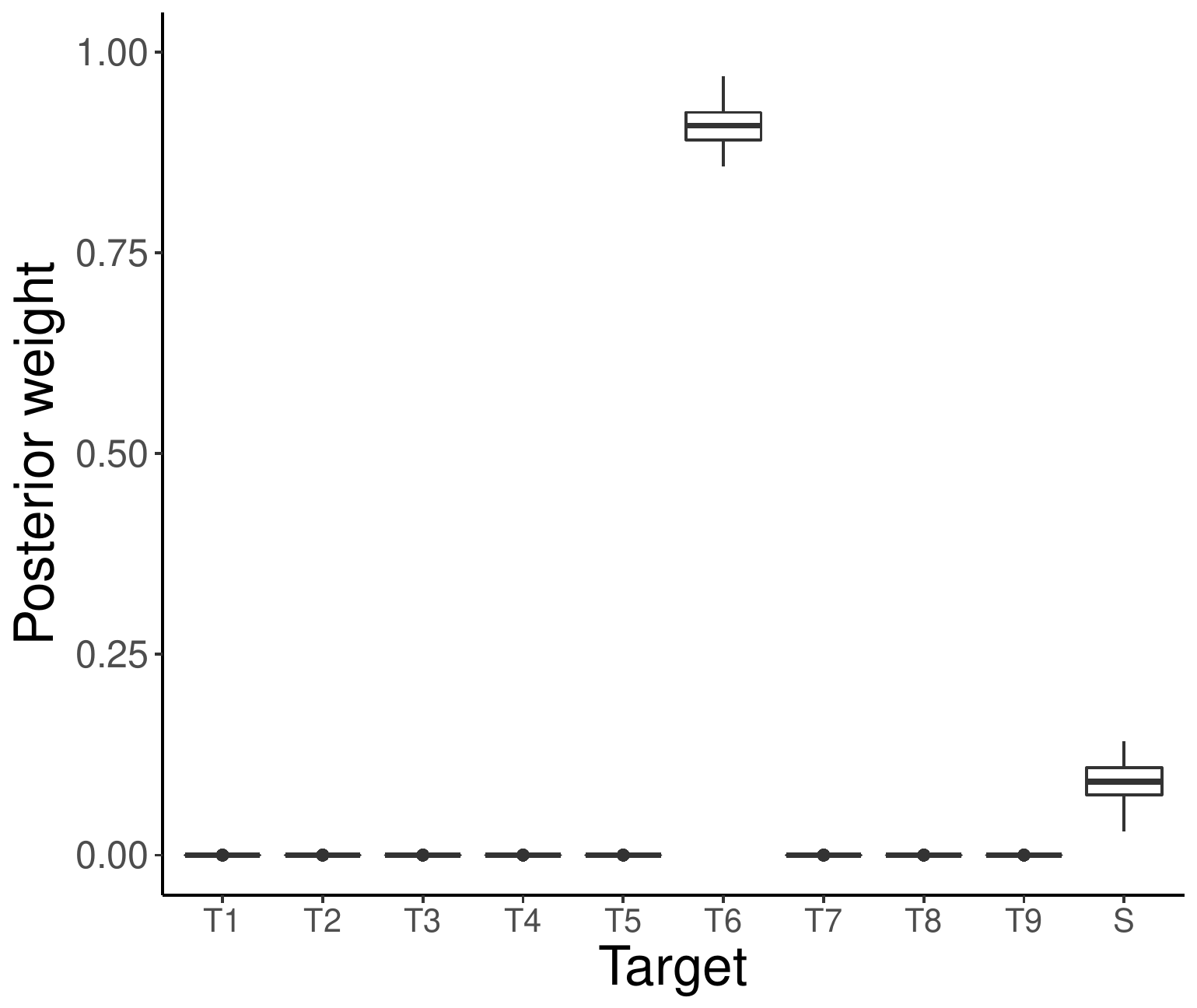}
  			\label{Fig:modelSim1:sub2w}
  		}
  \hfill\null
  \end{minipage}\hfill
  		\caption{Simulation results for scenarios 1 and 2 when $n=25$. Barplots display the PRIAL for each estimator and boxplots display target-specific posterior weights (see equation \eqref{Eq:TAS_weights}) of the TAS estimator. ST1,~\ldots,~ ST9 refer to the nine STS estimators, TAS to estimator \eqref{Eq:TAS}, AT1, \ldots, AT3 to the three estimators of \citet{touloumis2015} and CPC to the estimator of \citet{schafer2005}.}
		\label{Fig:modelSim1}
\end{figure}

\begin{figure}[ht]
 \begin{minipage}[c]{0.5\linewidth}
    \null\hfill
  		\subfigure[][Scenario 3: PRIAL]{
  			\includegraphics[width=0.8\textwidth]{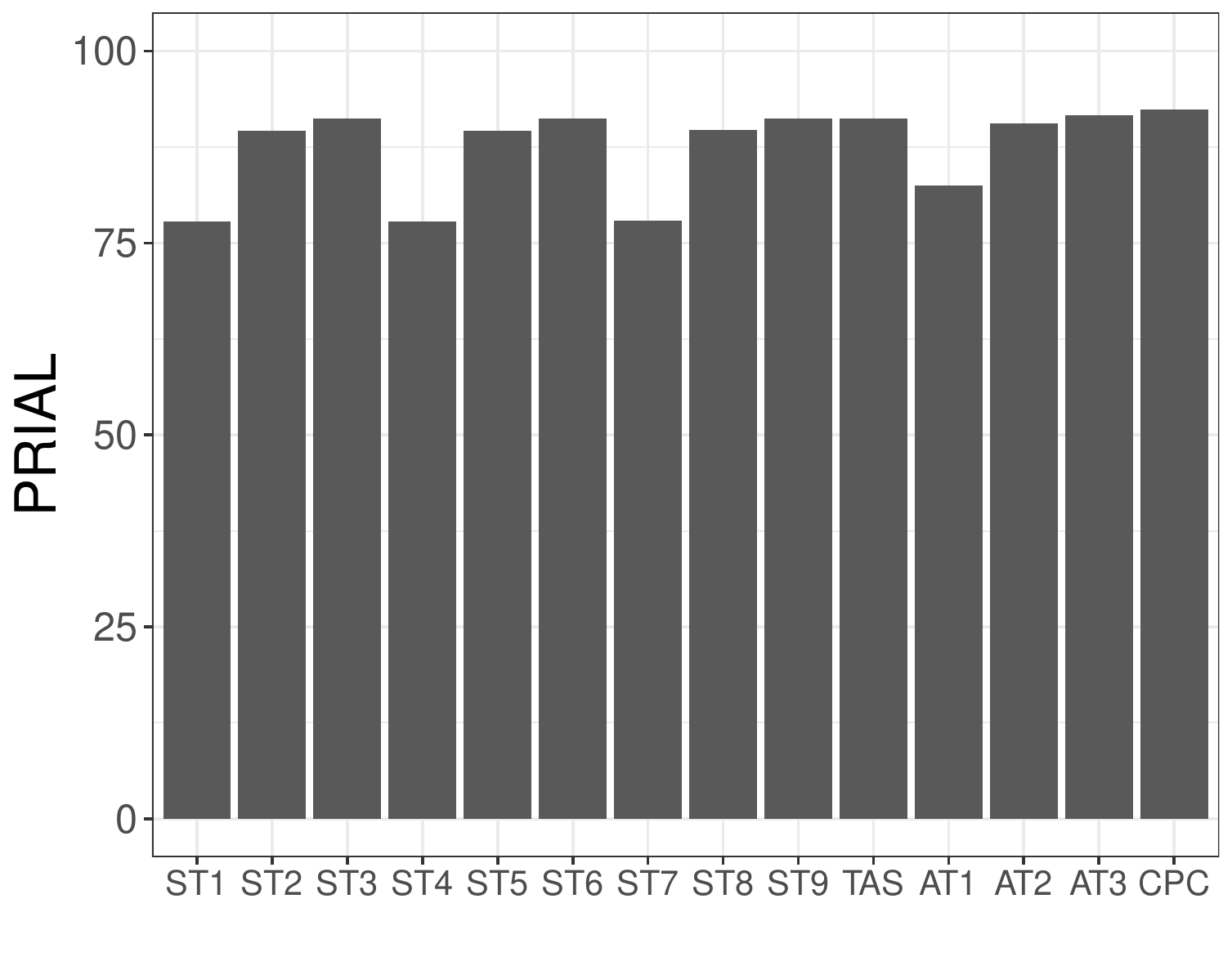}
  			\label{Fig:modelSim2:sub3}
  		}
  \end{minipage}\hfill
      	\begin{minipage}[c]{0.5\linewidth}
    		\subfigure[][Scenario 3: target-specific posterior weights]{
  			\includegraphics[width=0.8\textwidth]{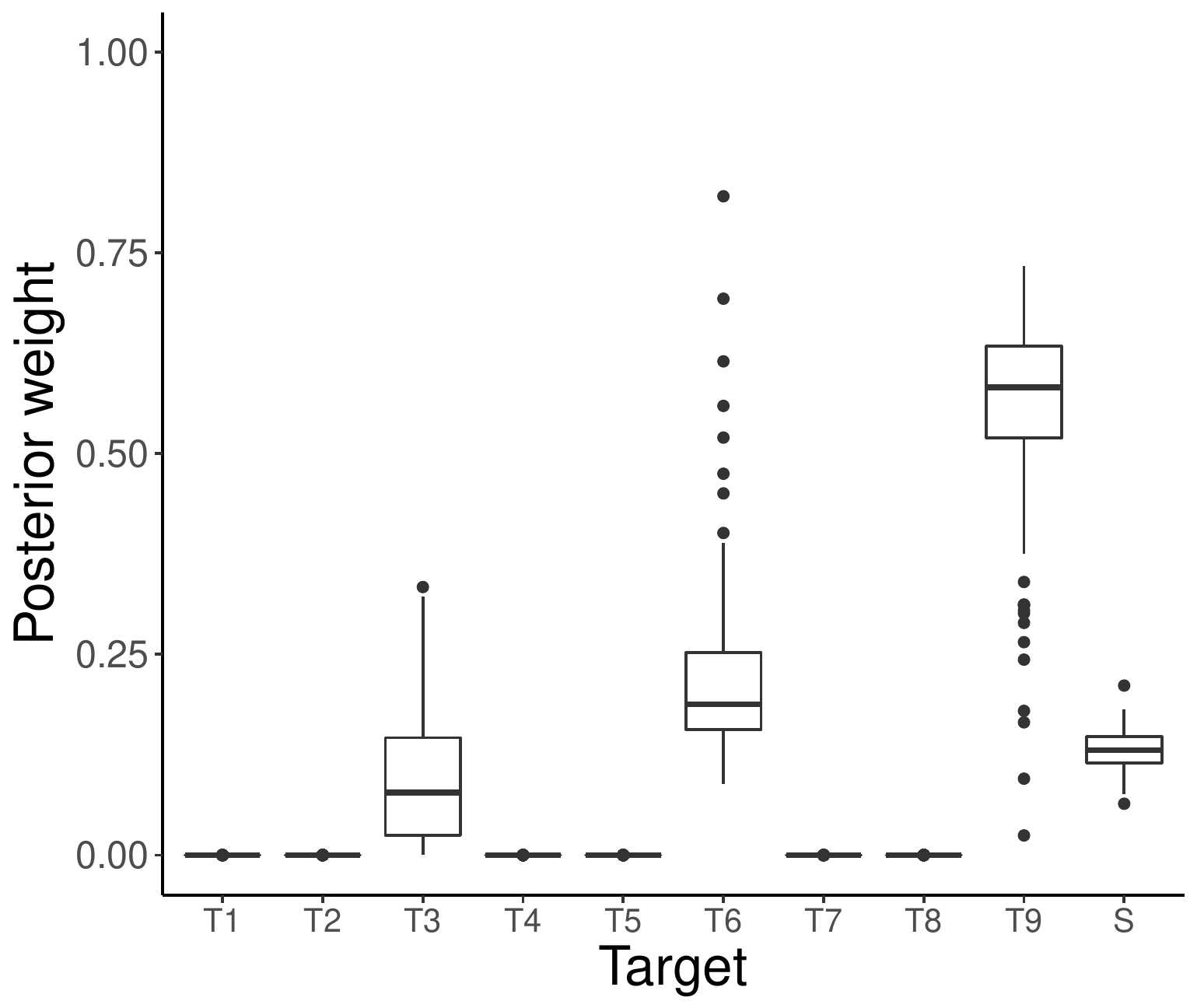}
  			\label{Fig:modelSim2:sub3w}
  		}
  \hfill\null
  \end{minipage}\hfill
  \begin{minipage}[c]{0.5\linewidth}
   \null\hfill
  		\subfigure[][Scenario 4: PRIAL]{
  			\includegraphics[width=0.8\textwidth]{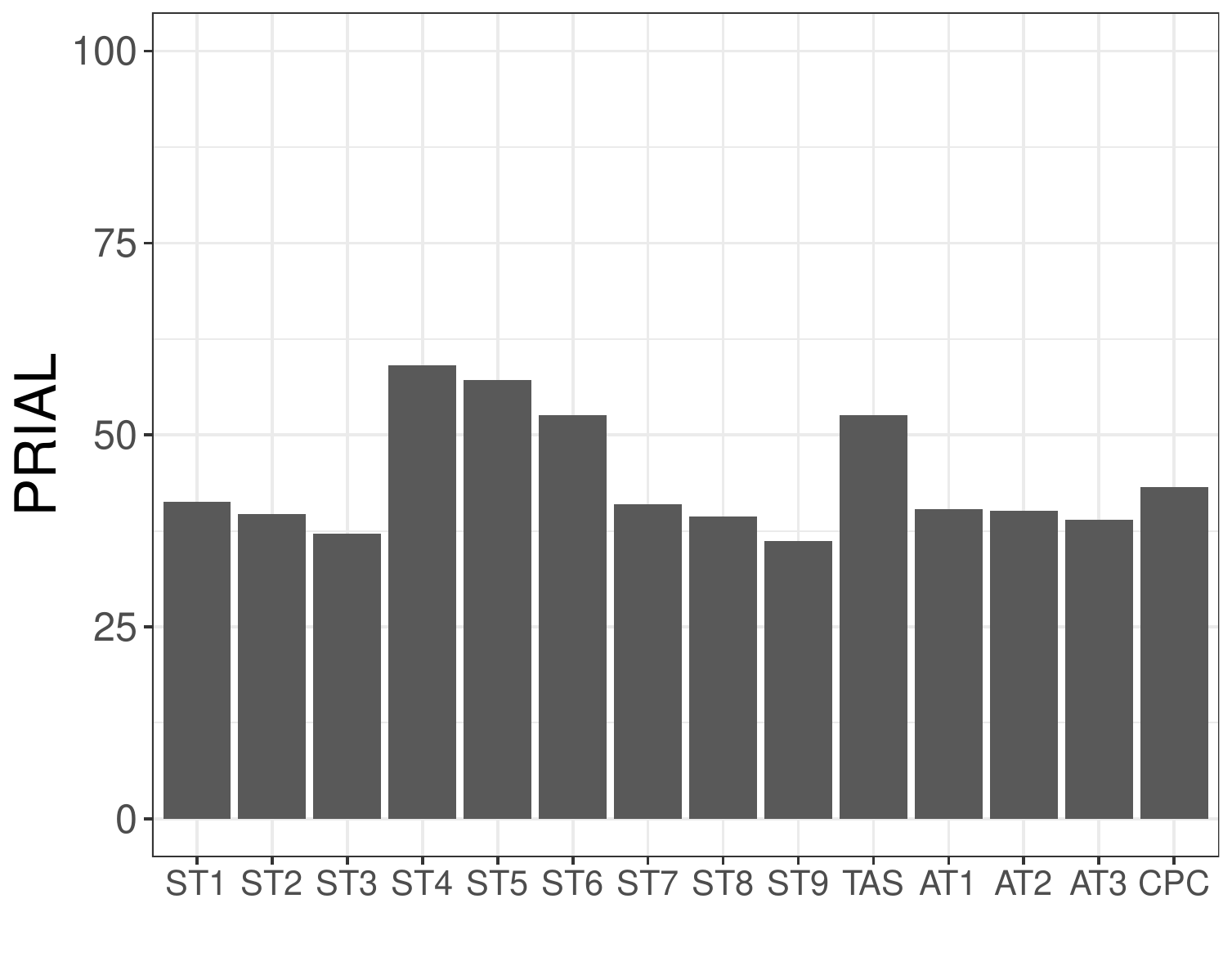}
  			\label{Fig:modelSim2:sub4}
  		}
  \end{minipage}\hfill
      	\begin{minipage}[c]{0.5\linewidth} 
    		\subfigure[][Scenario 4: target-specific posterior weights]{
  			\includegraphics[width=0.8\textwidth]{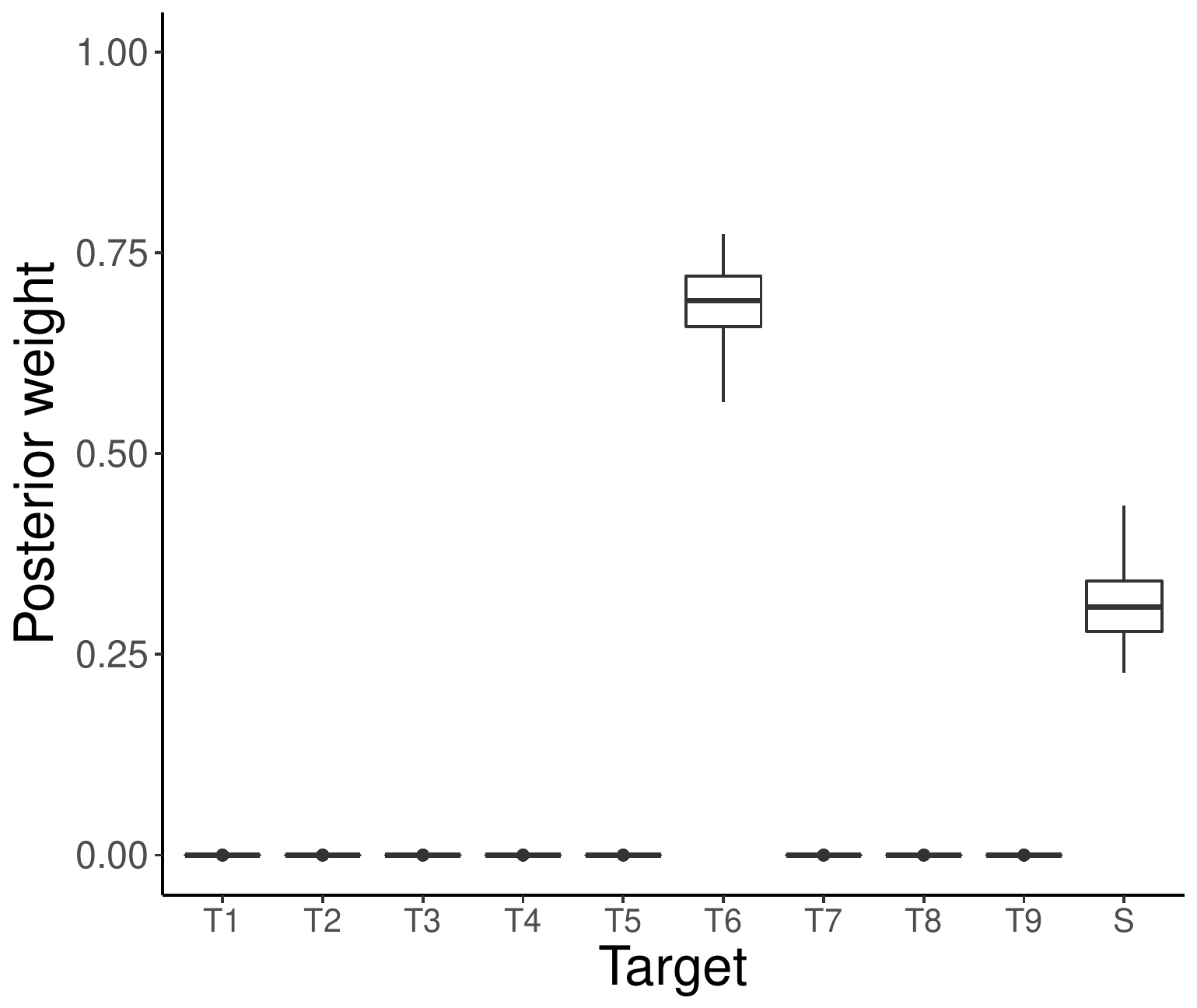}
  			\label{Fig:modelSim2:sub4w}
  		}
  \hfill\null
  \end{minipage}\hfill
  		\caption{Simulation results for scenarios 3 and 4 when $n=25$. Barplots display the PRIAL for each estimator and boxplots display target-specific posterior weights (see equation \eqref{Eq:TAS_weights}) of the TAS estimator. ST1,~\ldots,~ ST9 refer to the nine STS estimators, TAS to estimator \eqref{Eq:TAS}, AT1, \ldots, AT3 to the three estimators of \citet{touloumis2015} and CPC to the estimator of \citet{schafer2005}.}
		\label{Fig:modelSim2}
\end{figure}

Figures~\ref{Fig:modelSim1} and \ref{Fig:modelSim2} summarise the results obtained for $n=25$ (results for $n \in \{50, 75\}$, which are similar to that of $n=25$, are provided in Supplementary Material \ref{supp:model_sims}). Overall, we observe that the performance of STS estimators clearly varies across the different simulation scenarios, and that it may strongly depend on the choice of shrinkage target. Large PRIAL values are observed for STS estimators when the shrinkage target resembles the true covariance matrix (e.g.~$\boldsymbol{T}_4$ in scenario 2), whereas negative PRIAL values (indicating that the estimator performs worse than the sample covariance matrix) are observed in cases where the shrinkage target is \emph{misspecified} (see scenario 2). In contrast, the TAS estimator achieves a similar performance with respect to the best STS estimator without having to choose the correct shrinkage target, and this even when the target set does not contain the true underlying covariance structure (see scenario 4). This highlights a key strength of the proposed multi-target estimator, namely that it is less sensitive to misspecification of its targets. %\\[10pt]

As illustrated in the right panels of Figures~\ref{Fig:modelSim1} and \ref{Fig:modelSim2}, target-specific posterior weights (see equation \eqref{Eq:TAS_weights}) can also provide insights about the structure of the true covariance matrix $\boldsymbol{\Sigma}$. For example, in scenario 3, TAS allocates the highest posterior weight to shrinkage targets that match the underlying covariance structure of the data (i.e.~$\boldsymbol{T}_9$). A similar behaviour is observed in scenario 1 and 2, although this is less clear. Indeed, the shrinkage target $\boldsymbol{T}_6$ is assigned the largest weight in scenario 2, while it would be expected that $\boldsymbol{T}_4$ has the highest weight. Similarly, the shrinkage targets $\boldsymbol{T}_3$, $\boldsymbol{T}_6$ and $\boldsymbol{T}_9$ have high posterior weights in scenario 1 whereas it would be expected that $\boldsymbol{T}_2$ has the highest weight. However, closer inspection of the shrinkage targets (see Supplementary Figure~\ref{supp:Fig:modelSim5}) shows that $\boldsymbol{T}_6$ is almost equal to $\boldsymbol{T}_4$ in scenario 2, and that $\boldsymbol{T}_3$, $\boldsymbol{T}_6$ and $\boldsymbol{T}_9$ are almost equal to $\boldsymbol{T}_2$ in scenario 1. It is also observed that the distances (as measured by the Frobenius norm) between each of these targets to the true covariance matrix are almost equal (see Supplementary Figure~\ref{supp:Fig:modelSim5}). Additionally, in scenario 4, the highest posterior weight is assigned to the shrinkage target $\boldsymbol{T}_6$ that is the closest to the true covariance matrix, along with targets $\boldsymbol{T}_4$ and $\boldsymbol{T}_5$. Overall, these simulations suggest that shrinkage weights are capable to exclude (i.e.~the posterior weight is equal to zero) shrinkage targets whose shape is quite distinct to the true underlying covariance structure. These results also show that having very similar shrinkage targets in the target set $\mathcal{D}$ does not harm the performance of the TAS estimator, but that it may complicate the interpretation of the (posterior) shrinkage weights. Thus we would recommend that Frobenius distance between targets are systematically evaluated and considered together with the shrinkage weights.

The non-parametric estimators $\boldsymbol{\hat{\Sigma}}_{\text{AT1}}, \boldsymbol{\hat{\Sigma}}_{\text{AT2}}$ and $\boldsymbol{\hat{\Sigma}}_{\text{AT3}}$ perform in general better than their parametric counterparts $\boldsymbol{\hat{\Sigma}}_{\text{ST1}}, \boldsymbol{\hat{\Sigma}}_{\text{ST2}}$ and $\boldsymbol{\hat{\Sigma}}_{\text{ST3}}$. This suggests that, when using the same shrinkage target, improved performance can be obtained by relaxing distributional assumptions. However, alike the behaviour observed for $\boldsymbol{\hat{\Sigma}}_{\text{T1}}, \ldots, \boldsymbol{\hat{\Sigma}}_{\text{T9}}$, the performance of $\boldsymbol{\hat{\Sigma}}_{\text{AT1}}, \ldots, \boldsymbol{\hat{\Sigma}}_{\text{AT3}}$ can also be affected by the choice of shrinkage target (see scenarios 1 and 3). Finally, on average, we observe that the proposed multi-target TAS estimator performs similarly to $\boldsymbol{\hat{\Sigma}}_{\text{cpc}}$ (scenarios 1 and 3) or better (scenario 2 and 4, where the true covariance matrix has a more dense structure).

\newpage
\null

\section{Data-based simulation}
\label{Sec:datasim}

Here, we employ gene expression data from The Cancer Genome Atlas (TCGA) and a data partitioning strategy to assess the performance of estimator~\eqref{Eq:TAS} and evaluate the benefits of incorporating external information into the target set $\mathcal{D}$. We retrieved, using the R package \emph{cgdsr} \citep{jacobson2015}, all TCGA level 3 normalised gene expression data that were measured using the Agilent 244K Custom gene Expression G4502A{\_}07 array. The data span 10 cancer types. However, for reasons that will become clearer below, we consider the following two low-dimensional extracts:
\begin{itemize}
\item \textbf{Data set 1:} p53 pathway in breast cancer ($p=68$ genes in $N=529$ samples)
\item \textbf{Data set 2:} apoptosis pathway in ovarian cancer ($p=86$ genes in $N=558$ samples)
\end{itemize}

As the true covariance structures between genes in these two data sets are unknown, we use a data partitioning strategy \citep{vanDeWiel2013, leday2017} to assess the performance of estimators $\boldsymbol{\hat{\Sigma}}_{\text{TAS}}$, $\boldsymbol{\hat{\Sigma}}_{\text{AT1}}$, $\boldsymbol{\hat{\Sigma}}_{\text{AT2}}$, $\boldsymbol{\hat{\Sigma}}_{\text{AT3}}$ and $\boldsymbol{\hat{\Sigma}}_{\text{cpc}}$ (in light of the results shown in Section \ref{Sec:modelsim}, $\boldsymbol{\hat{\Sigma}}_{\text{ST1}}, \ldots , \boldsymbol{\hat{\Sigma}}_{\text{ST9}}$ are excluded from this comparison). The strategy is illustrated in Supplementary Material~\ref{supp:data_sims}. For a given data set, the strategy consists on randomly splitting the full data matrix ($p\times N$) into a small sample size ($p\times n$) and a large sample size ($p \times (N-n)$) data matrix, for $n \in \{p/4, p/2, 3p/4\}$. Given this partition, all estimators are computed using the small sample size data matrix, whereas the sample covariance matrix obtained from the large sample size data matrix is used as a proxy for the true covariance when calculating the PRIAL (see \eqref{prial}). This procedure is repeated $1,000$ times for data sets 1 and 2, and for the three different values of $n$ investigated.

To illustrate the benefits of incorporating external information into the target set, we also consider the multi-target shrinkage estimator $\boldsymbol{\hat{\Sigma}}_{\text{TAS-info}}$ with target set $\mathcal{D}_{\text{info}}=\mathcal{D} \cup \boldsymbol{\hat{\Sigma}}_{\text{ext}}$, where $\boldsymbol{\hat{\Sigma}}_{\text{ext}}$ is an estimate of the covariance between genes that is obtained from independent data. For data sets 1 and 2, we obtain such estimate by pooling the TCGA gene expression data from the nine other cancer types for which expression levels were measured using the Agilent platform. To ensure that $\boldsymbol{\hat{\Sigma}}_{\text{ext}}$ is positive definite and well-conditioned, we use a regularised estimate (obtained using~\eqref{Eq:TAS}) instead of the pooled sample covariance.

\begin{figure}[ht]
\centering
\begin{minipage}[c]{0.5\linewidth}
    \centering
    \subfigure[][P53 pathway - breast cancer]{
  		\includegraphics[width=0.99\textwidth]{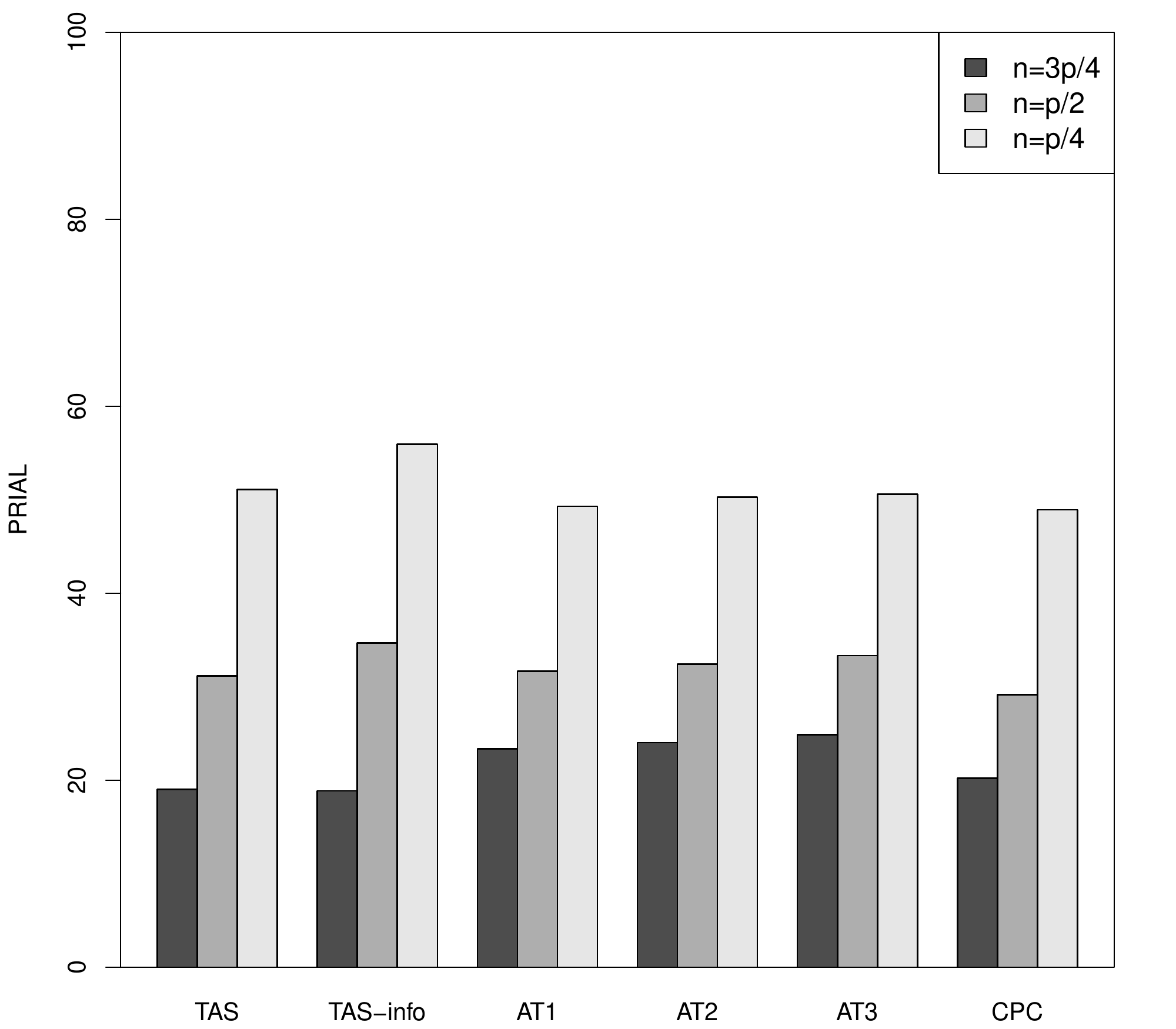}
  		\label{Fig:dataSim1:sub1}
  	}
  \end{minipage}\hfill
	\begin{minipage}[c]{0.5\linewidth}
    \centering
    \subfigure[][Apoptosis pathway - ovarian cancer]{
  		\includegraphics[width=0.99\textwidth]{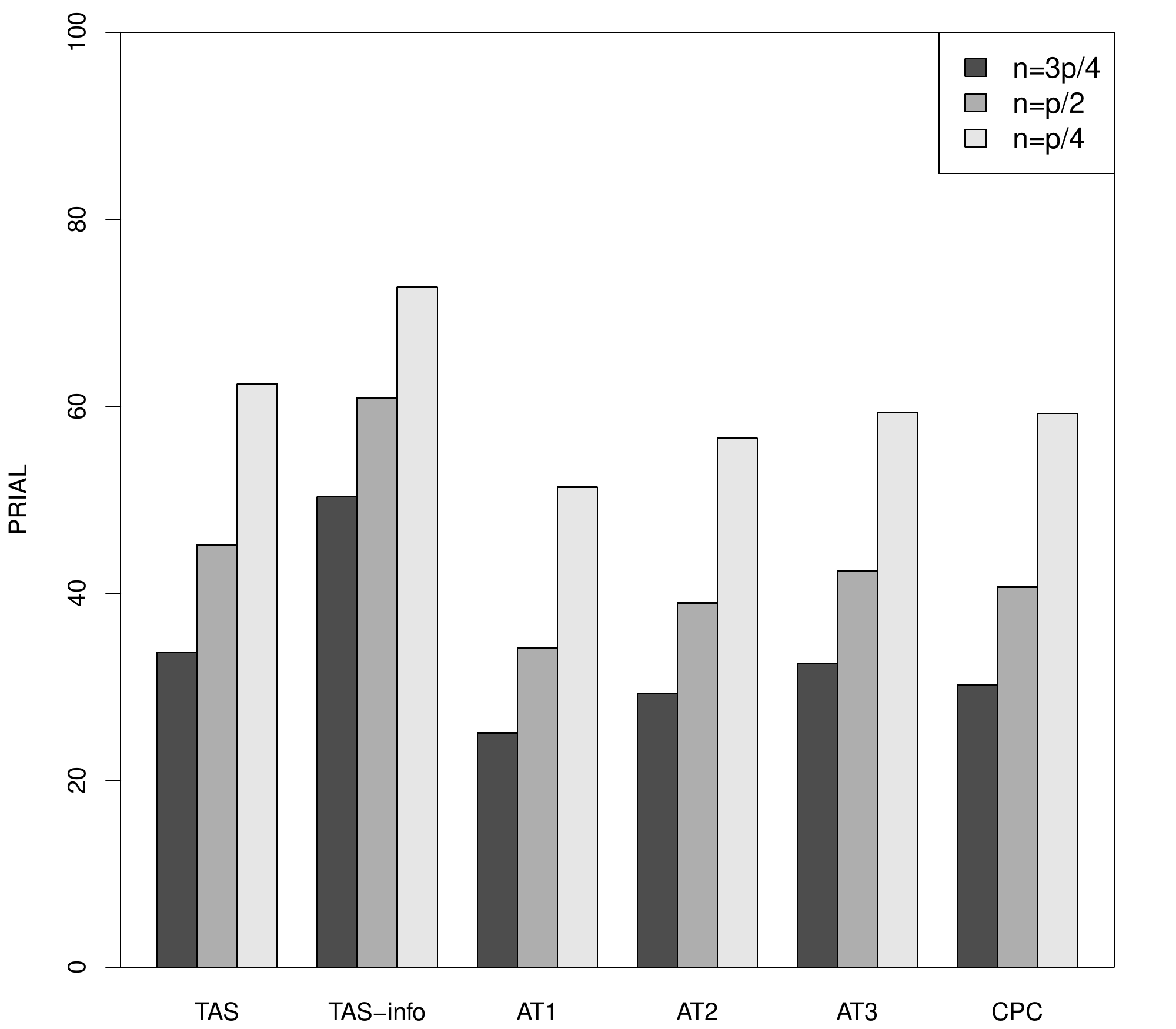}
  		\label{Fig:dataSim1:sub2}
  	}
  \end{minipage}\hfill
\caption{Results of the TCGA gene expression data-based simulation. Barplots display the PRIAL calculated for each estimator for (a) data set 1 (p53 pathway, breast cancer samples) and, (b) data set 2 (apoptosis pathway, ovarian cancer samples).}
\label{Fig:dataSim1}
\end{figure}

Figure~\ref{Fig:dataSim1} summarises the results for the experiment described above. Overall, for data set 1, all estimators achieve a similar PRIAL regardless of $n/p$ ratios (Figure~\ref{Fig:dataSim1:sub1}). For data set 2, however, we observe that $\boldsymbol{\hat{\Sigma}}_{\text{TAS-info}}$ (and to a lesser extent $\boldsymbol{\hat{\Sigma}}_{\text{TAS}}$) outperforms all other estimators. This highlights another key strength of the TAS estimator: its ability to incorporate external information within the target set can substantially improve performance.\\

\null

\begin{figure}[ht]
  	\begin{minipage}[c]{0.5\linewidth}
    \centering
    \subfigure[][$\boldsymbol{\hat{\Sigma}}_{\text{TAS}}$]{
  	\includegraphics[width=0.9\textwidth]{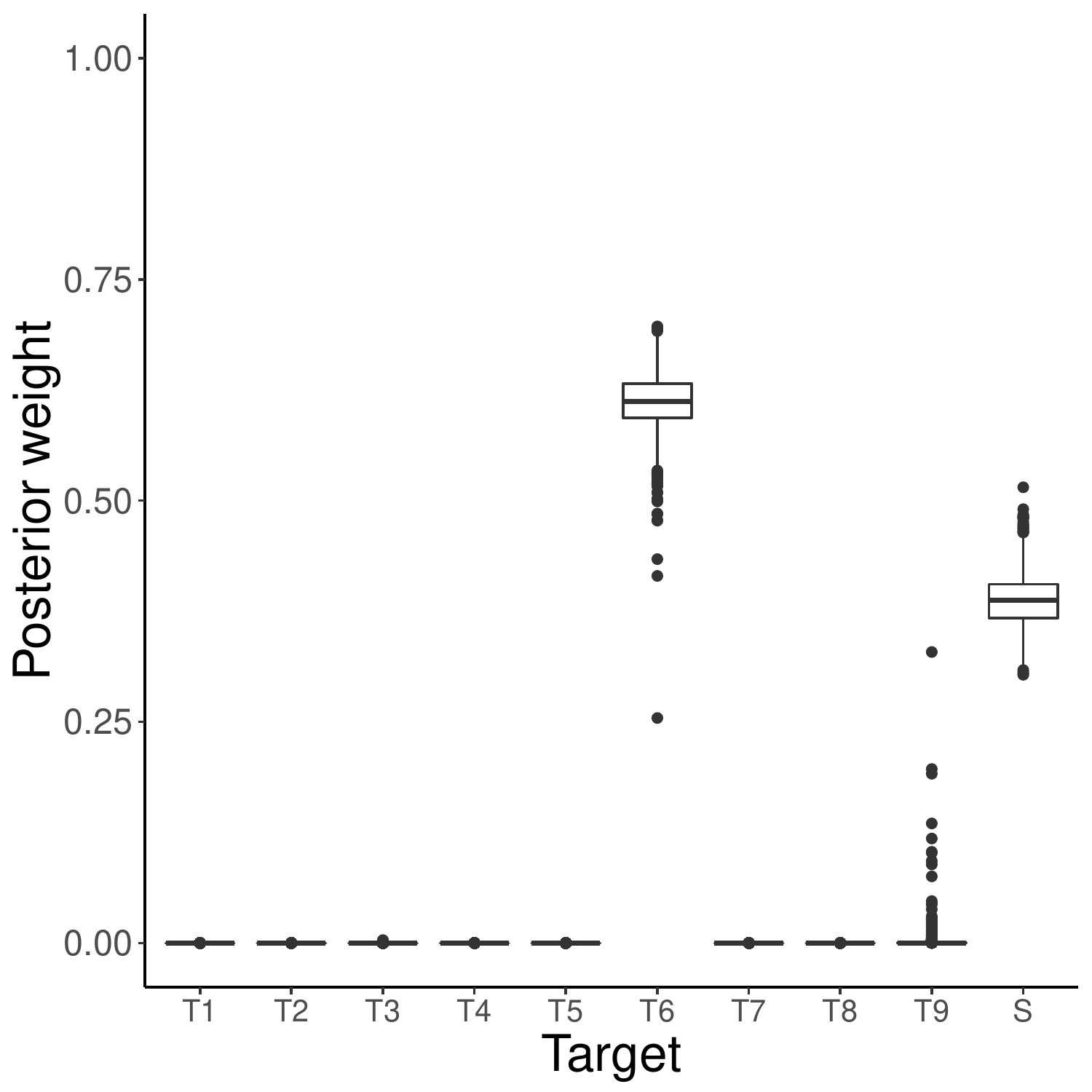}
  	\label{Fig:dataSim2:sub1}
  	}
  	\end{minipage}\hfill
  	\begin{minipage}[c]{0.5\linewidth}
    \centering
    \subfigure[][$\boldsymbol{\hat{\Sigma}}_{\text{TAS-info}}$]{
  	\includegraphics[width=0.9\textwidth]{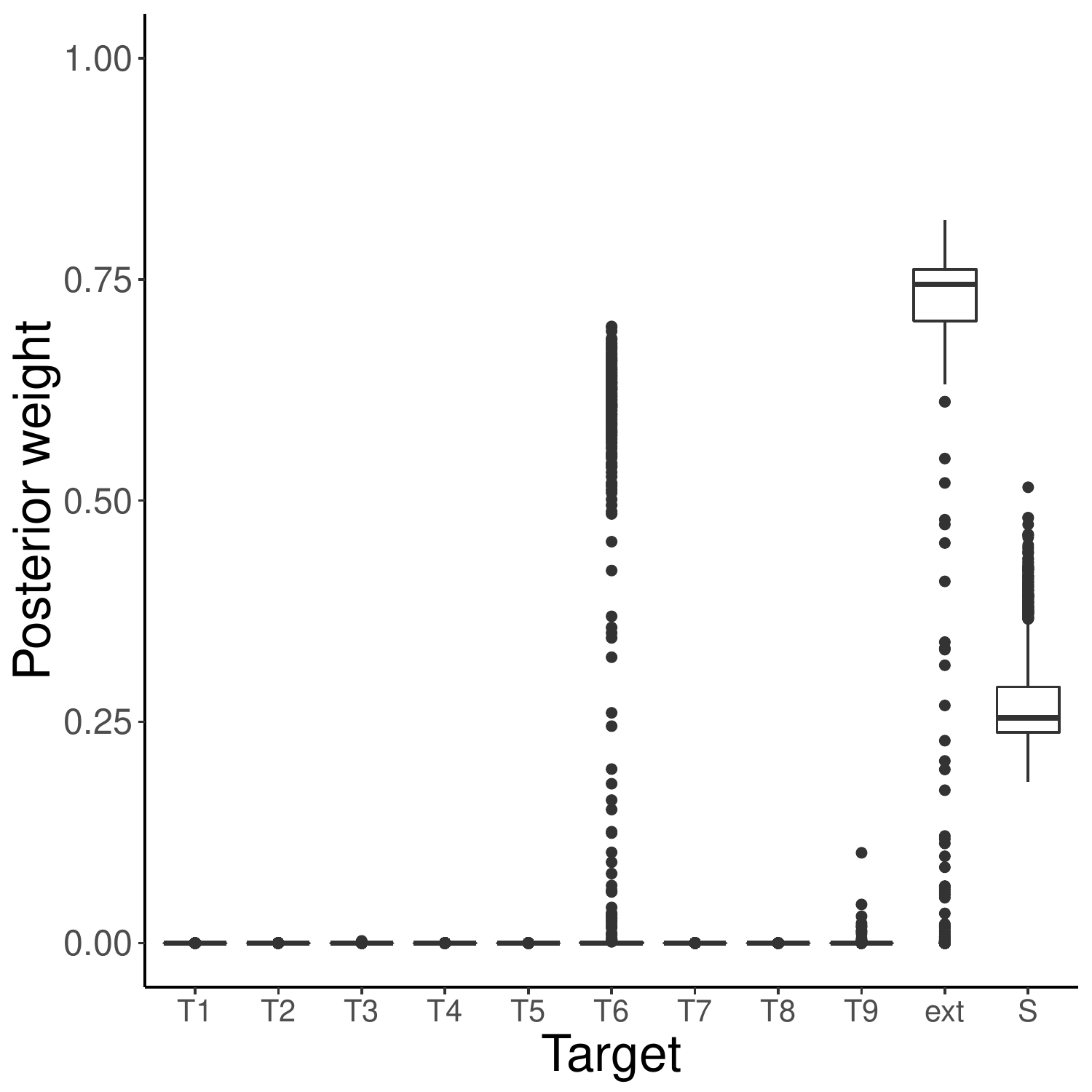}
  	\label{Fig:dataSim2:sub2}
  	}
  	\end{minipage}\hfill
\caption{Target-specific posterior weights (see equation \eqref{Eq:TAS_weights}) obtained for estimators $\boldsymbol{\hat{\Sigma}}_{\text{TAS}}$ and $\boldsymbol{\hat{\Sigma}}_{\text{TAS-info}}$ across the $1,000$ random data partitions of the ovarian cancer data set when $n=p/2$. The target ``ext'' in $\boldsymbol{\hat{\Sigma}}_{\text{TAS-info}}$ stands for the shrinkage target $\boldsymbol{\hat{\Sigma}}_{\text{ext}}$ estimated from external data. }
\label{Fig:dataSim2}
\end{figure}

Figure \ref{Fig:dataSim2} shows the ditribution of target-specific posterior weights (see equation \eqref{Eq:TAS_weights}) in estimators $\boldsymbol{\hat{\Sigma}}_{\text{TAS}}$ and $\boldsymbol{\hat{\Sigma}}_{\text{TAS-info}}$ across the $1,000$ random data partitions performed for data set 2 . We observe in Figure~\ref{Fig:dataSim2:sub1} that the shrinkage target $\boldsymbol{T}_6$ (constant correlation and unequal variances) is assigned the largest weight in estimator $\boldsymbol{\hat{\Sigma}}_{\text{TAS}}$, among all targets. This may be due to the fact that genes within the apoptosis pathway are expected to have high correlations between each other. Therefore, the shrinkage estimation of the covariance matrix may benefit from a shrinkage target whose off-diagonal elements are not equal to zero. On the other hand, we observe in Figure~\ref{Fig:dataSim2:sub2} that the shrinkage target $\boldsymbol{\hat{\Sigma}}_{\text{ext}}$, derived from external data, is assigned the largest weight in estimator $\boldsymbol{\hat{\Sigma}}_{\text{TAS-info}}$. This is in line with the results shown in Figure~\ref{Fig:dataSim1}, where the incorporation of external information substantially improved performance in data set 2. Supplementary Figures~\ref{supp:Fig:dataSim1} and~\ref{supp:Fig:dataSim2} show that $\boldsymbol{\hat{\Sigma}}_{\text{TAS-info}}$ also puts more weight on the shrinkage target $\boldsymbol{\hat{\Sigma}}_{\text{ext}}$ for data set 1, which results only in a small improvement in PRIAL (see figure~\ref{Fig:dataSim1:sub1}). Overall, complementary results in Supplementary Material~\ref{supp:data_sims} for others $n/p$ ratios show, as expected, that when $n$ increases, both $\boldsymbol{\hat{\Sigma}}_{\text{TAS}}$ and $\boldsymbol{\hat{\Sigma}}_{\text{TAS-info}}$ put more weight on the sample covariance matrix in both data sets.

Finally, while the multivariate normal assumption does not seem to be supported by these two gene expression data sets (see Supplementary Material~\ref{supp:norm}), it is found that the non-parametric estimators of \citet{touloumis2015} do not generally outperform the TAS estimator, which assumes multivariate normality. In fact, the opposite can occur for specific choices of target matrices (e.g.~when external information is included). This may suggest that accounting for multiple shrinkage target matrices may be more critical than flexible distributional assumptions.

\section{Application to protein expression data}
\label{Sec:app}

In this section, we apply our method to protein expression data from The Cancer Proteome Atlas (\href{tcpaportal.org/tcpa} {tcpaportal.org/tcpa}). In particular, we consider the PANCAN32 data set, focusing on level 4 normalised expression levels of 209 proteins that were measured on 7,694 samples across 32 cancer types. Supplementary Table~\ref{supp:Tab:cancer} provides for each cancer type its acronym and the number of samples.

We first use the TAS estimator to estimate the covariance between the 209 proteins separately for three histologically different cancers, namely cholangiocarcinoma (CHOL), liver hepatocellular carcinoma  (LIHC) and rectum adenocarcinoma (READ). For each of these three data sets, the target set of the TAS estimator includes the nine targets of Table~\ref{Tab:targets} (denoted $\boldsymbol{T}_1$, $\dots$, $\boldsymbol{T}_9$), 31 targets derived from each of the other cancer types (which we will refer to by their acronyms in Supplementary Table~\ref{supp:Tab:cancer}) and one target obtained by pooling the data from the 31 cancer types (referred to as PANCAN). To ensure that shrinkage targets derived from independent data sets are positive definite and well-conditioned, we use the TAS estimate using the nine targets of Table~\ref{Tab:targets} instead of the sample covariance matrix (however any other regularisation technique may be used instead). 

\newpage
\begin{figure}[ht]
  	\begin{minipage}[c]{0.95\linewidth}
    \centering
    \subfigure[][Cholangiocarcinoma (CHOL)]{
  	\includegraphics[width=0.8\textwidth]{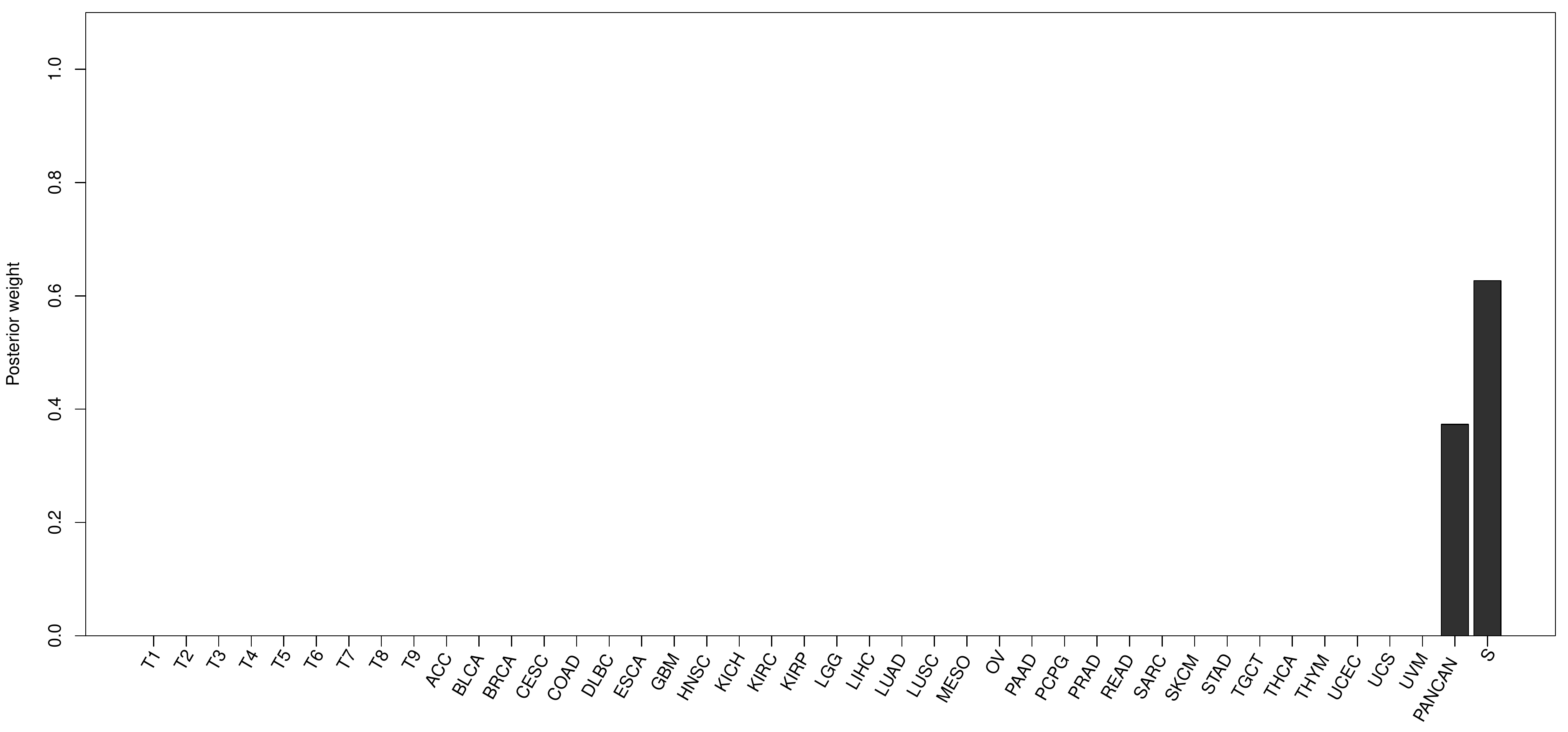}
  	\label{tcpatargetweights:sub1}
  	}
  	\end{minipage}\hfill
  	\begin{minipage}[c]{0.95\linewidth}
    \centering
    \subfigure[][Liver hepatocellular carcinoma (LIHC)]{
 	\includegraphics[width=0.8\textwidth]{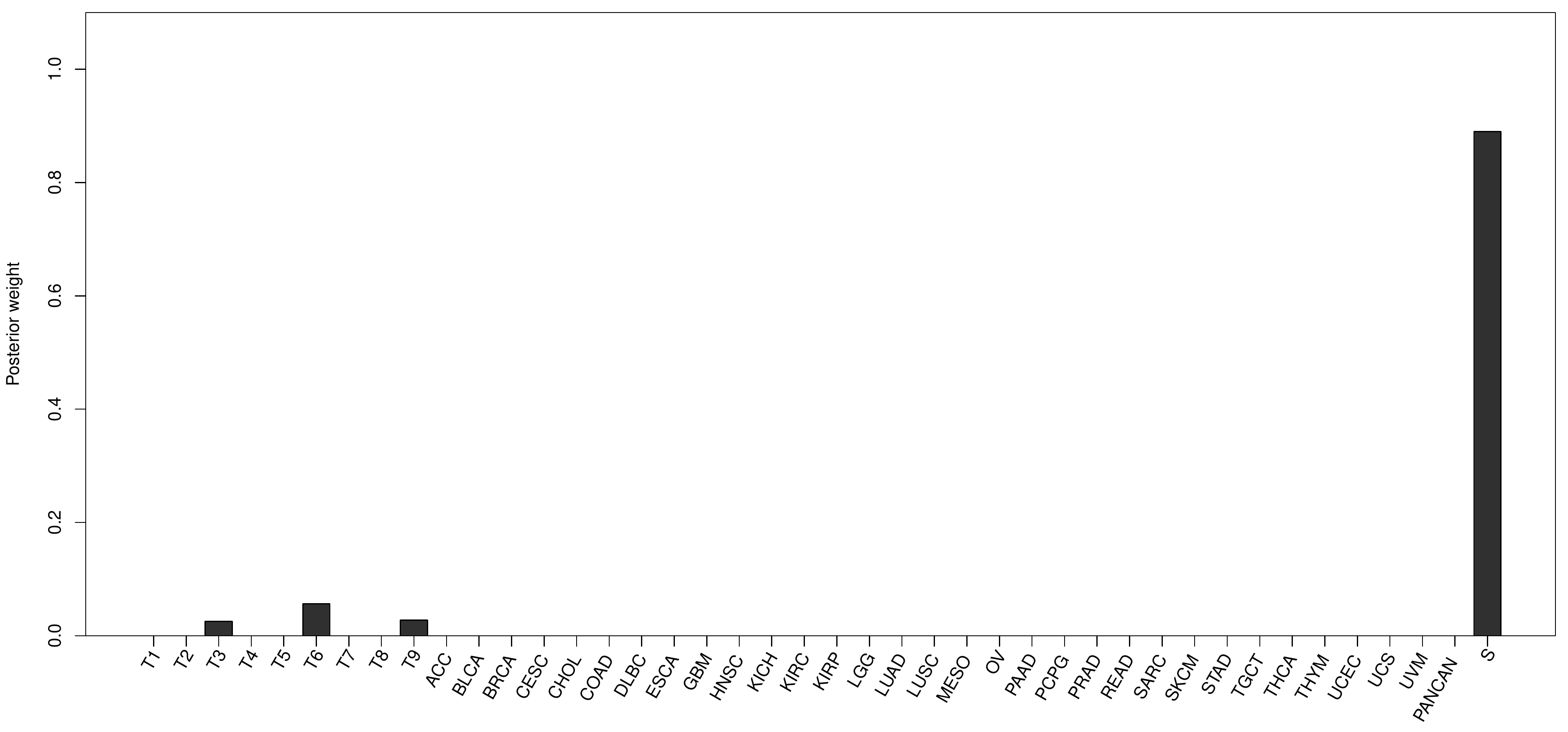}
  	\label{tcpatargetweights:sub2}
  	}
  	\end{minipage}\hfill
  	\begin{minipage}[c]{0.95\linewidth}
    \centering
    \subfigure[][Rectum adenocarcinoma (READ)]{
  	\includegraphics[width=0.8\textwidth]{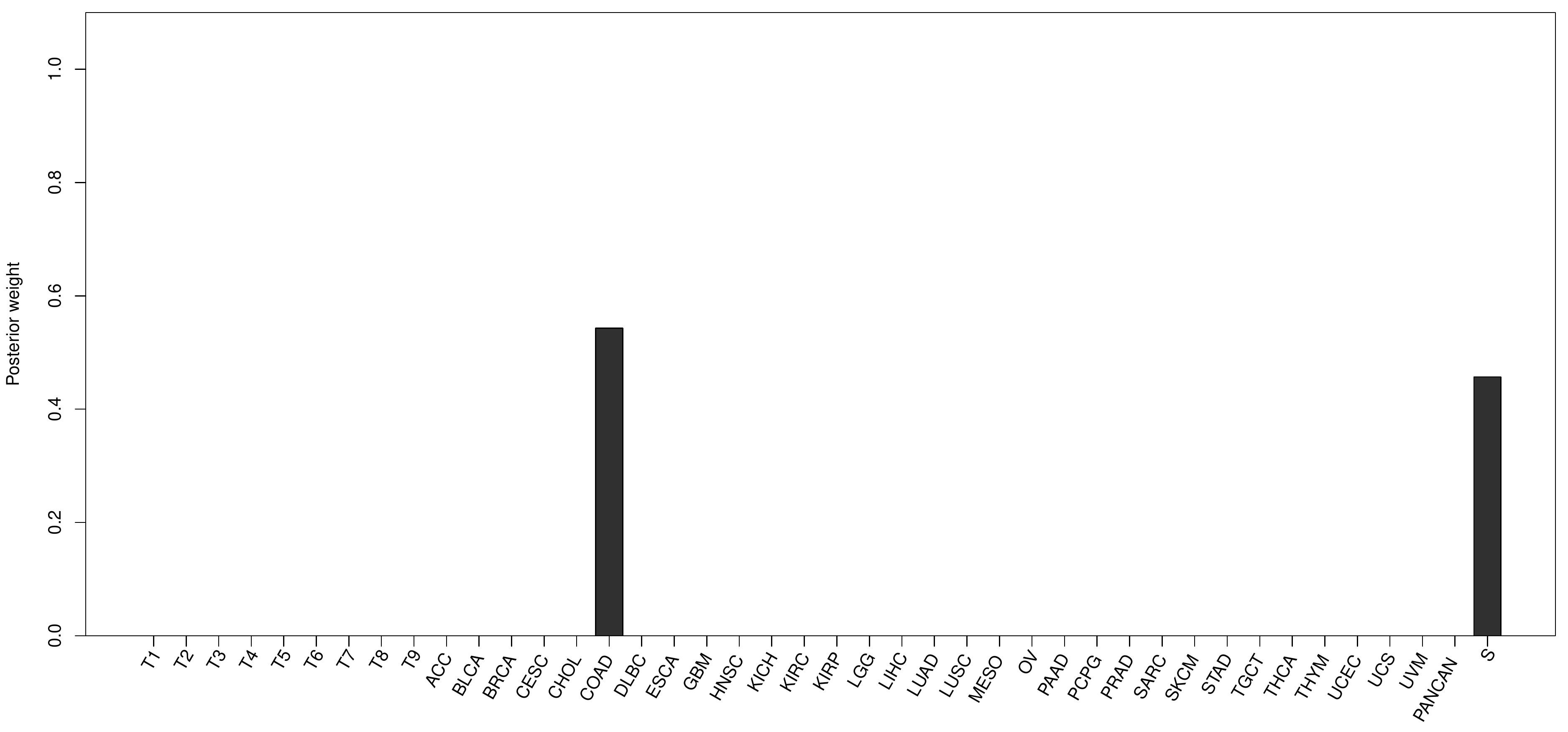}
  	\label{tcpatargetweights:sub3}
  	}
  	\end{minipage}\hfill
\caption{Target-specific posterior weights in estimators $\boldsymbol{\hat{\Sigma}}_{\text{TAS}}$ when analysing a) the cholangiocarcinoma, b) liver hepatocellular carcinoma and, c) rectum adenocarcinoma proteomic data sets.}
\label{Fig:tcpa3}
\end{figure}
\null 

Figure~\ref{Fig:tcpa3} reports target-specific posterior weights (see equation \eqref{Eq:TAS_weights}) of the TAS estimator obtained for each of these three data sets. This shows that the TAS estimator assigns large weights to different types of shrinkage targets across these datasets. For example, the PANCAN shrinkage target (that pools data from the 31 remaining cancers) is assigned a large weight in the Cholangiocarcinoma (CHOL) data set but not in the other two data sets. Virtually no weight is attributed to any of the targets derived from external data in the Liver hepatocellular carcinoma (LIHC) data set, whereas in the Rectum adenocarcinoma (READ) data set a large weight is assigned to the shrinkage target derived from the colon adenocarcinoma (COAD) cancer data. For the latter, it is biologically plausible that the dependence structure between proteins in rectum and colon adenocarcinoma samples are similar because both tumours are histologically related. Overall, these observations support the conclusions that covariance estimation may or may not benefit from the incorporation of external information and that, when it does, estimation can benefit both from generic (e.g. the PANCAN shrinkage target) and specific (e.g. the COAD shrinkage target) prior information.

\begin{figure}[ht]
	\centering
	\includegraphics[width=0.9\textwidth]{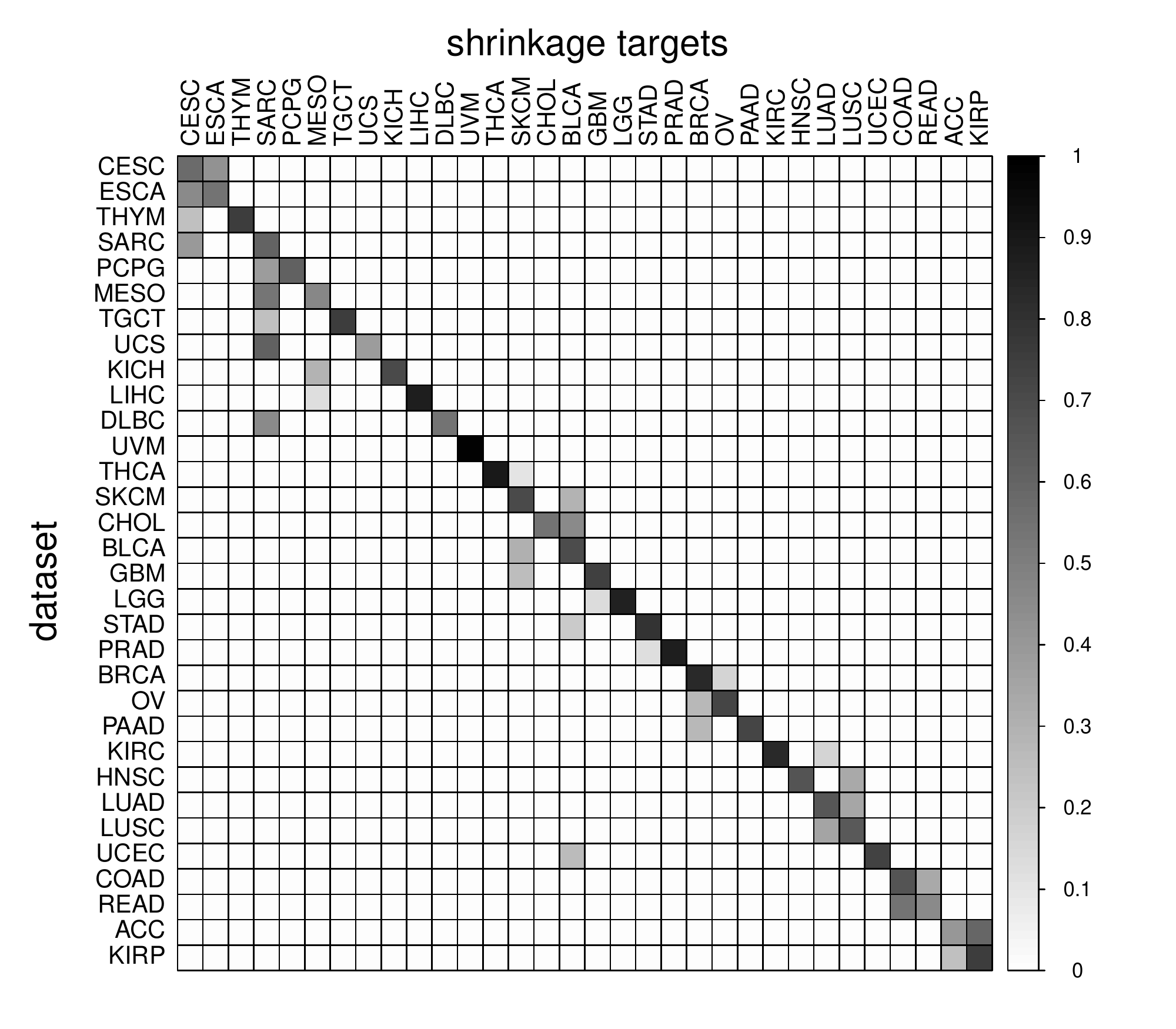}
\caption{Posterior shrinkage weights obtained by the TAS estimator.  Columns represent the shrinkage targets comprised in the target set of the TAS estimator. Elements on the diagonal represent shrinkage weights associated with the sample covariance of the data set. The per-row sum is equal to one.}
\label{Fig:tcpacancerweights}
\end{figure}

We now illustrate that the TAS estimator can provide insights regarding the relationship between the 32 cancer types shown in Table~\ref{Tab:targets}. For each of the 32 cancer data sets, we consider the TAS estimator with target set comprising of 31 shrinkage targets derived from the other 31 cancer types. We use the same strategy as above to make sure the shrinkage targets are positive definite and well-conditioned. Figure~\ref{Fig:tcpacancerweights} displays the posterior shrinkage weights obtained by the TAS estimator for each of the 32 cancer data sets. Our results suggest that high posterior weights might indicate similarity between cancers in terms of covariance structures. In particular, the target-specific posterior weights suggest a relatively high similarity between cancers with known putative biological similarity: (a) lung adenocarcinoma (LUAD) and lung squamous cell carcinoma (LUSC), both subtypes of non-small cell lung cancer \citep{ettinger2017}; (b) COAD and READ, both colorectal cancers \citep{cancer2012} and (c) breast invasive carcinoma (BRCA) and ovarian serous cystadenocarcinoma (OV), with known common susceptibility genes \citep{king2003}. These pairs of cancers have been also shown to be similar by pancancer analyses of previous releases of the TCPA dataset \citep[e.g.~][]{senbabaouglu2016}. Additionally, our results suggest a high similarity between esophageal carcinoma (ESCA) and cervical squamous cell carcinoma and endocervical adenocarcinoma (CESC). Both of these cancers have been found to be linked to human papillomavirus \citep{walboomers1999, ludmir2015}.

Figure~\ref{Fig:tcpacancerweights} also suggests that covariance estimation for cancers with small sample size can benefit from shrinkage towards cancer types with a large number of samples. Examples include adrenocortical carcinoma (ACC; $n = 46$) with kidney renal papillary cell carcinoma (KIRP; $n = 208$), uterine carcinosarcoma (UCS; $n = 48$) with sarcoma (SARC; $n = 221$), as well as cholangiocarcinoma (CHOL; $n = 30$) with bladder urothelial carcinoma (BLCA; $n = 344$). Despite this, no posterior weight was allocated to other cancer types in the case of Uveal melanoma (UVM; $n = 12$). This could be a consequence of its very small sample size, or it may suggest that protein interactions in UVM are unrelated to that of the other cancers. Future releases of TCPA, in which more samples are available, could enable us to confirm this.

\section{Discussion}
\label{Sec:disc}

We proposed a flexible, yet computationally simple, Bayesian covariance estimator that can accommodate an arbitrary number of shrinkage target matrices. The estimator is particularly useful in high-dimensional settings ($n << p$), where shrinkage is most important, and when external information is available. For these reasons, the present work is particularly relevant in the context of high-throughput genomic experiments due to (i) the central role that covariance estimation plays in multivariate data analyses, (ii) the high-dimensionality of the data and, (iii) the increasing availability of large open data repositories (e.g.~TCGA) which can provide relevant external information for specific studies. 

To the best of our knowledge, only \cite{bartz2014} and \cite{lancewicki2014} have proposed multi-target linear shrinkage estimators for covariance estimation. Unfortunately, numerical comparison with these methods was not performed due to the lack of available software. The methods of \cite{bartz2014} and \cite{lancewicki2014} are conceptually different to the TAS estimator. Firstly, both methods attempt to simultaneously estimate the weights that produce an optimal linear combination in the mean square sense. Secondly, these methods require analytical derivations that are tied to a particular target set. In contrast, our estimator weights individual targets using a fully probabilistic framework which enables analytical expressions for any arbitrary target set.

We envisage two main extensions for our work. Firstly, much like the performance of STS estimators depends on the choice of a target matrix, the performance of the proposed TAS estimator depends on the choice of a target set. In the absence of relevant prior information, we constructed a \emph{default} target set using shrinkage target matrices that are popular in the STS literature. However, further research is required to determine a more comprehensive and generic default target set. Such a target set would ideally cover a wide range of structures, to ensure that there is enough flexibility in the shrinkage. The latter must also take into account that, if the chosen target set contains shrinkage targets with overlapping shape, shrinkage weights need to be interpreted together with the pairwise Frobenius distance between targets, and their distance to the empirical covariance. Finally, it would be useful to extend the present work to non-Gaussian settings to allow for example the analysis of count data obtained from RNA sequencing experiments. These experiments provide greater specificity with higher throughput than array-based technologies. Potential avenues include hierarchical latent representations \citep{aitchison1989, gallopin2013} and data transformation strategies  \citep{cloonan2008, jia2017,zhang2017}. Nonetheless, as normal approximations can have good performance in RNA sequencing data \citep[e.g.~][]{law2014voom}, we foresee that the current TAS estimator might have practical utility in such context.

%% ACKNOWLEDGEMENTS
\section*{Acknowledgements}

The authors would like to thank Lorenz Wernisch and Paul Kirk for valuable discussions. H.G. was supported by a PhD scholarship funded by The Wellcome Trust grant number 105362/Z/14/Z, G.G.R.L. was supported by the Medical Research Council grant number MR/M004421 and C.A.V. was supported by The Alan Turing Institute (under the EPSRC grant number EP/N510129/1) and by a Chancellor's Fellowship provided by The University of Edinburgh. S.R. was funded by MRC grant MC$\_$UP$\_$0801/1. No conflict of interest declared.

%% BIBLIOGRAPHY
%\newpage
\bibliographystyle{apalike}
\bibliography{TAS}

%% Supplementary Material
\newpage
\setcounter{table}{0}
\renewcommand{\thetable}{S.\arabic{table}}
\setcounter{figure}{0}
\renewcommand{\thefigure}{S.\arabic{figure}}
\setcounter{section}{0}
\renewcommand{\thesection}{\arabic{section}}

\begin{center}
{\huge \textsc{Supplementary Material}}
\end{center}

\section{Maximum likelihood estimation of $\boldsymbol{\Sigma}$}
\label{supp:MLE}

\begin{figure}[ht]
	\centering
	\includegraphics[width=0.49\textwidth]{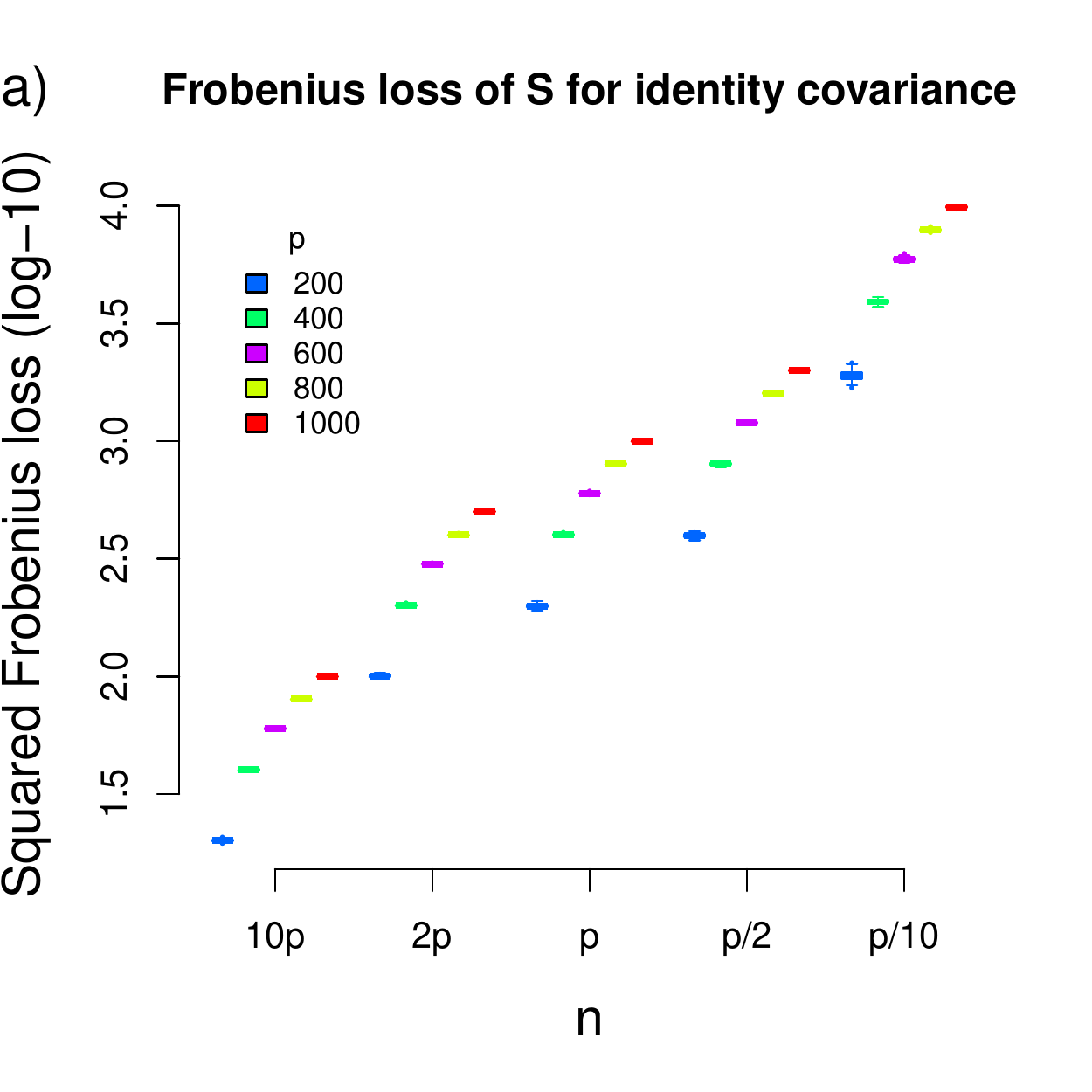}
	\includegraphics[width=0.49\textwidth]{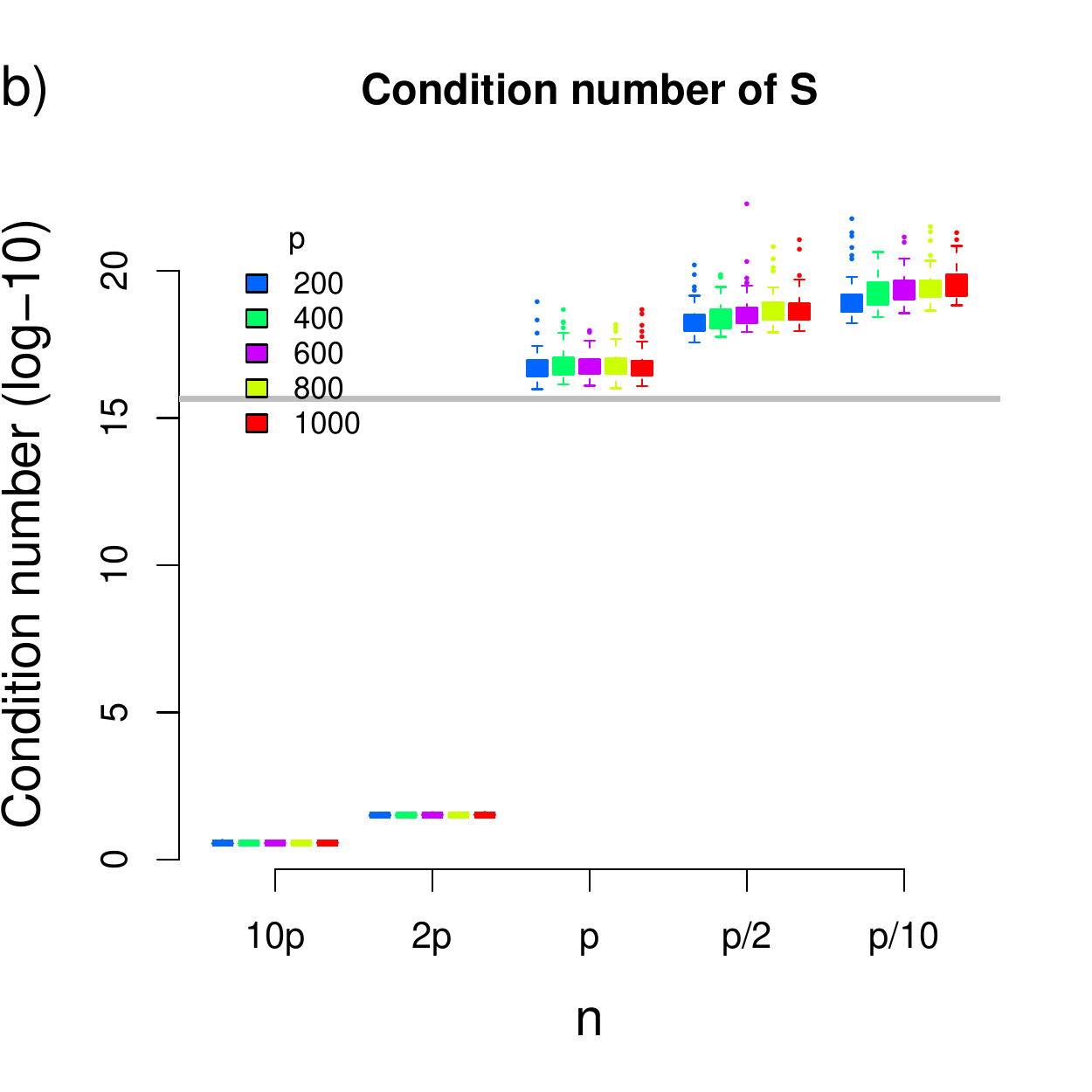}
\caption{Maximum likelihood estimation (MLE) for $\boldsymbol{\Sigma}$. Sub-figure (a) shows the squared Frobenius distance between the MLE $\boldsymbol{S} = \boldsymbol{XX}^\top/n$ and $\boldsymbol{\Sigma}$ whereas (b) shows the condition number of $\boldsymbol{S}$. The grey line represents the condition number for which matrices are declared numerically singular by the \textit{solve()} function in R.}
\label{supp:Fig:s}
\end{figure}

Here, we assess the behaviour of the MLE $\boldsymbol{S}$ of $\boldsymbol{\Sigma}$. For different combinations of the number of variables ($p \in \{200, 400, 600, 800, 1000\}$) and number of observations ($n \in \{10p, 2p, p, p/2, p/10\}$), we generate 100 data sets from a multivariate normal distribution $\mathcal{N}_p(\boldsymbol{0}, \boldsymbol{\Sigma})$, where $\boldsymbol{\Sigma}=\boldsymbol{I}_{p\times p}$. For each generated data set $\boldsymbol{X}$, we compute: (i) the MLE $\boldsymbol{S}=\boldsymbol{X}\boldsymbol{X}^{\intercal}/n$, (ii) the associated (squared) Frobenius distance between $\boldsymbol{S}$ and $\boldsymbol{\Sigma}$, $\|\boldsymbol{\Sigma}-\boldsymbol{S}\|_{F}^2=\sum_i^p\sum_j^p(\Sigma_{ij}-S_{ij})^2$, and (iii) the condition number of $\boldsymbol{S}$. The latter is defined as $\lambda_{\text{max}}/\lambda_{\text{min}}$, where $\lambda_{\text{max}}$ and $\lambda_{\text{min}}$ are the largest and smallest eigenvalues of $\boldsymbol{S}$, respectively. These results are summarised in Figure~\ref{supp:Fig:s}. As expected, we observe a higher estimation error (reflected in larger Frobenius distances) when the ratio $p/n$ increases. Moreover, we observe that $\boldsymbol{S}$ is singular whenever $n \leq p$.

\section{Parametrisation for the inverse Wishart distribution}\label{supp:reparametrisation}

A $p\times p$ random matrix $\boldsymbol{\Sigma}$ with probability density
\begin{equation*}
	2^{-\frac{\nu p}{2}} \Gamma_P^{-1}\left( \frac{\nu}{2} \right) \mid \boldsymbol{\Psi} \mid^{\frac{\nu}{2}} \mid \boldsymbol{\Sigma} \mid^{-\frac{\nu + p + 1}{2}} \exp\left\lbrace -\frac{1}{2} \text{tr}\left( \boldsymbol{\Psi} \boldsymbol{\Sigma}^{-1} \right)\right\rbrace
\end{equation*}
is said to follow an inverse Wishart distribution with scale matrix $\boldsymbol{\Psi}$ and degree of freedom $\nu$. Instead, we adopt the mean-centred parametrisation of the Inverse Wishart distribution used by \cite{hannart2014}, which facilitates interpretation in the context of STS estimators. The new parameterisation is obtained through the following bijective transformation:
\begin{equation*}
	(\alpha, \boldsymbol{\Delta})=\bigg(\frac{\nu-p-1}{n+\nu-p-1}, \frac{\boldsymbol{\Psi}}{\nu-p-1}\bigg)\quad \Leftrightarrow \quad (\nu, \boldsymbol{\Psi}) = \bigg(\frac{\alpha n}{1-\alpha}+p+1, \frac{\alpha n}{1-\alpha}\boldsymbol{\Delta}\bigg),
\end{equation*}
which yields $\mathbb{E}(\boldsymbol{\Sigma} | \alpha, \boldsymbol{\Delta}) = \boldsymbol{\Delta}$ and $\mathbb{E}(\boldsymbol{\Sigma} | \boldsymbol{X}, \alpha, \boldsymbol{\Delta})=\alpha \boldsymbol{\Delta}+(1-\alpha)\boldsymbol{S}$, given the data matrix $\boldsymbol{X}$. The new hyperparameters $\alpha$ and $\boldsymbol{\Delta}$ respectively correspond to the shrinkage intensity and target of an STS estimator.

\section{Uncertainty around the empirical Bayes estimate of $\alpha$} \label{supp:eb}

\begin{figure}[ht]
	\centering
	\includegraphics[width=0.55\textwidth]{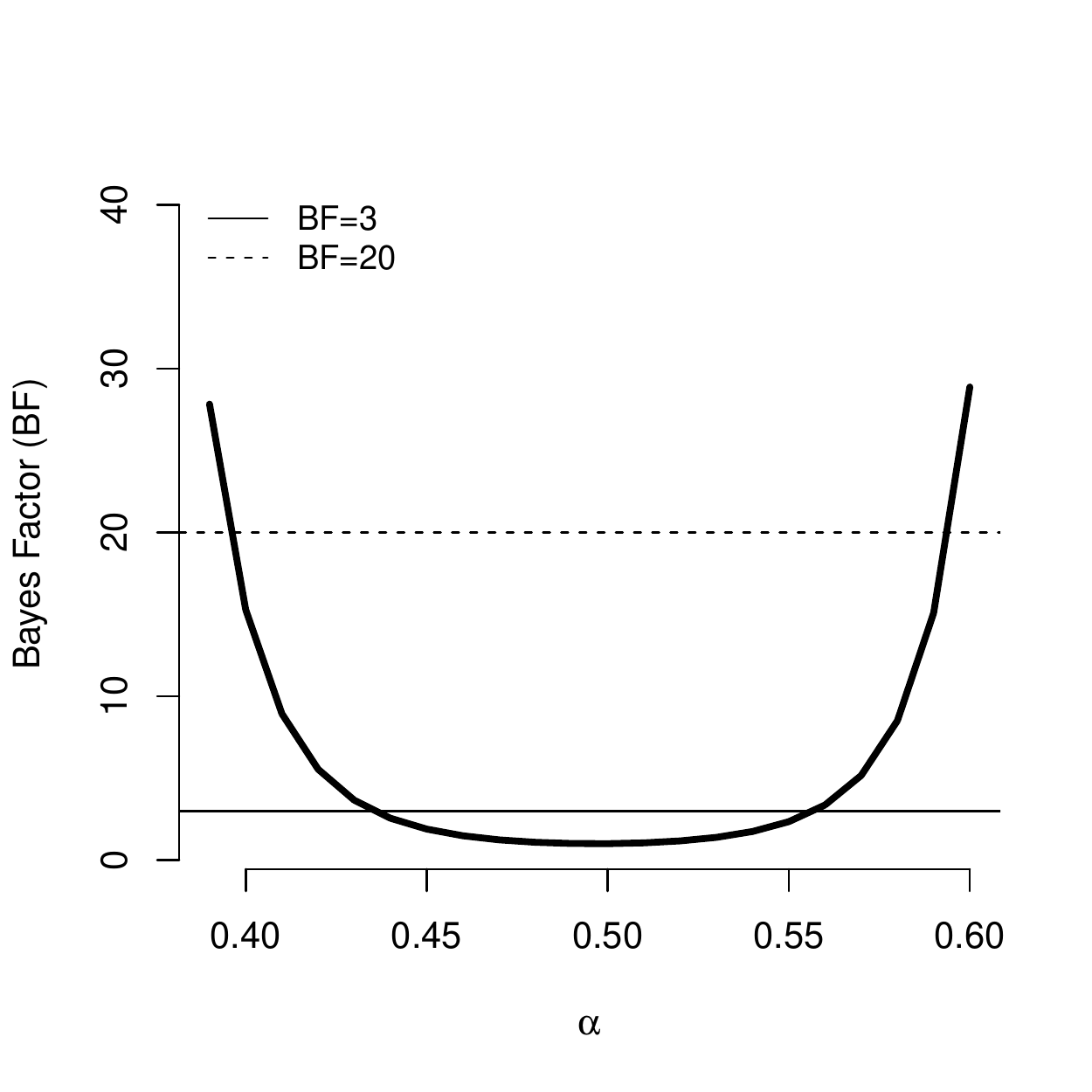}
\caption{Bayes factor quantifying the strength of support for $\alpha^*$ (the empirical Bayes estimate for $\alpha$) compared to alternative values of $\alpha \in (0,1)$. Horizontal lines correspond to the heuristic rules of Kass and Raftery (1995) for which $\text{BF} < 3$ is ``not worth more than a bare mention'' and $\text{BF} < 20$ provides ``less than strong evidence''. }
\label{supp:Fig:uncertainty}
\end{figure}

Here, we illustrate the statistical uncertainty surrounding the empirical Bayes estimate $\alpha^\ast$, defined as the value of $\alpha$ that maximises the marginal likelihood defined in \ref{supp:ml} for fixed $\boldsymbol{\Delta}$. We generate a data matrix $\boldsymbol{X}=(\boldsymbol{x}_1, \ldots, \boldsymbol{x}_n)$ using $\boldsymbol{x}_i\sim \mathcal{N}_p(\boldsymbol{0}, 2*\boldsymbol{I}_{p\times p})$, $p=200$ and $n=20$. For the generated data set, we observe that a range of values of $\alpha$ lead to similar marginal likelihood values. Figure~\ref{supp:Fig:uncertainty} displays the Bayes factor
\begin{equation*}
\text{BF}(\alpha) = \frac{\text{p}(\boldsymbol{X}|\alpha^*, \boldsymbol{I}_{p\times p})}{\text{p}(\boldsymbol{X}|\alpha, \boldsymbol{I}_{p\times p})},
\end{equation*}
which quantifies evidence in favour of $\alpha^*$ when compared to alternative values of $\alpha \in (0,1)$.

\section{Marginal likelihood of the Gaussian conjugate model}\label{supp:ml}

The marginal likelihood of the Gaussian conjugate model, which is required to calculate the posterior probabilities introduced in ~\eqref{Eq:margexp}, can be found in closed-form as
\begin{equation*}
\begin{split}
	\text{p}(\boldsymbol{X}|\alpha, \boldsymbol{\Delta})&=\int\text{p}(\boldsymbol{X}|\boldsymbol{\Sigma})\text{p}(\boldsymbol{\Sigma}|\alpha, \boldsymbol{\Delta})\text{d}\boldsymbol{\Sigma}
	=\frac{\Gamma_p\{\frac{1}{2}( \frac{n}{1-\alpha}+p+1)\}|\frac{\alpha}{1-\alpha}\boldsymbol{\Delta}|^{ \frac{\alpha n}{1-\alpha} + p + 1} }{(n\pi)^{\frac{np}{2}}\Gamma_p\{\frac{1}{2}( \frac{\alpha n}{1-\alpha}+p+1)\}|\boldsymbol{S}+\frac{\alpha}{1-\alpha}\boldsymbol{\Delta}|^{ \frac{n}{1-\alpha} + p + 1}}.
	\end{split}
\end{equation*}

\medskip

\section{Cardinality for the support of $\alpha$}
\label{supp:card}

\begin{figure}[ht]
	\centering
	\includegraphics[width=0.55\textwidth]{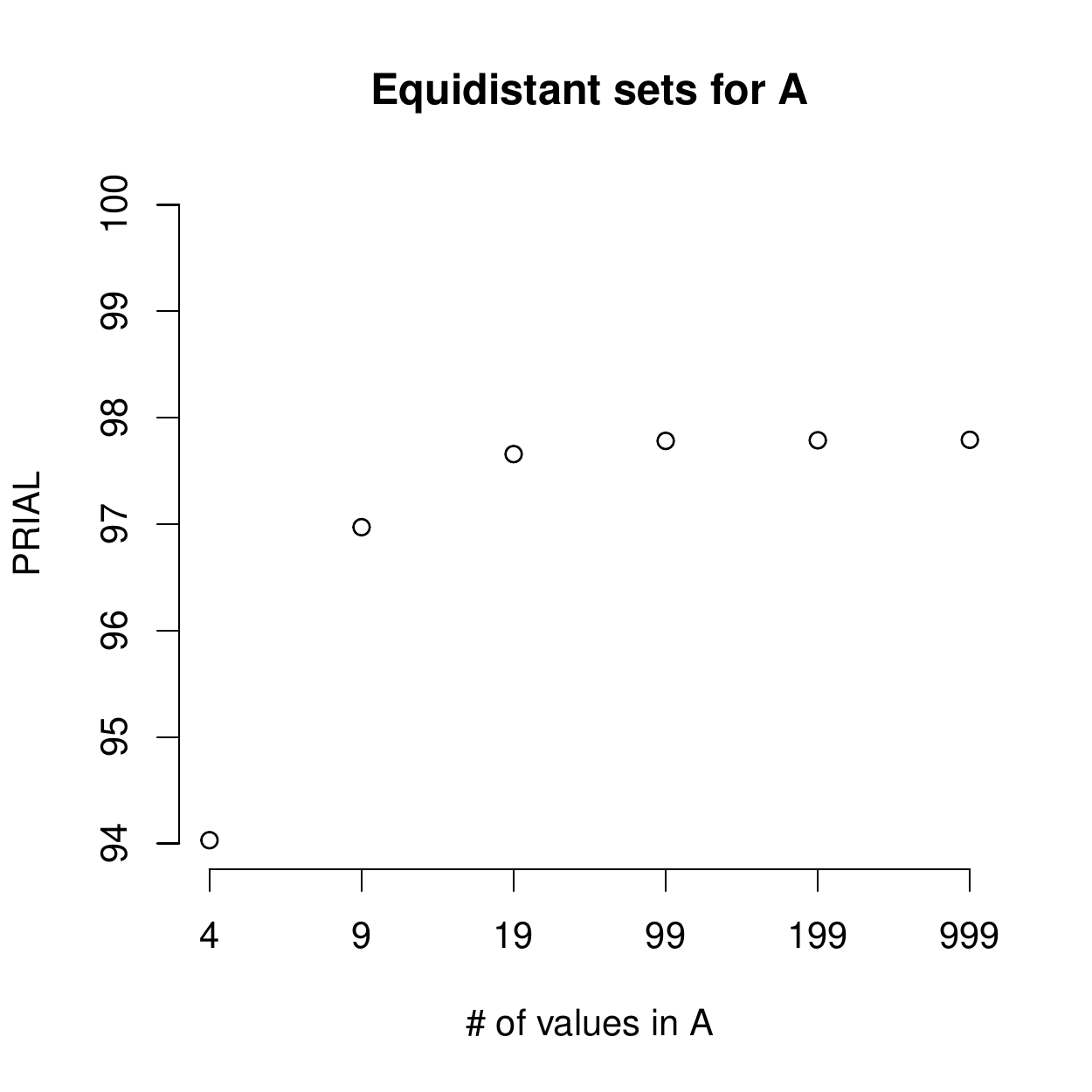}
\caption{PRIAL associated to the TAS estimator ($\mathcal{D}=\{\boldsymbol{T}_1, \ldots, \boldsymbol{T}_9\}$, see Table \ref{Tab:targets}) across different  cardinalities of $\mathcal{A}$. Results are associated to the simulation setup described in Section \ref{supp:card}. }
\label{supp:Fig:AvsPRIAL}
\end{figure}

Our approach assigns a discrete prior distribution with support $\mathcal{A}$ to the shrinkage intensity parameter $\alpha$. As $0 < \alpha < 1$, a natural support for this prior is an equidistant grid of values within the $(0,1)$ interval. Here, we study the stability of the multi-target estimate for different choices of support. We generate $100$ data sets of size $n=25$ from $\mathcal{N}_{100}(\boldsymbol{0}, \boldsymbol{\Sigma})$ with $\boldsymbol{\Sigma}=4*\boldsymbol{I}_{100\times 100}$. Subsequently, we compute \eqref{Eq:TAS} using $\mathcal{D}=\{\boldsymbol{T}_1, \ldots, \boldsymbol{T}_9\}$ (see Table 1 in main text) and $d \in \{0.2, 0.1, 0.05, 0.01, 0.005, 0.001\}$, where $d$ denotes the distance between consecutive elements of $\mathcal{A}$. Figure \ref{supp:Fig:AvsPRIAL} shows the PRIAL of estimator \eqref{Eq:TAS} as a function of the cardinality $\text{card}_d(\mathcal{A})=d^{-1}-1$ of $\mathcal{A}$ and shows that for sufficiently small values of $d$, i.e. large values of $\text{card}_d(\mathcal{A})$, minimal improvement is obtained beyond $d = 0.01$ ($\text{card}(\mathcal{A})=99$).\\

\newpage

\section{Model-based simulation: additional results}
\label{supp:model_sims}

\begin{figure}[ht]
	\begin{minipage}[c]{0.5\linewidth}
    \centering
  		\subfigure[][Scenario 1: PRIAL]{
  			\includegraphics[width=0.9\textwidth]{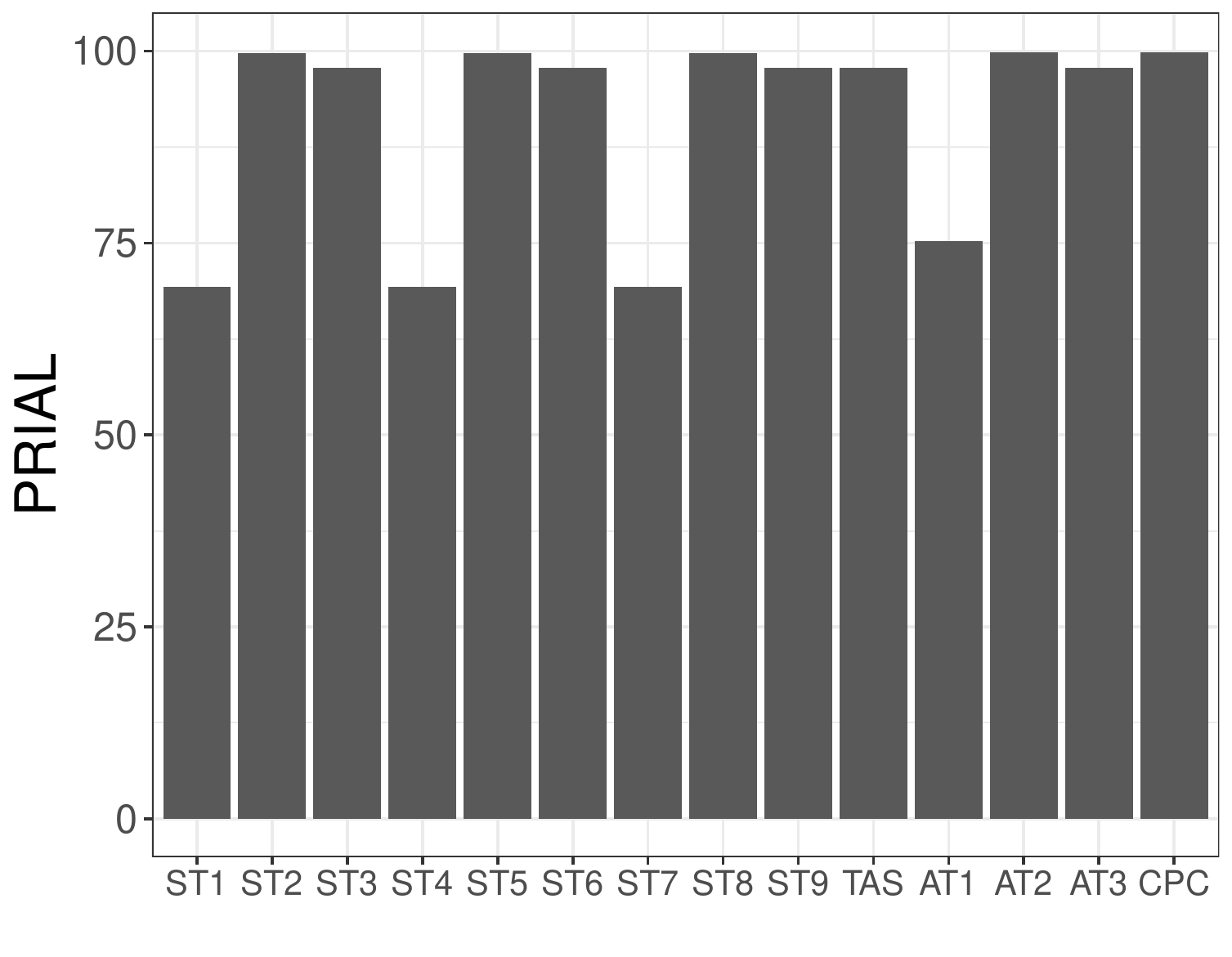}
  			\label{supp:Fig:modelSim1:sub1}
  		}
  \end{minipage}\hfill
  	\begin{minipage}[c]{0.5\linewidth}
    \centering 
    		\subfigure[][Scenario 1: target-specific posterior weights]{
  			\includegraphics[width=0.9\textwidth]{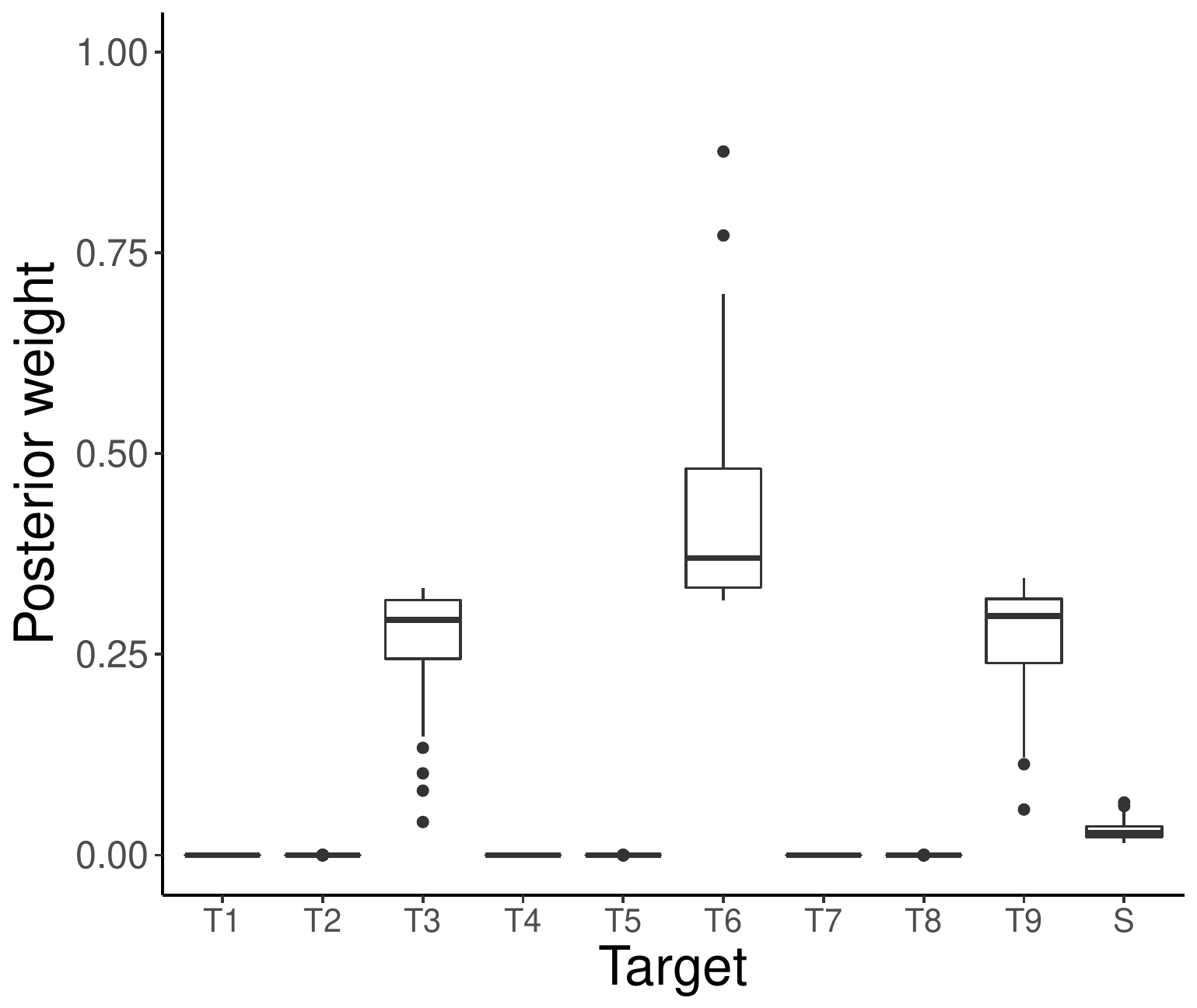}
  			\label{supp:Fig:modelSim1:sub1w}
  		}
  \end{minipage}\hfill
	\begin{minipage}[c]{0.5\linewidth}
    \centering
  		\subfigure[][Scenario 2: PRIAL]{
  			\includegraphics[width=0.9\textwidth]{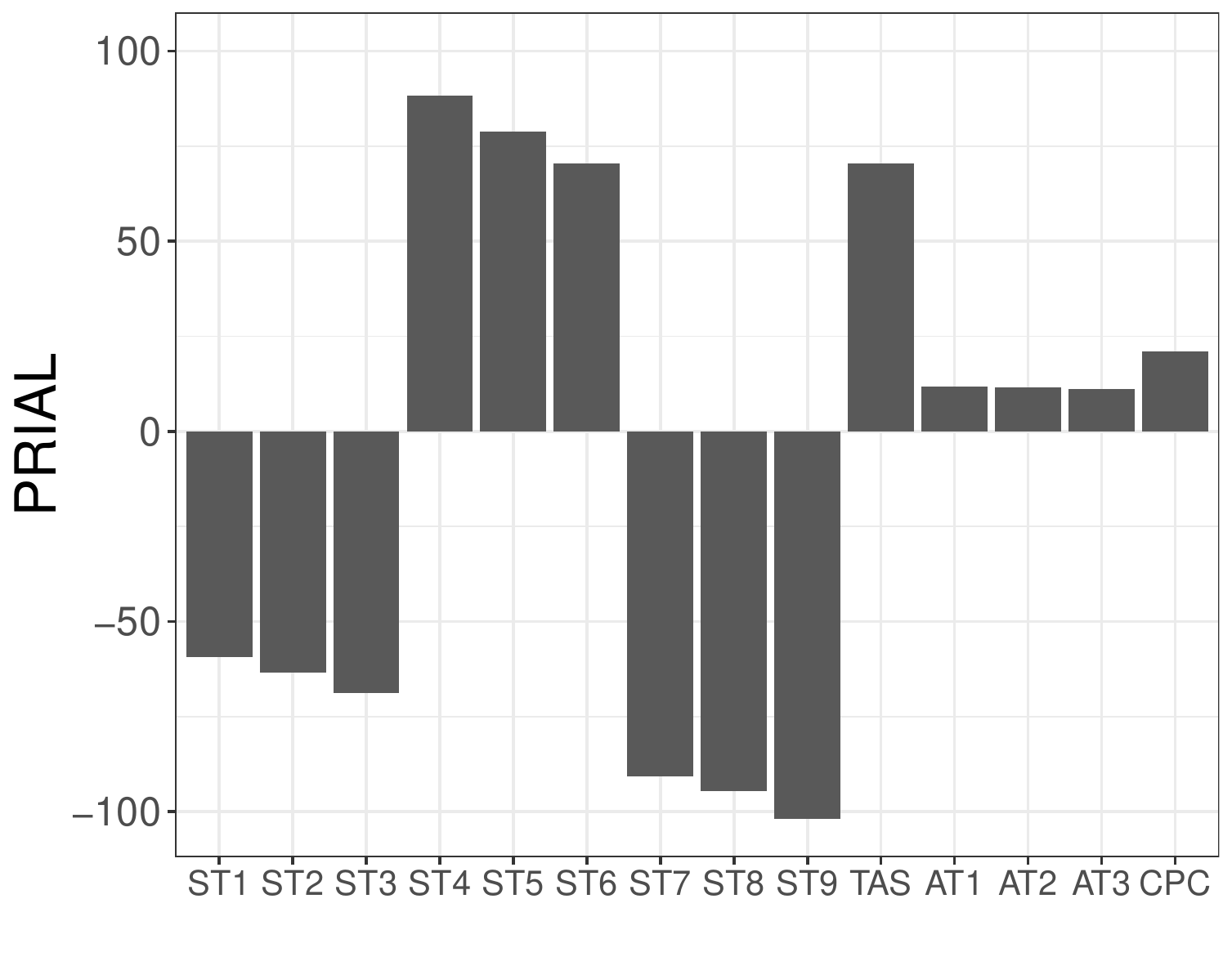}
  			\label{supp:Fig:modelSim1:sub2}
  		}
  \end{minipage}\hfill
    	\begin{minipage}[c]{0.5\linewidth}
    \centering 
    		\subfigure[][Scenario 2: target-specific posterior weights]{
  			\includegraphics[width=0.9\textwidth]{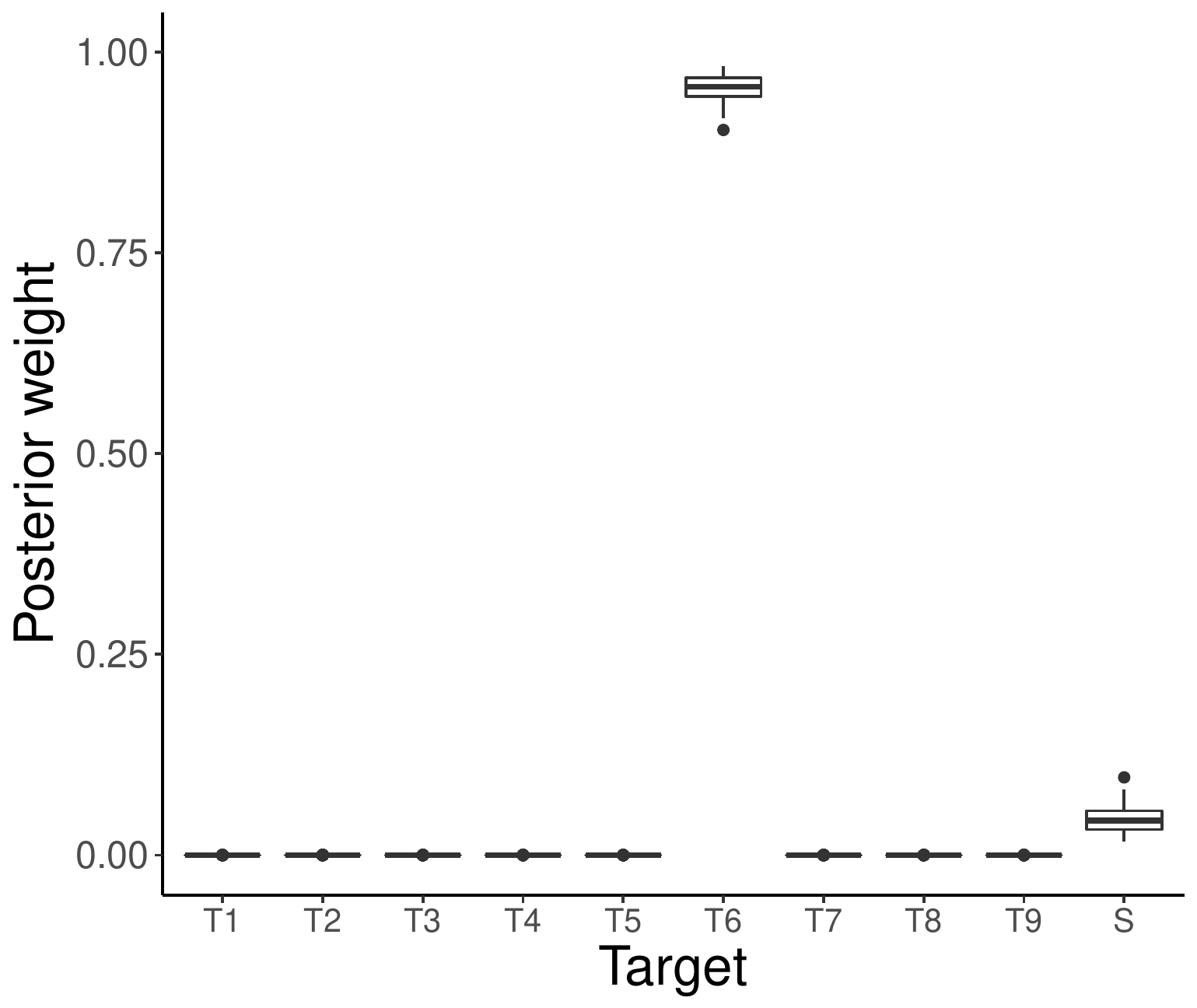}
  			\label{supp:Fig:modelSim1:sub2w}
  		}
  \end{minipage}\hfill
		\caption{Simulation results for scenarios 1 and 2 when $n=50$. Barplots display the PRIAL for each estimator and boxplots display target-specific posterior weights (see equation \eqref{Eq:TAS_weights}) of the TAS estimator. ST1,~\ldots,~ ST9 refer to the nine STS estimators, TAS to estimator \eqref{Eq:TAS}, AT1, \ldots, AT3 to the three estimators of \citet{touloumis2015} and CPC to the estimator of \citet{schafer2005}.}
		\label{supp:Fig:modelSim1}
\end{figure}

\begin{figure}[ht]
  \begin{minipage}[c]{0.5\linewidth}
    \centering
  		\subfigure[][Scenario 3: PRIAL]{
  			\includegraphics[width=0.9\textwidth]{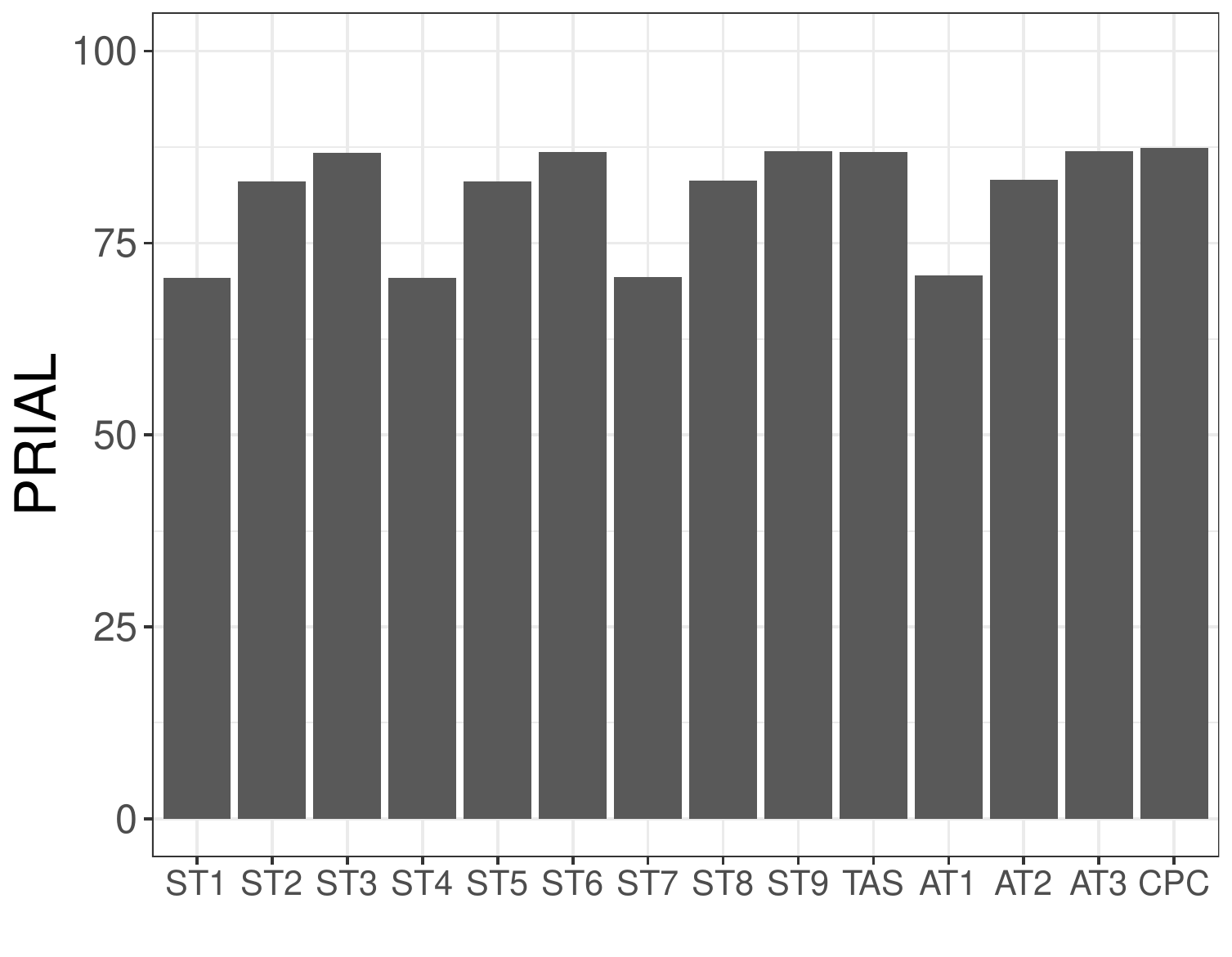}
  			\label{supp:Fig:modelSim2:sub3}
  		}
  \end{minipage}\hfill
      	\begin{minipage}[c]{0.5\linewidth}
    \centering 
    		\subfigure[][Scenario 3: target-specific posterior weights]{
  			\includegraphics[width=0.9\textwidth]{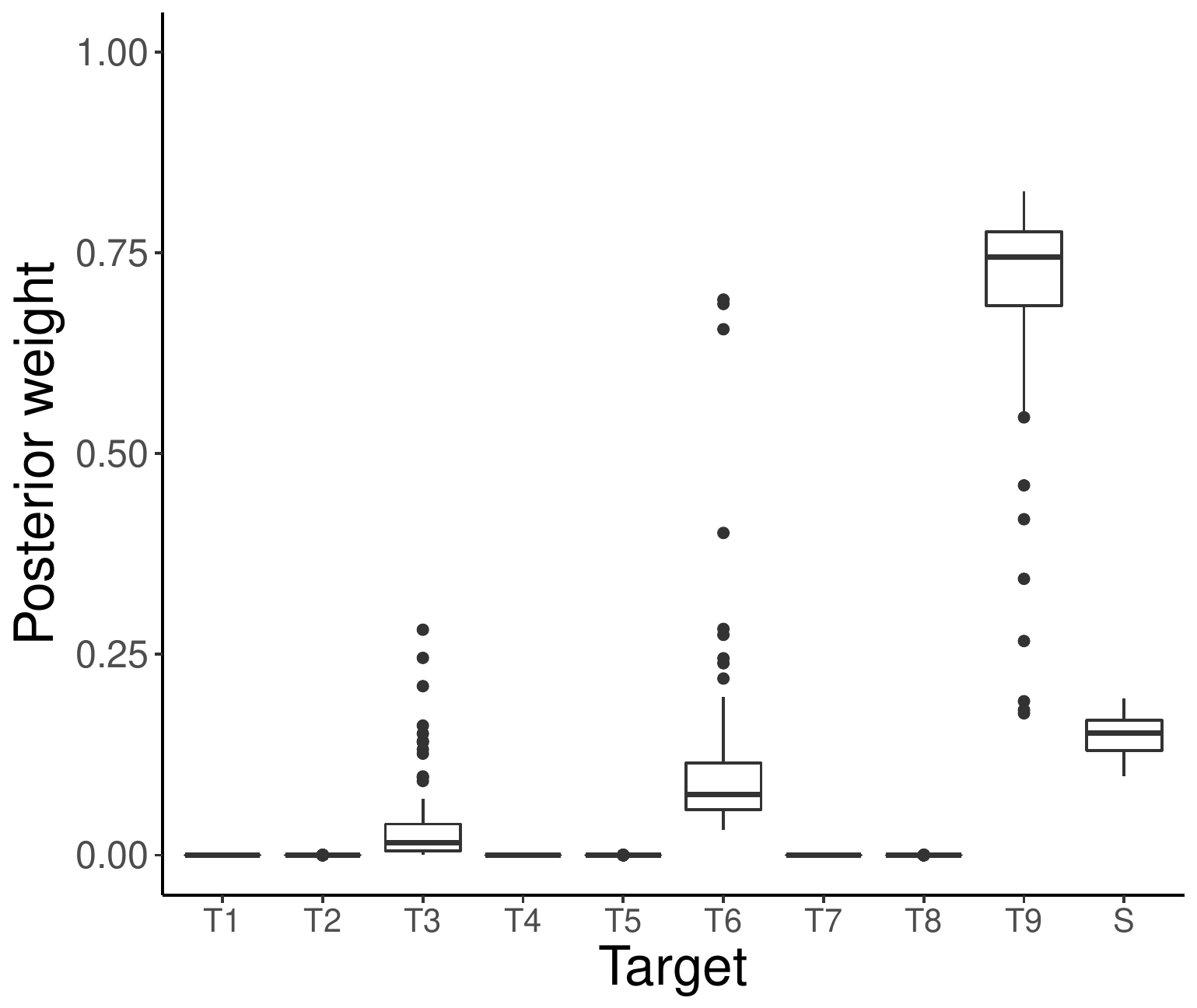}
  			\label{supp:Fig:modelSim2:sub3w}
  		}
  \end{minipage}\hfill
  \begin{minipage}[c]{0.5\linewidth}
    \centering
  		\subfigure[][Scenario 4: PRIAL]{
  			\includegraphics[width=0.9\textwidth]{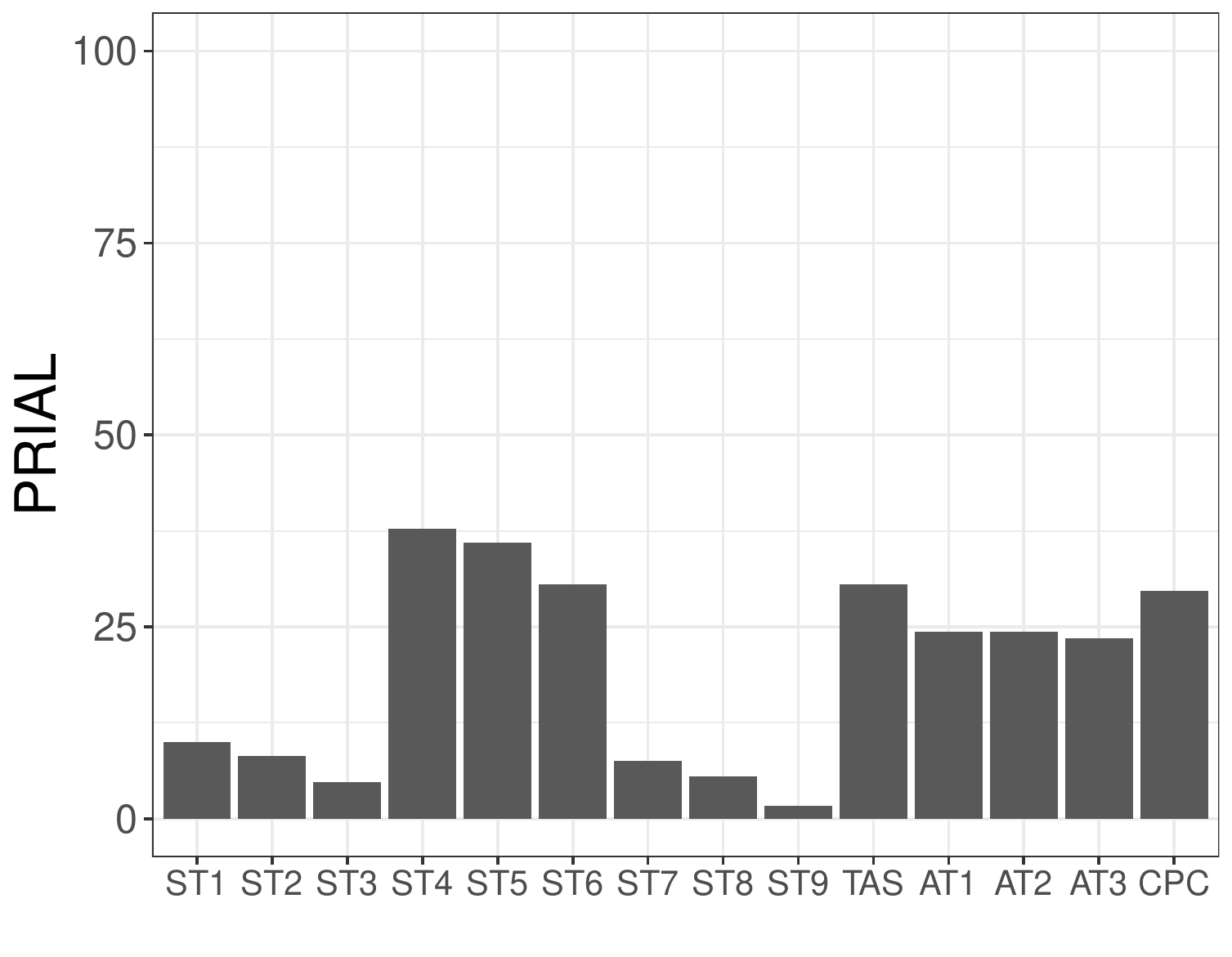}
  			\label{supp:Fig:modelSim2:sub4}
  		}
  \end{minipage}\hfill
      	\begin{minipage}[c]{0.5\linewidth}
    \centering 
    		\subfigure[][Scenario 4: target-specific posterior weights]{
  			\includegraphics[width=0.9\textwidth]{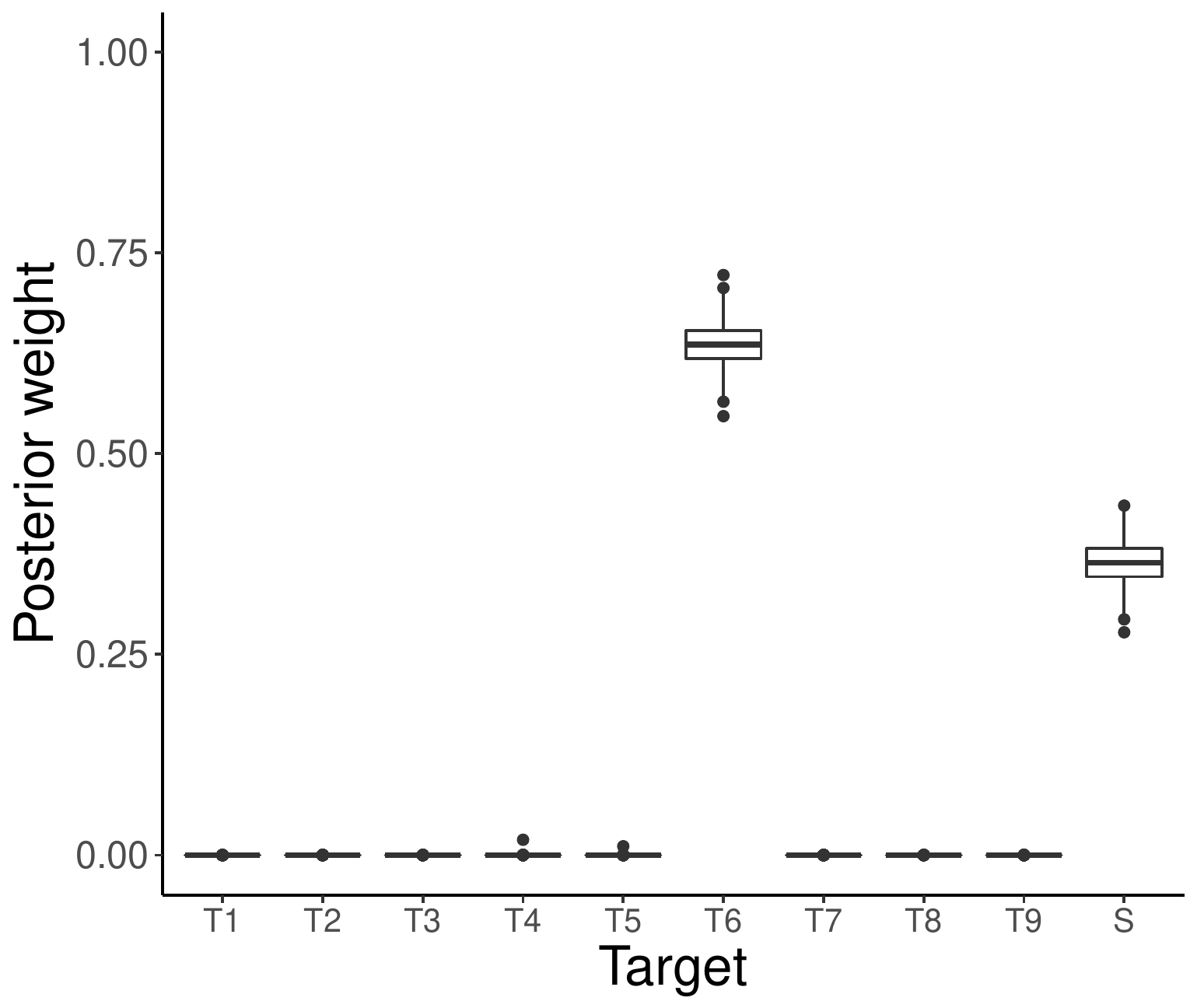}
  			\label{supp:Fig:modelSim2:sub4w}
  		}
  \end{minipage}\hfill
		\caption{Simulation results for scenarios 3 and 4 when $n=50$. Barplots display the PRIAL for each estimator and boxplots display target-specific posterior weights (see equation \eqref{Eq:TAS_weights}) of the TAS estimator. ST1,~\ldots,~ ST9 refer to the nine STS estimators, TAS to estimator \eqref{Eq:TAS}, AT1, \ldots, AT3 to the three estimators of \citet{touloumis2015} and CPC to the estimator of \citet{schafer2005}.}
		\label{supp:Fig:modelSim2}
\end{figure}

\begin{figure}[ht]
	\begin{minipage}[c]{0.5\linewidth}
    \centering
  		\subfigure[][Scenario 1: PRIAL]{
  			\includegraphics[width=0.9\textwidth]{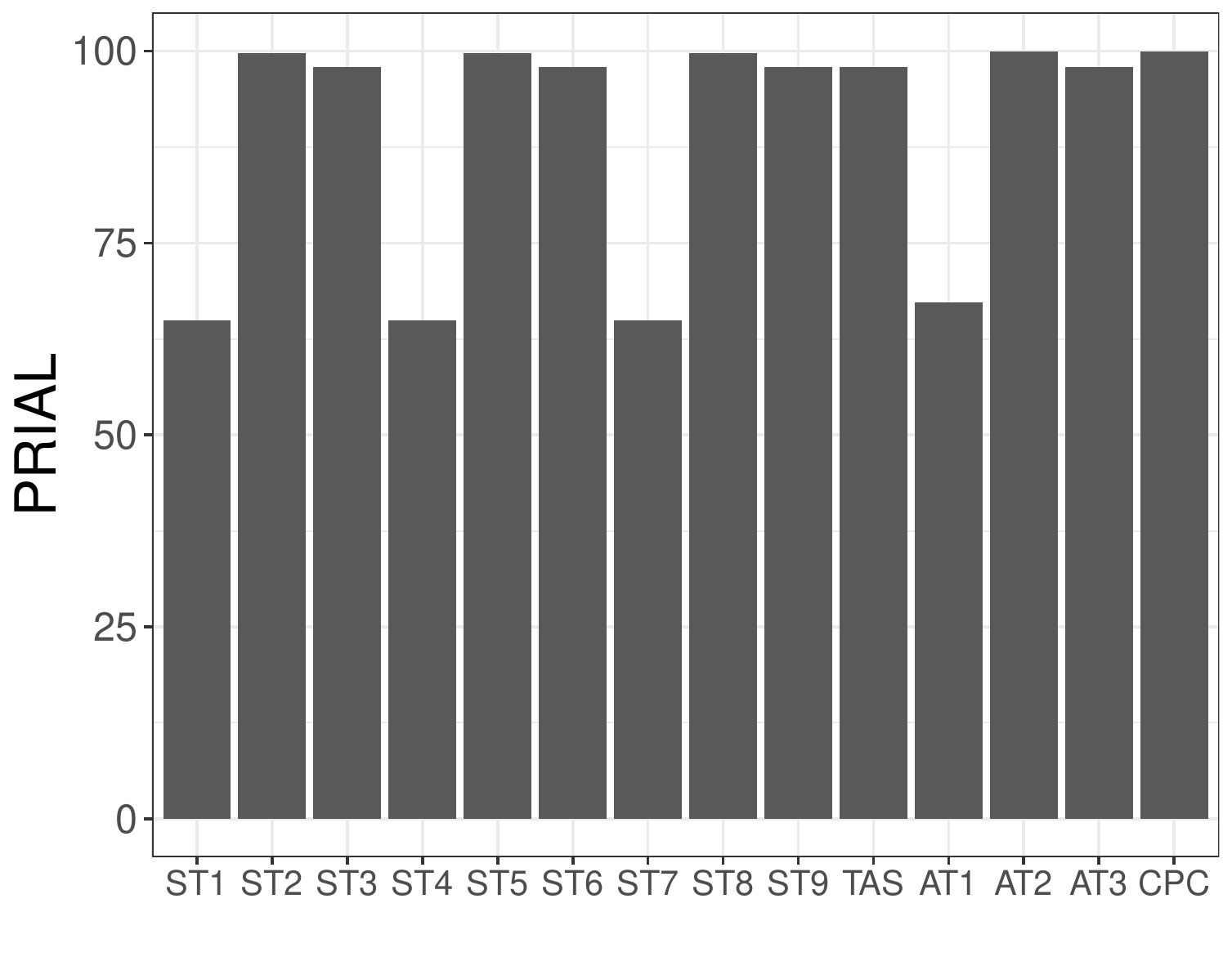}
  			\label{supp:Fig:modelSim3:sub1}
  		}
  \end{minipage}\hfill
  	\begin{minipage}[c]{0.5\linewidth}
    \centering 
    		\subfigure[][Scenario 1: target-specific posterior weights]{
  			\includegraphics[width=0.9\textwidth]{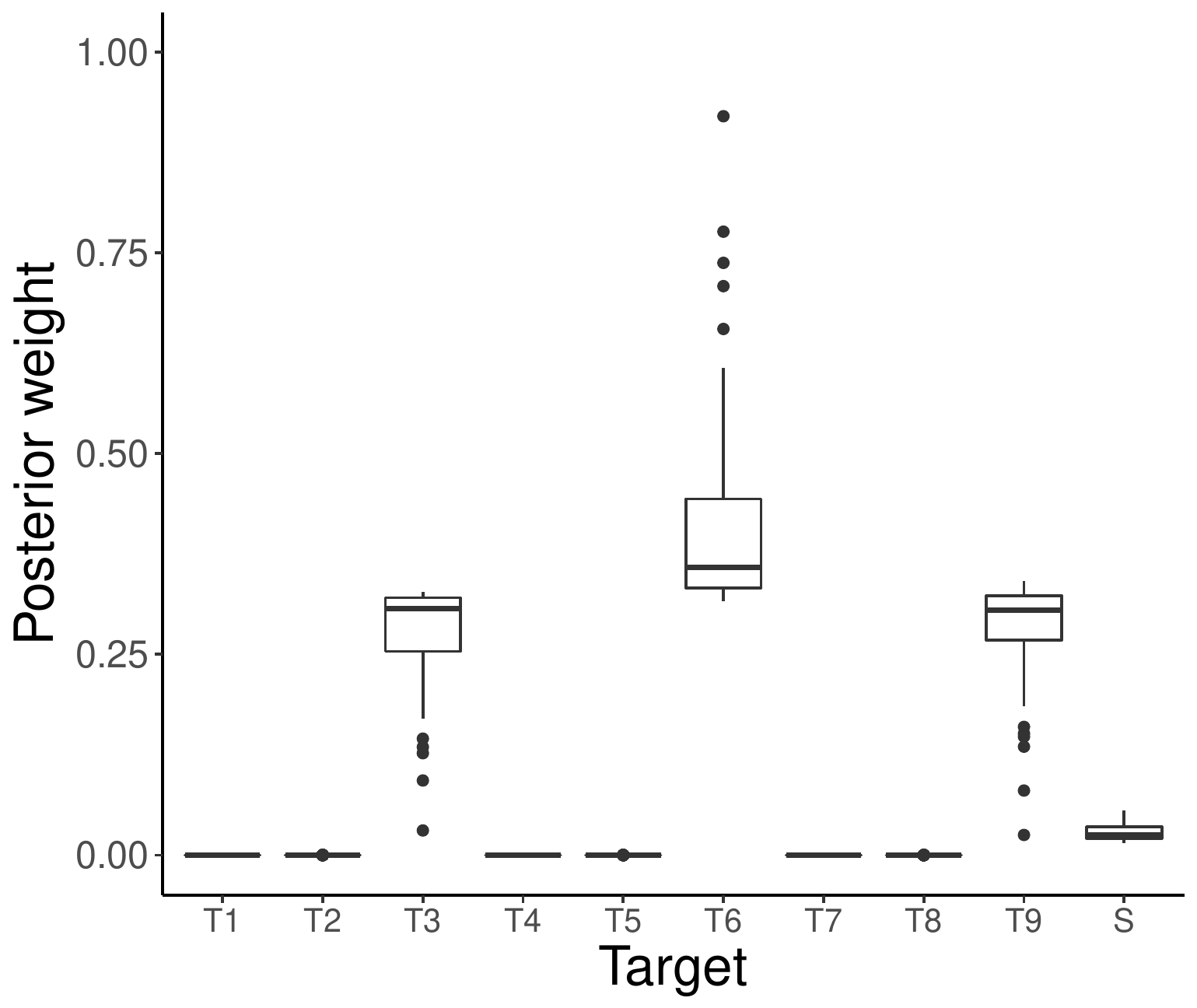}
  			\label{supp:Fig:modelSim3:sub1w}
  		}
  \end{minipage}\hfill
	\begin{minipage}[c]{0.5\linewidth}
    \centering
  		\subfigure[][Scenario 2: PRIAL]{
  			\includegraphics[width=0.9\textwidth]{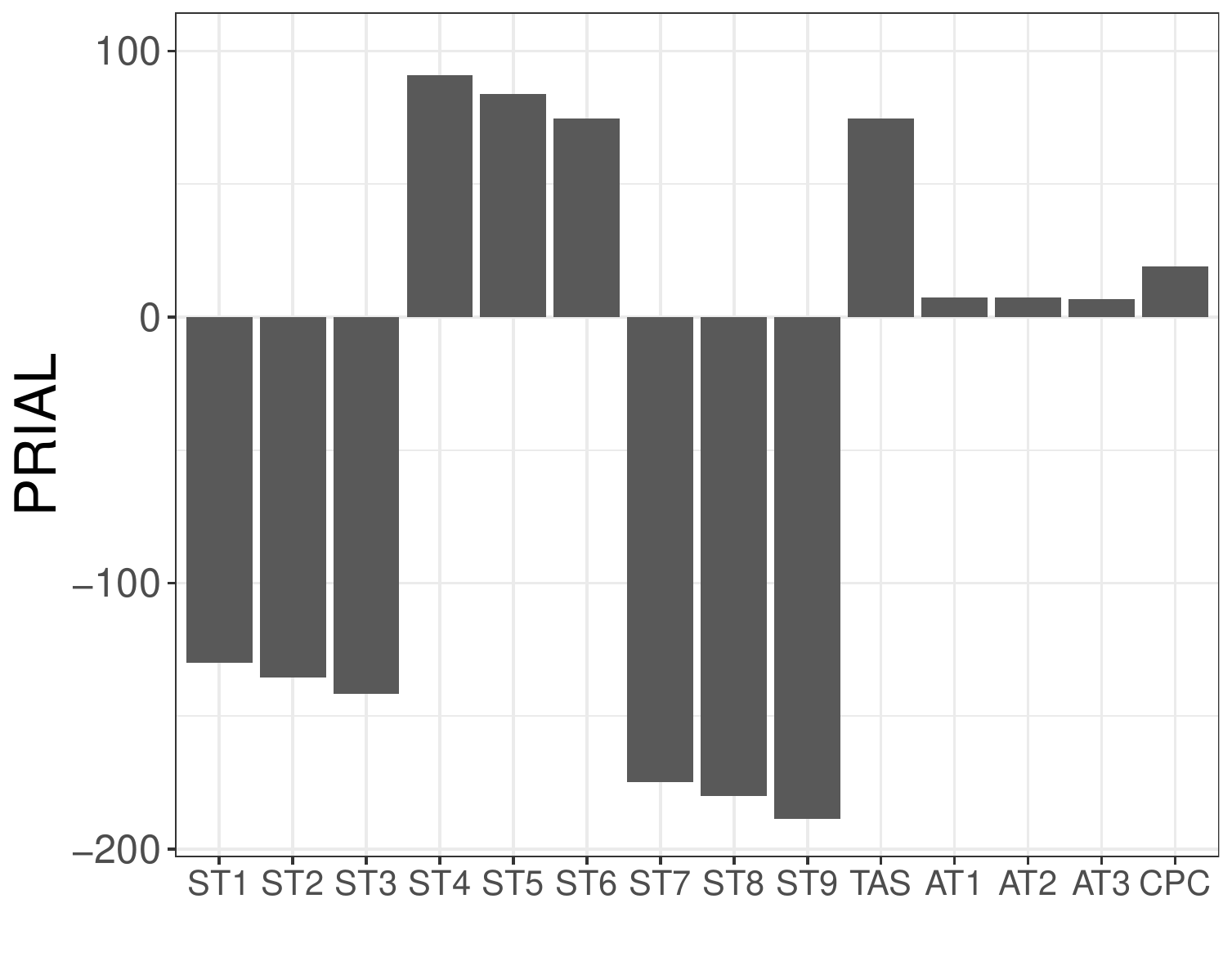}
  			\label{supp:Fig:modelSim3:sub2}
  		}
  \end{minipage}\hfill
    	\begin{minipage}[c]{0.5\linewidth}
    \centering 
    		\subfigure[][Scenario 2: target-specific posterior weights]{
  			\includegraphics[width=0.9\textwidth]{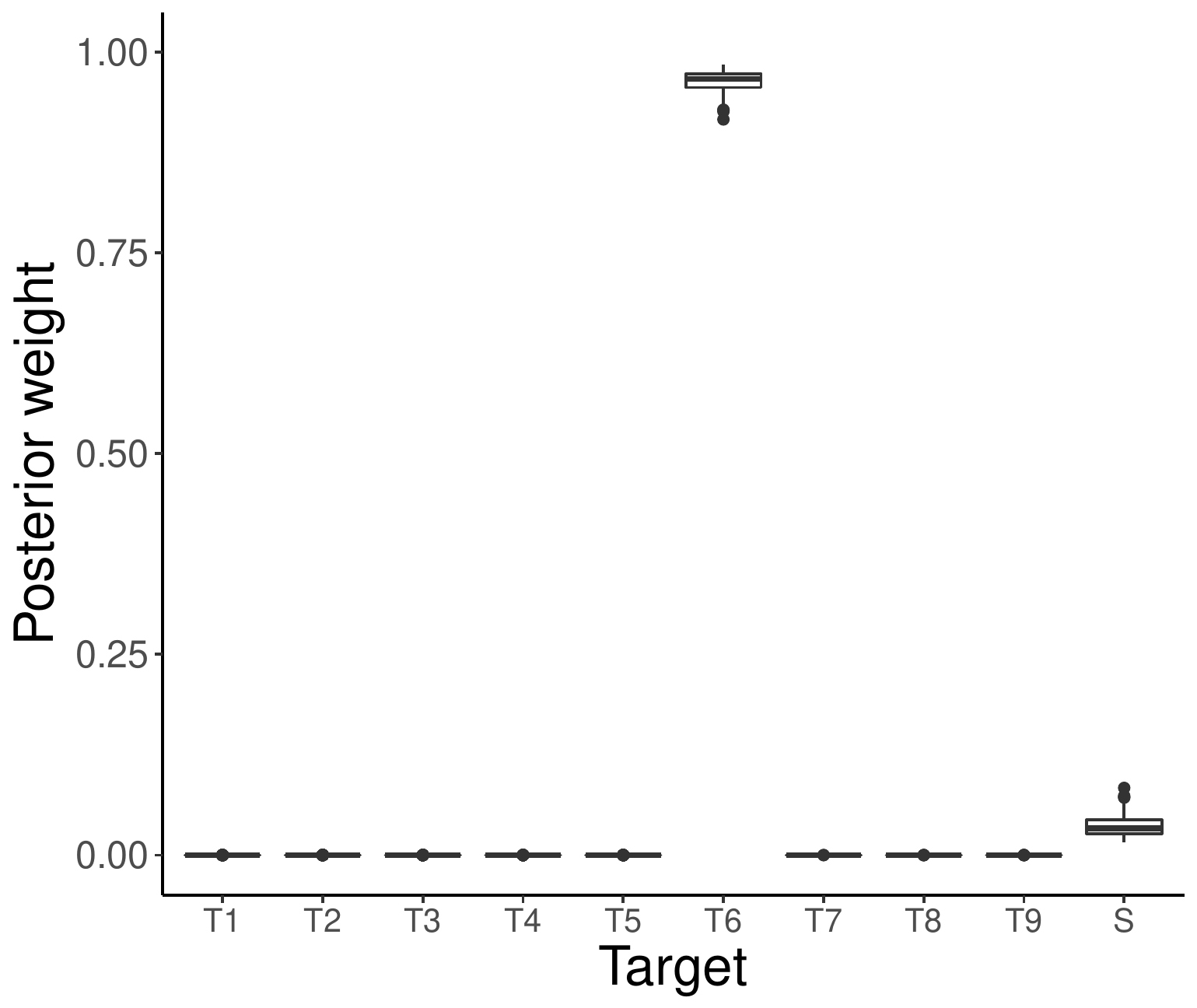}
  			\label{supp:Fig:modelSim3:sub2w}
  		}
  \end{minipage}\hfill
		\caption{Simulation results for scenarios 1 and 2 when $n=75$. Barplots display the PRIAL for each estimator and boxplots display target-specific posterior weights (see equation \eqref{Eq:TAS_weights}) of the TAS estimator. ST1,~\ldots,~ ST9 refer to the nine STS estimators, TAS to estimator \eqref{Eq:TAS}, AT1, \ldots, AT3 to the three estimators of \citet{touloumis2015} and CPC to the estimator of \citet{schafer2005}.}
	\label{supp:Fig:modelSim3}
\end{figure}

\begin{figure}[ht]
  \begin{minipage}[c]{0.5\linewidth}
    \centering
  		\subfigure[][Scenario 3: PRIAL]{
  			\includegraphics[width=0.9\textwidth]{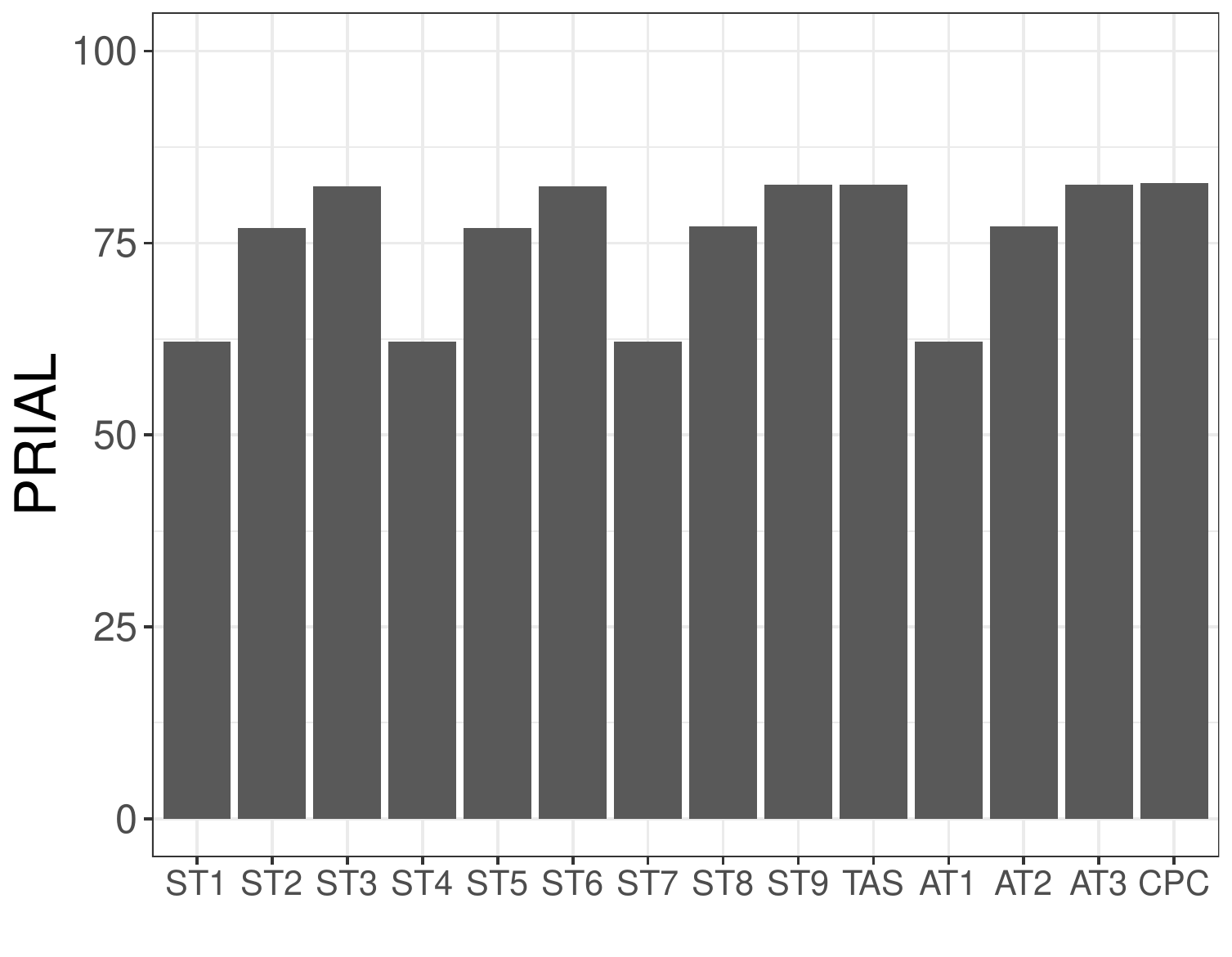}
  			\label{supp:Fig:modelSim4:sub3}
  		}
  \end{minipage}\hfill
      	\begin{minipage}[c]{0.5\linewidth}
    \centering 
    		\subfigure[][Scenario 3: target-specific posterior weights]{
  			\includegraphics[width=0.9\textwidth]{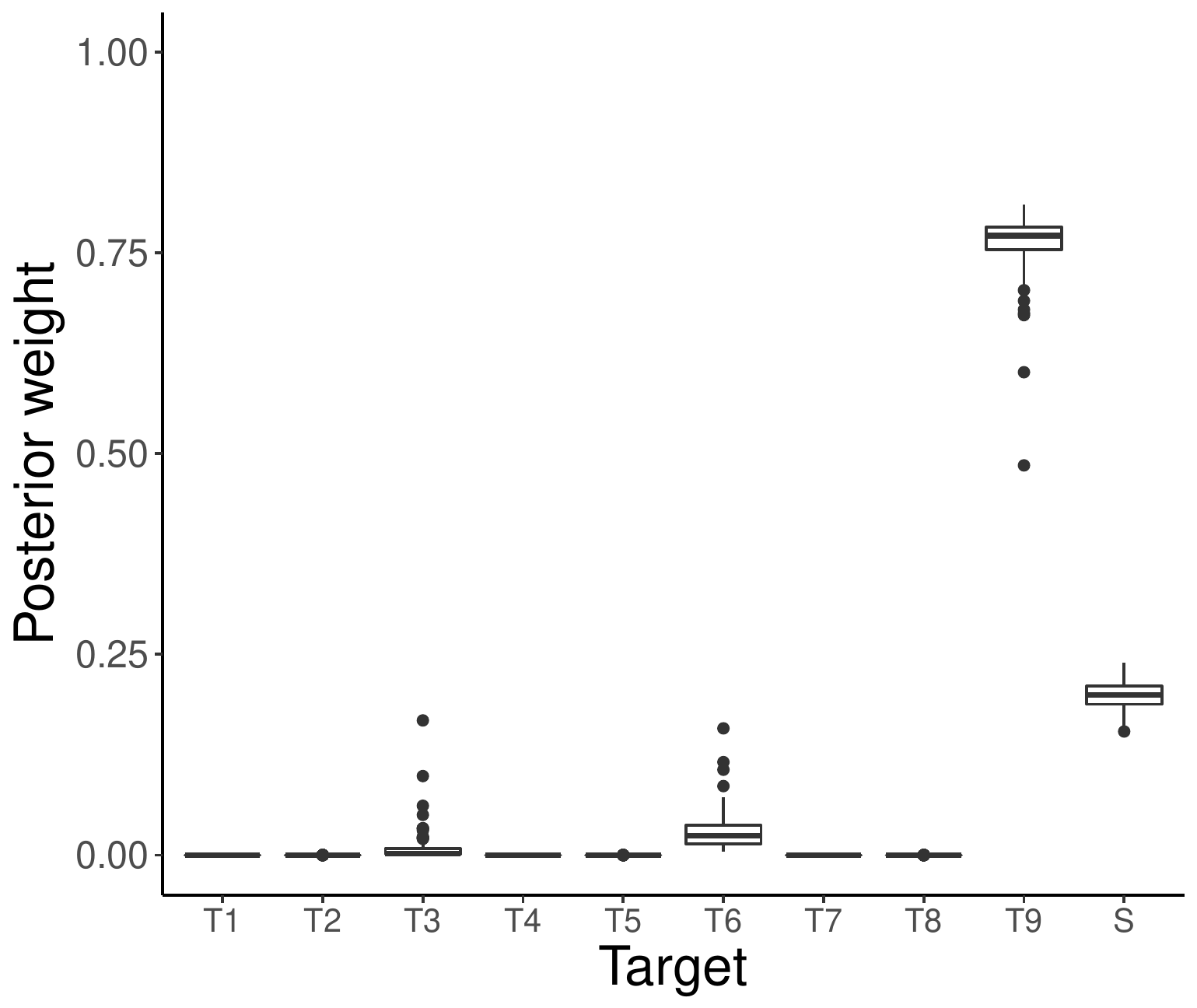}
  			\label{supp:Fig:modelSim4:sub3w}
  		}
  \end{minipage}\hfill
  \begin{minipage}[c]{0.5\linewidth}
    \centering
  		\subfigure[][Scenario 4: PRIAL]{
  			\includegraphics[width=0.9\textwidth]{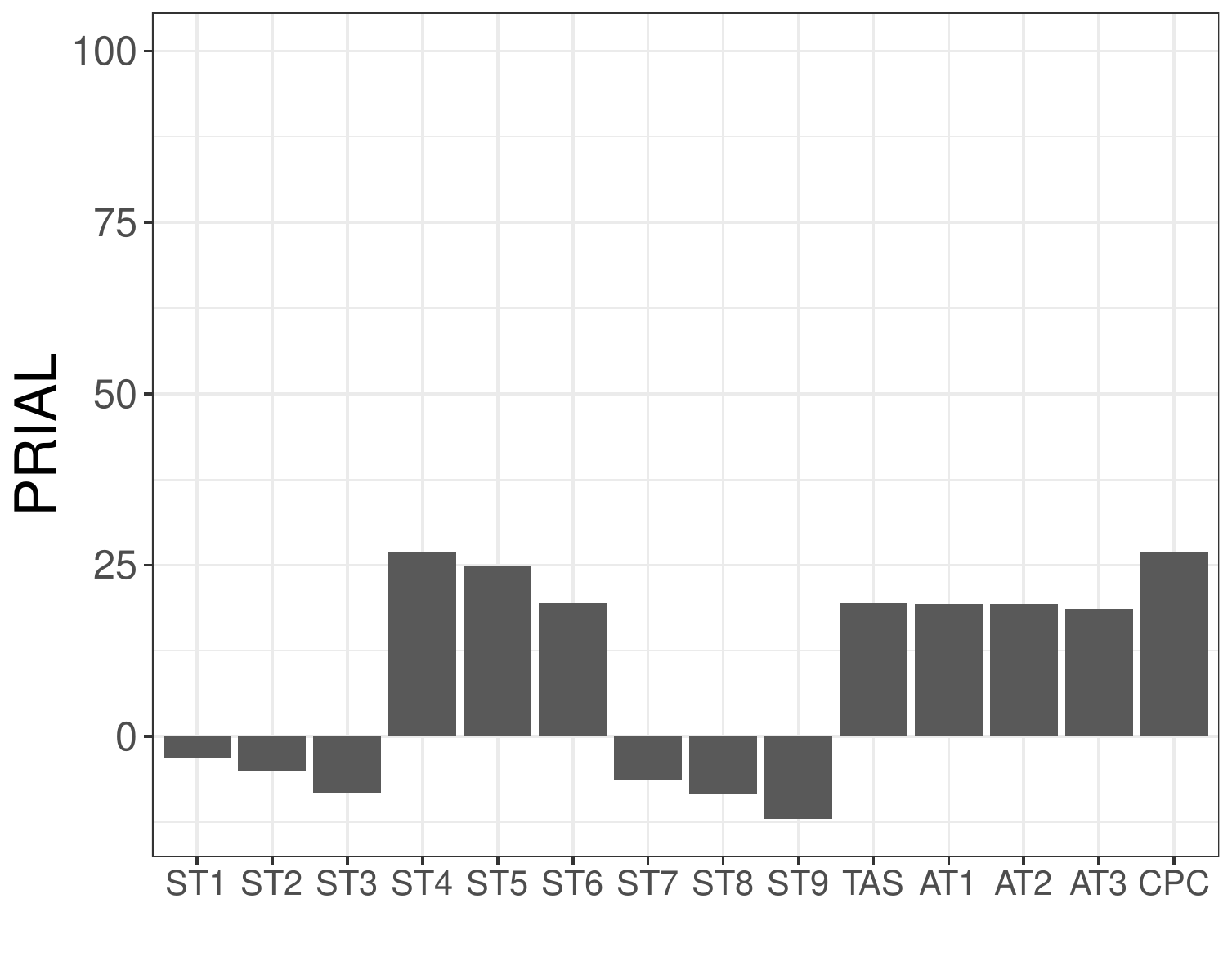}
  			\label{supp:Fig:modelSim4:sub4}
  		}
  \end{minipage}\hfill
      	\begin{minipage}[c]{0.5\linewidth}
    \centering 
    		\subfigure[][Scenario 4: target-specific posterior weights]{
  			\includegraphics[width=0.9\textwidth]{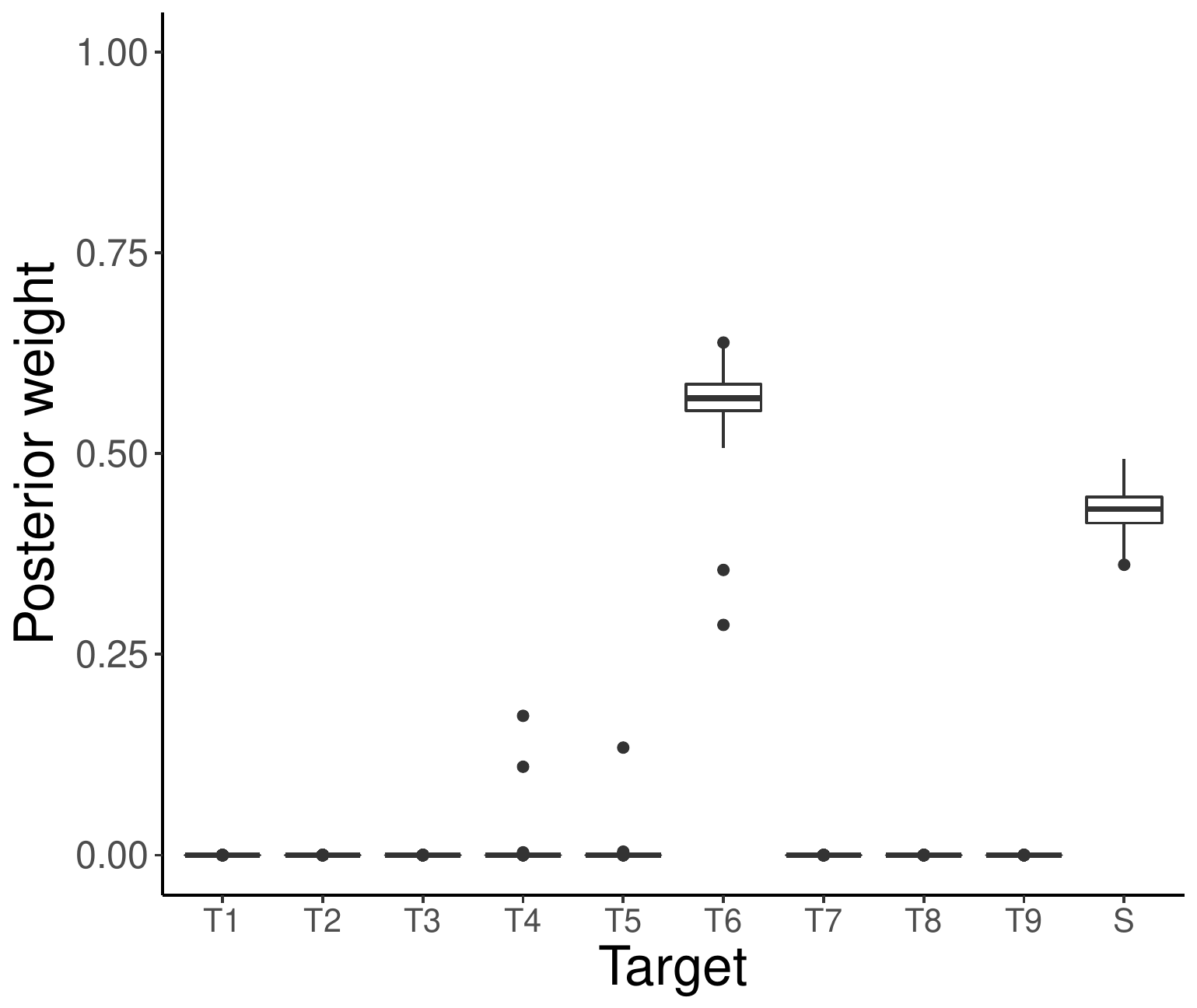}
  			\label{supp:Fig:modelSim4:sub4w}
  		}
  \end{minipage}\hfill
		\caption{Simulation results for scenarios 3 and 4 when $n=75$. Barplots display the PRIAL for each estimator and boxplots display target-specific posterior weights (see equation \eqref{Eq:TAS_weights}) of the TAS estimator. ST1,~\ldots,~ ST9 refer to the nine STS estimators, TAS to estimator \eqref{Eq:TAS}, AT1, \ldots, AT3 to the three estimators of \citet{touloumis2015} and CPC to the estimator of \citet{schafer2005}.}
		\label{supp:Fig:modelSim4}
\end{figure}

Figures \ref{supp:Fig:modelSim1}, \ref{supp:Fig:modelSim2}, \ref{supp:Fig:modelSim3} and \ref{supp:Fig:modelSim4} complement Figures \ref{Fig:modelSim1} and \ref{Fig:modelSim2} in Section \ref{Sec:modelsim} by providing results for $n \in \{50, 75\}$.\\

\newpage
\null\vfill
\vfill\null

\newpage
\null\vfill
\vfill\null

\newpage
\null\vfill
\vfill\null

\newpage

\null\vfill
\begin{figure}[ht]
  \begin{minipage}[c]{0.5\linewidth}
    \centering
  		\subfigure[][Scenario 1]{
  			\includegraphics[width=1.0\textwidth]{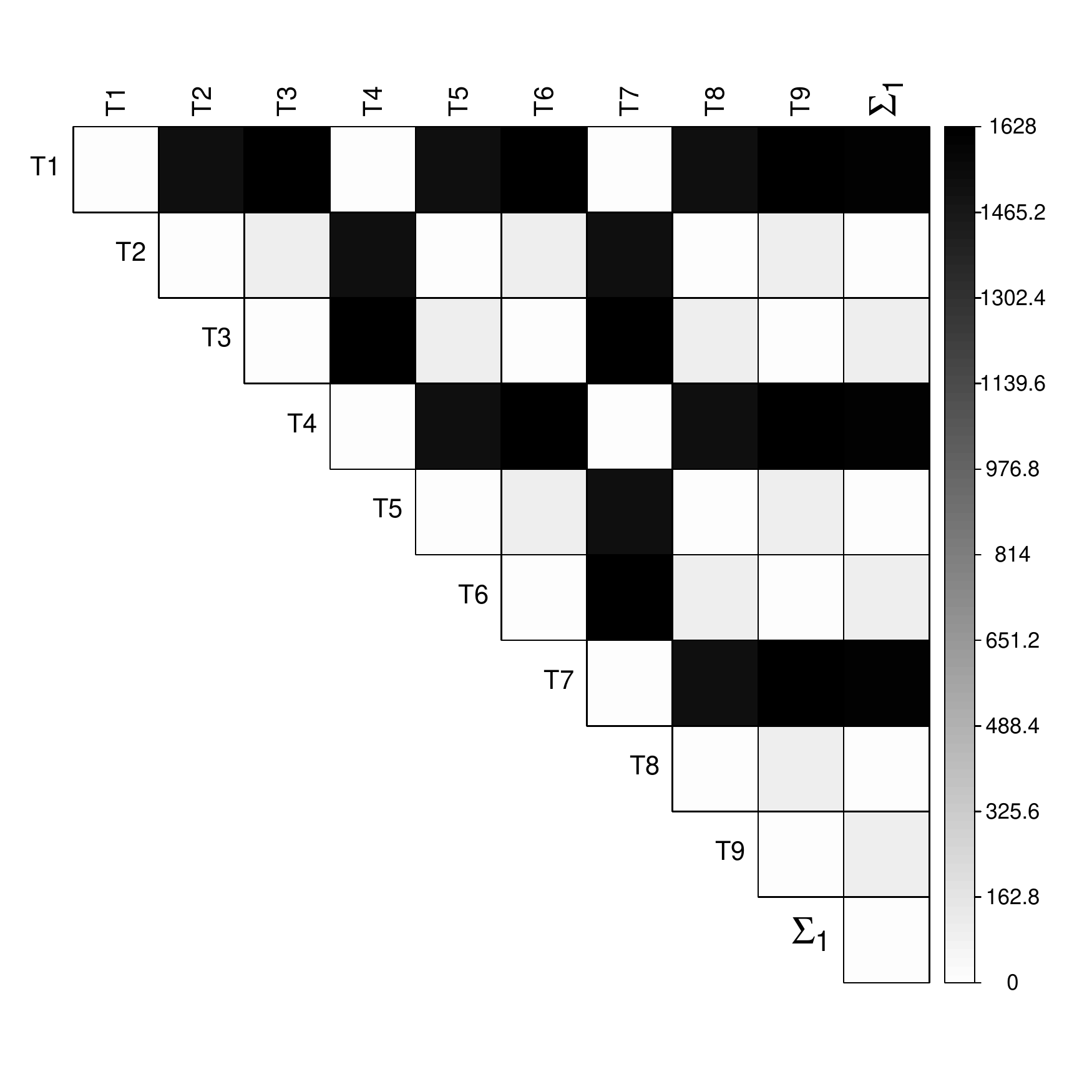}
  			\label{supp:Fig:modelSim5:sub1}
  		}
  \end{minipage}\hfill
      	\begin{minipage}[c]{0.5\linewidth}
    \centering 
    		\subfigure[][Scenario 2]{
  			\includegraphics[width=1.0\textwidth]{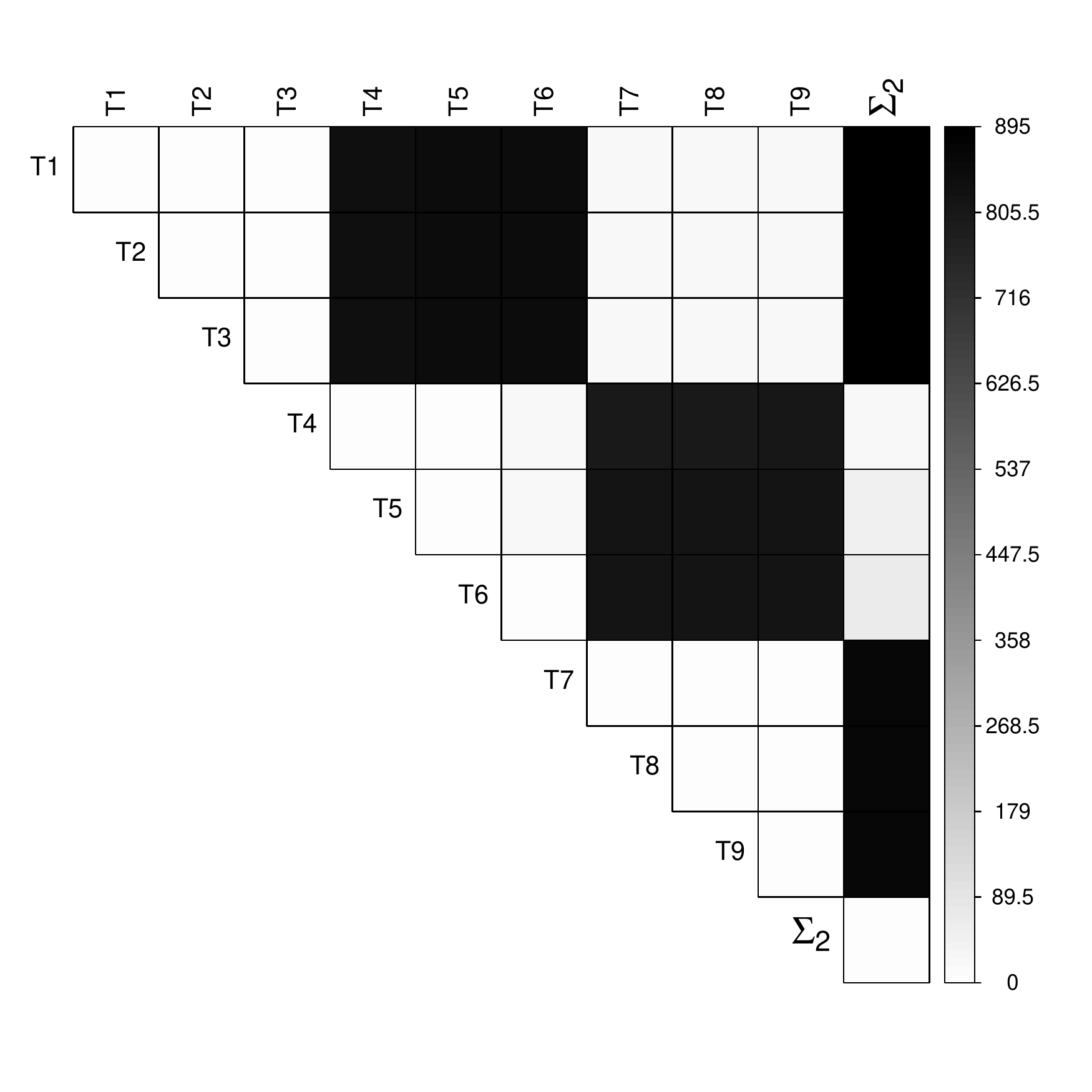}
  			\label{supp:Fig:modelSim5:sub2}
  		}
  \end{minipage}\hfill
  \begin{minipage}[c]{0.5\linewidth}
    \centering
  		\subfigure[][Scenario 3]{
  			\includegraphics[width=1.0\textwidth]{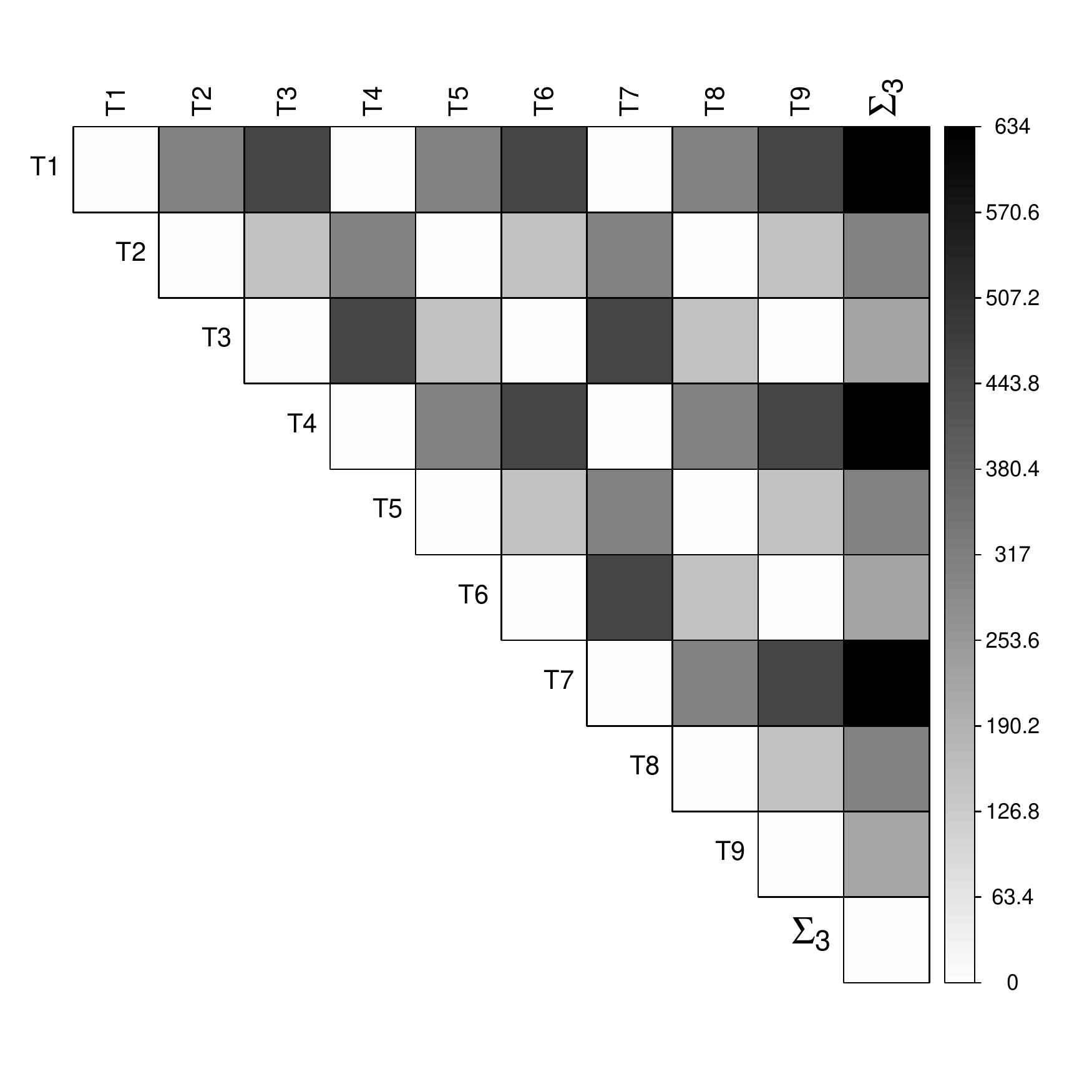}
  			\label{supp:Fig:modelSim5:sub3}
  		}
  \end{minipage}\hfill
      	\begin{minipage}[c]{0.5\linewidth}
    \centering 
    		\subfigure[][Scenario 4]{
  			\includegraphics[width=1.0\textwidth]{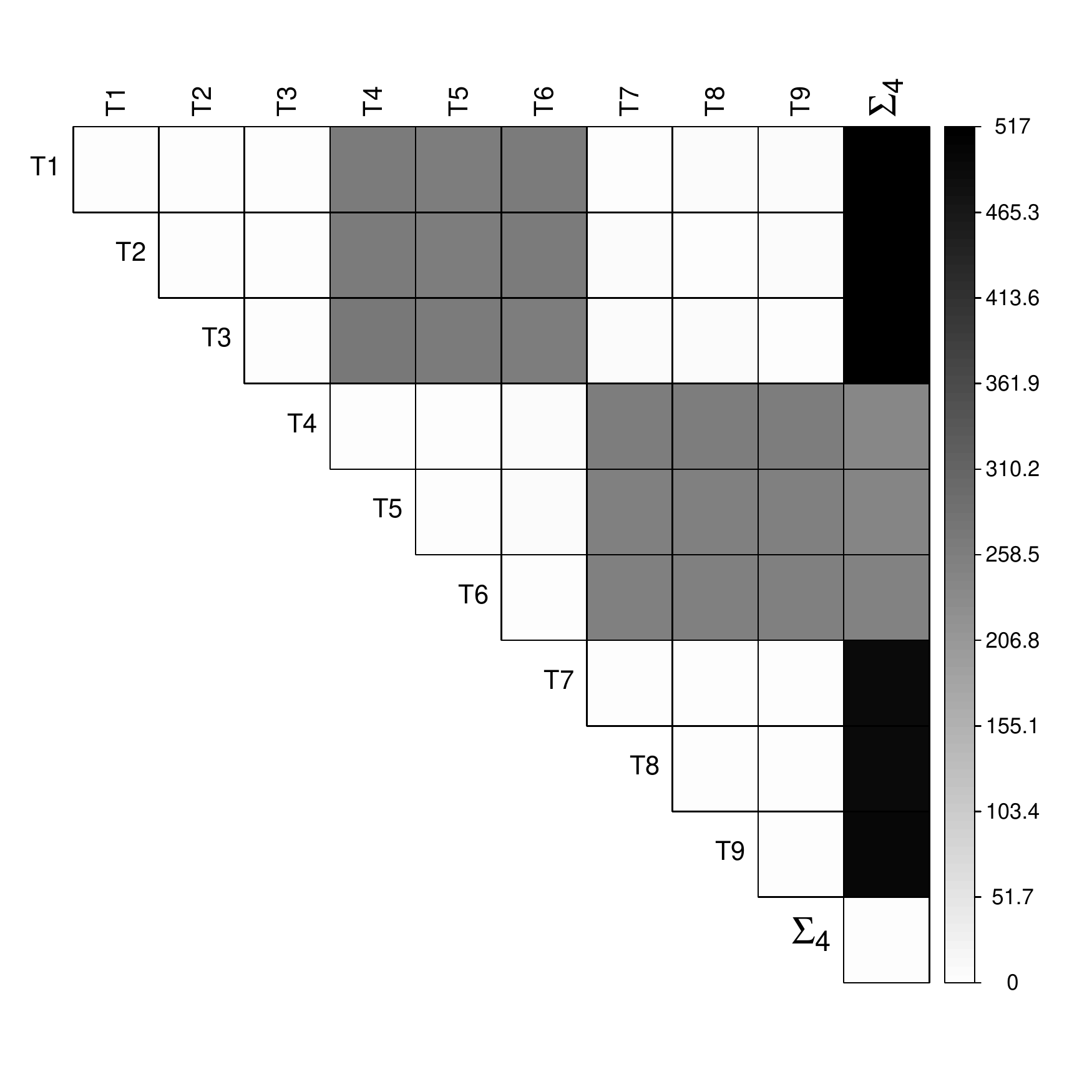}
  			\label{supp:Fig:modelSim5:sub4}
  		}
  \end{minipage}\hfill
		\caption{Heatmaps displaying the average Frobenius norm (over the 100 simulated data sets) between all pairs of shrinkage targets in Table~\ref{Tab:targets} for simulation scenarios 1, 2, 3 and 4 when $n=50$ (results are omitted for $n\in\{ 25,75 \}$ as they are identical). The true covariance matrices $\Sigma_1, \dots, \Sigma_4$ were also included in the comparison. Light (dark) colors indicate that the shrinkage targets are (dis-)similar.}
		\label{supp:Fig:modelSim5}
\end{figure}
\vfill\null

\newpage

\section{Data-based simulation strategy}
\label{supp:data_sims}

\begin{figure}[ht]
\centering
	\centering
	\includegraphics[width=0.6\textwidth]{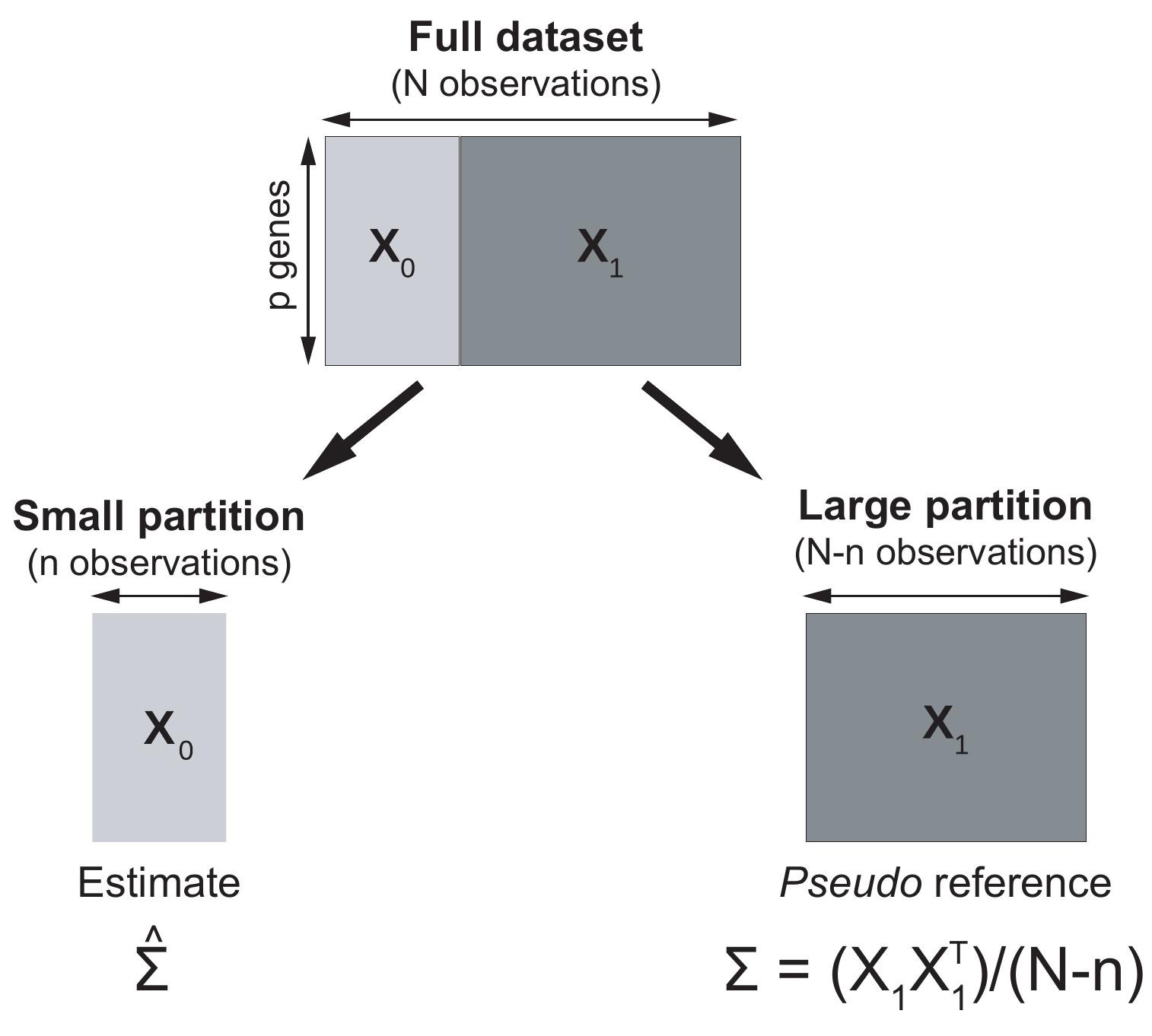}
\caption{Illustration of the data-partition strategy used in Sections \ref{Sec:datasim}.}
\label{supp:Fig:protocol}
\end{figure}

Figure~\ref{supp:Fig:protocol} illustrates the data partition strategy adopted in Section~\ref{Sec:datasim} when evaluating the performance of multiple covariance estimators. Figure~\ref{Fig:tcgatargetweights2} complements  Figure~\ref{Fig:dataSim2} by providing results for $n=p/4$ and $n=3p/4$.

\begin{figure}[ht]
  	 \begin{minipage}[c]{0.5\linewidth}
    \centering
    \subfigure[][$\boldsymbol{\hat{\Sigma}}_{\text{TAS}}$ ($n = p/2$)]{
  	\includegraphics[width=0.8\textwidth]{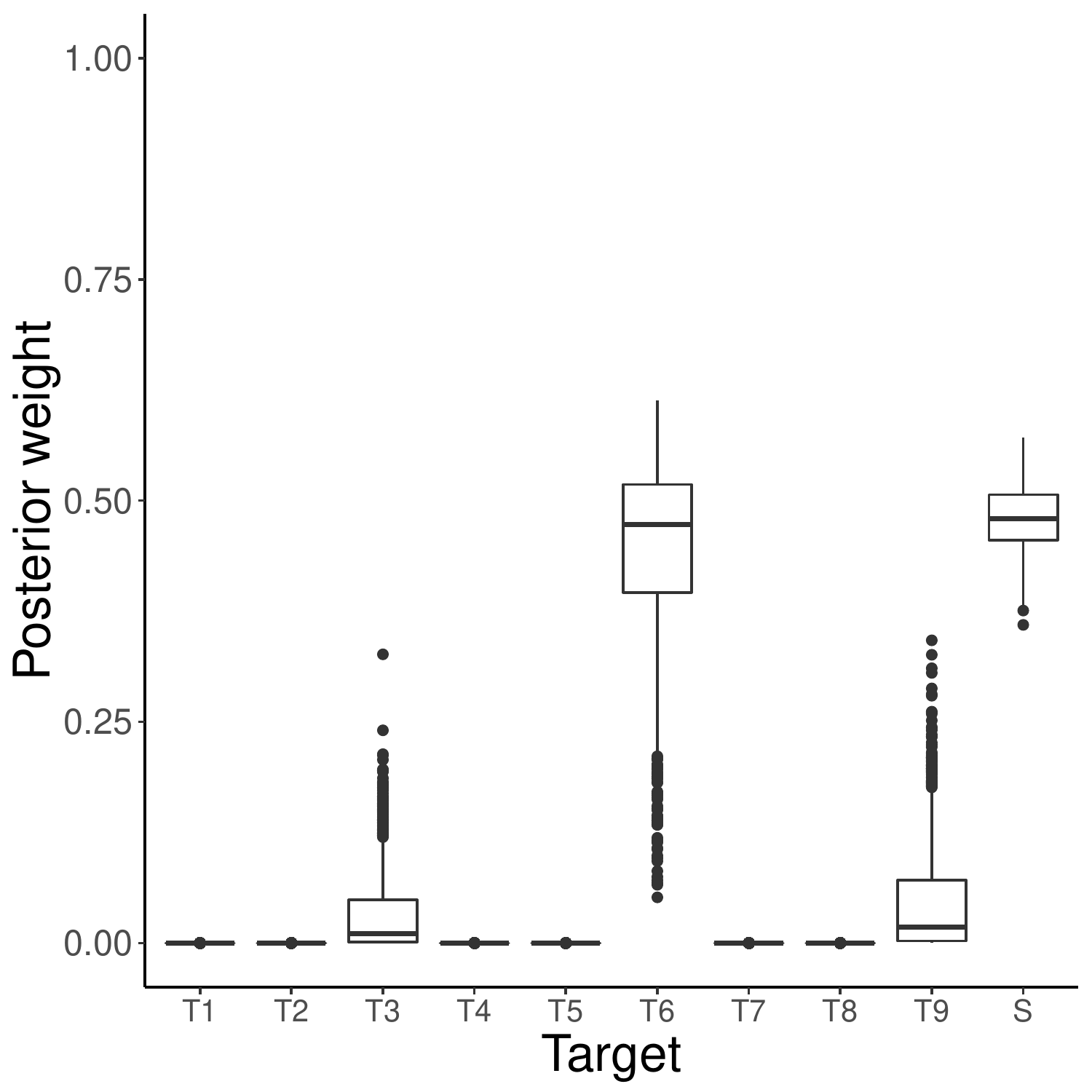}
  	\label{supp:Fig:dataSim1:sub1}
  	}
  	\end{minipage}\hfill
  	\begin{minipage}[c]{0.5\linewidth}
    \centering
    \subfigure[][$\boldsymbol{\hat{\Sigma}}_{\text{TAS-info}}$  ($n = p/2$)]{
  	\includegraphics[width=0.8\textwidth]{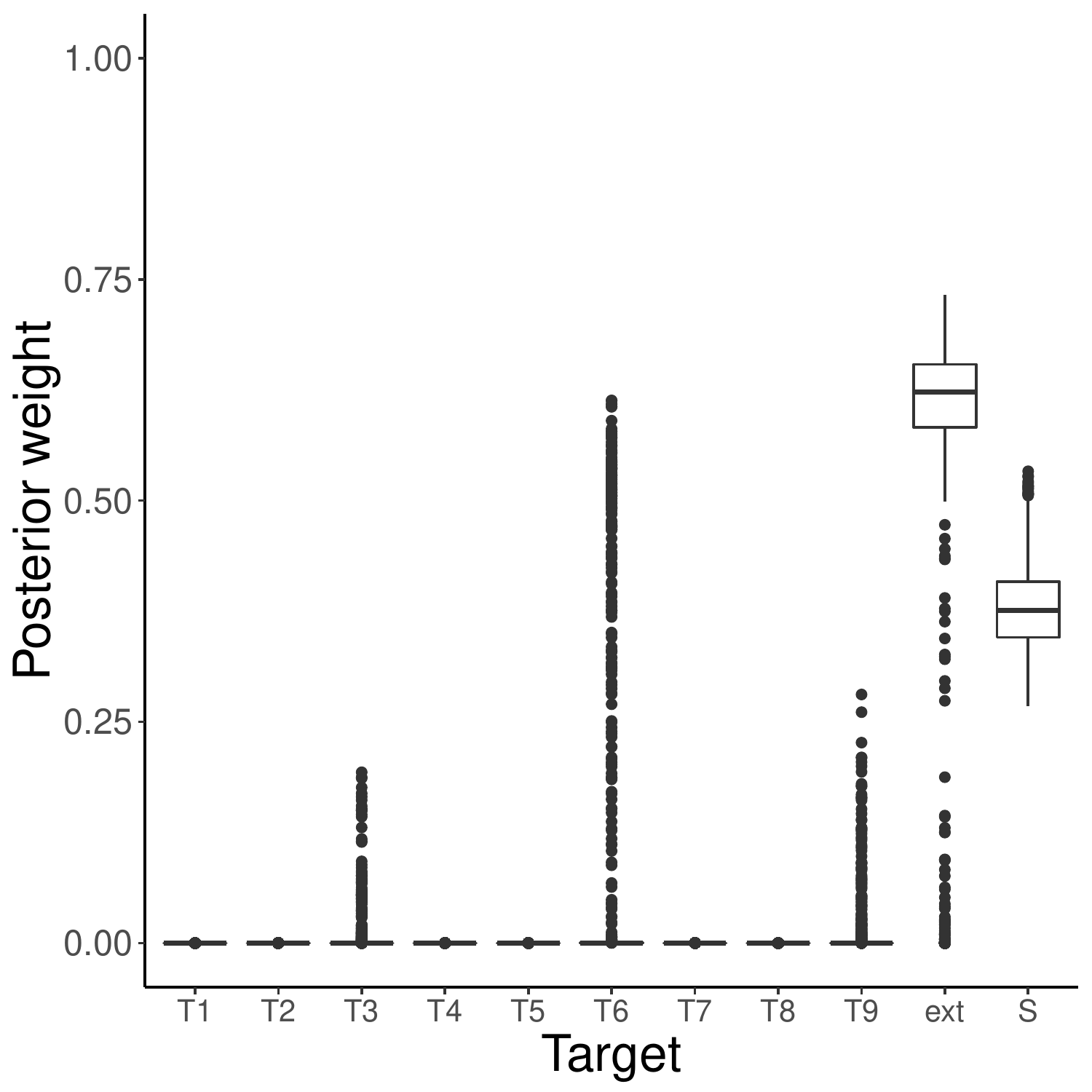}
  	\label{supp:Fig:dataSim1:sub2}
  	}
  	\end{minipage}\hfill
  	\caption{Target-specific posterior weights (see equation \eqref{Eq:TAS_weights}) obtained for estimators $\boldsymbol{\hat{\Sigma}}_{\text{TAS}}$ and $\boldsymbol{\hat{\Sigma}}_{\text{TAS-info}}$ across the $1,000$ random data partitions of the breast cancer data set when $n \in \{ p/2 \}$. The target ``ext'' in $\boldsymbol{\hat{\Sigma}}_{\text{TAS-info}}$ stands for the shrinkage target $\boldsymbol{\hat{\Sigma}}_{\text{ext}}$ estimated from external data.}
\label{supp:Fig:dataSim1}
\end{figure}

\newpage
\null\vfill
\begin{figure}[ht]
  	\begin{minipage}[c]{0.5\linewidth}
    \centering
    \subfigure[][$\boldsymbol{\hat{\Sigma}}_{\text{TAS}}$ ($n = p/4$)]{
  	\includegraphics[width=0.8\textwidth]{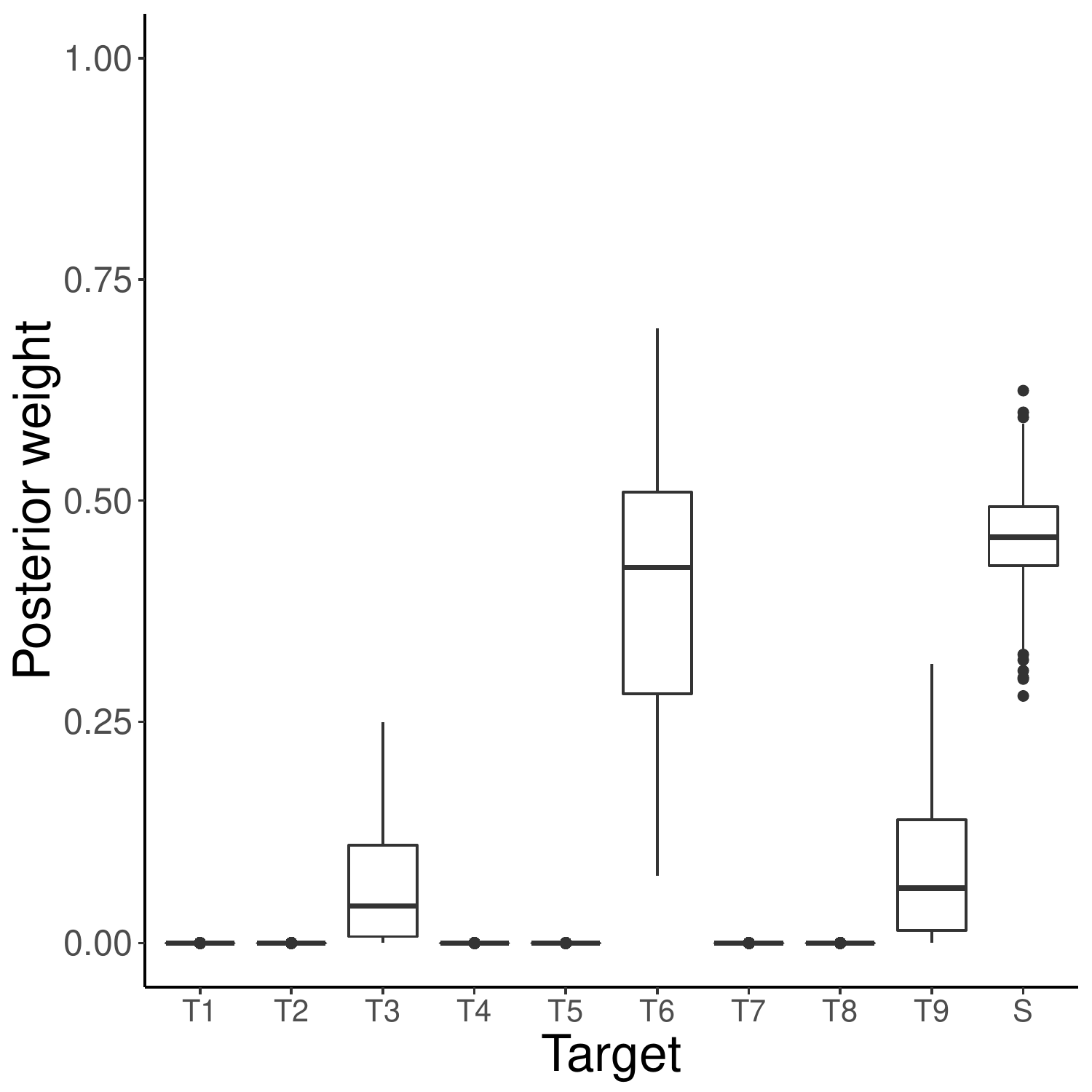}
  	\label{supp:Fig:dataSim2:sub1}
  	}
  	\end{minipage}\hfill
  	\begin{minipage}[c]{0.5\linewidth}
    \centering
    \subfigure[][$\boldsymbol{\hat{\Sigma}}_{\text{TAS-info}}$ ($n = p/4$)]{
  	\includegraphics[width=0.8\textwidth]{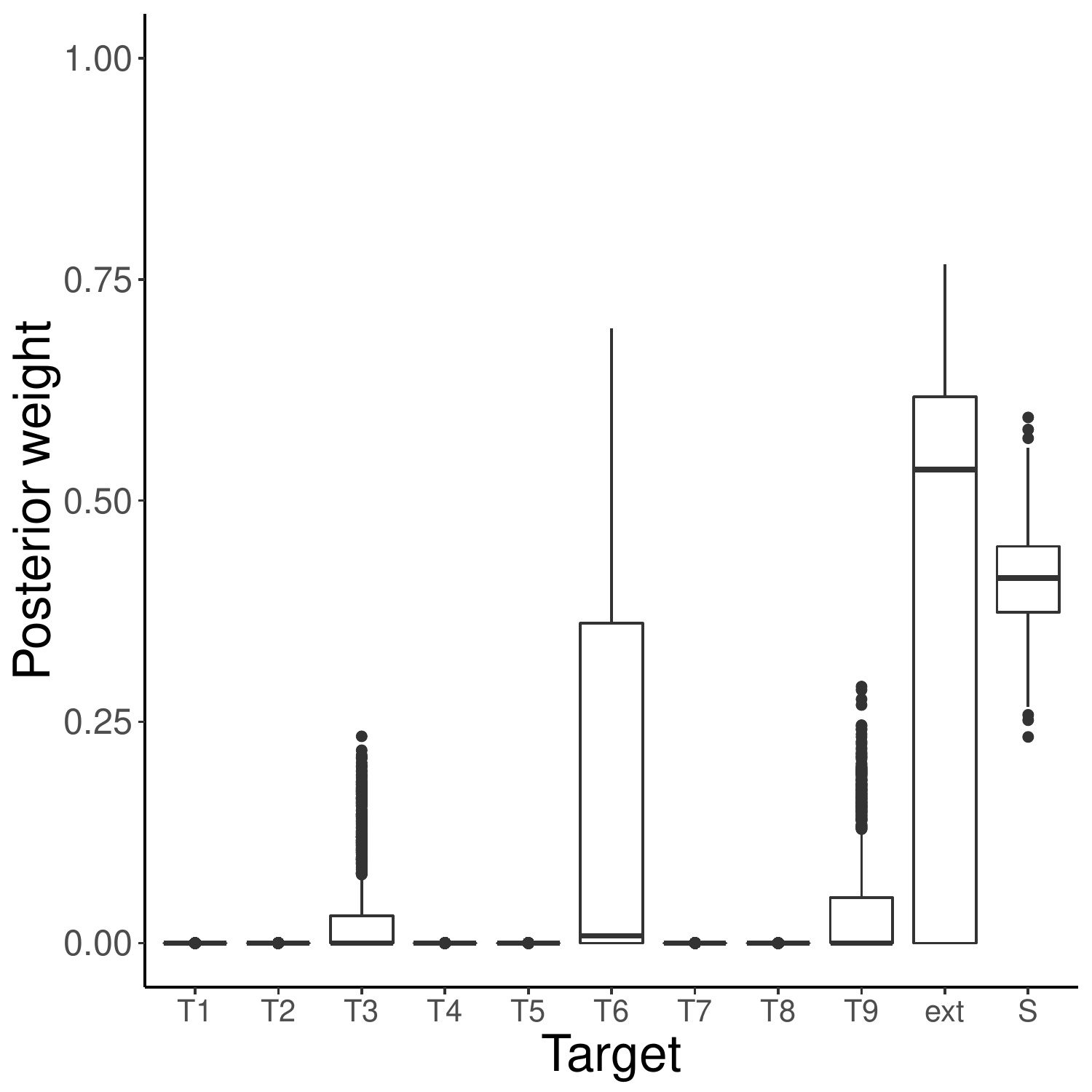}
  	\label{supp:Fig:dataSim2:sub2}
  	}
  	\end{minipage}\hfill
  	 \begin{minipage}[c]{0.5\linewidth}
    \centering
    \subfigure[][$\boldsymbol{\hat{\Sigma}}_{\text{TAS}}$ ($n = 3p/4$)]{
  	\includegraphics[width=0.8\textwidth]{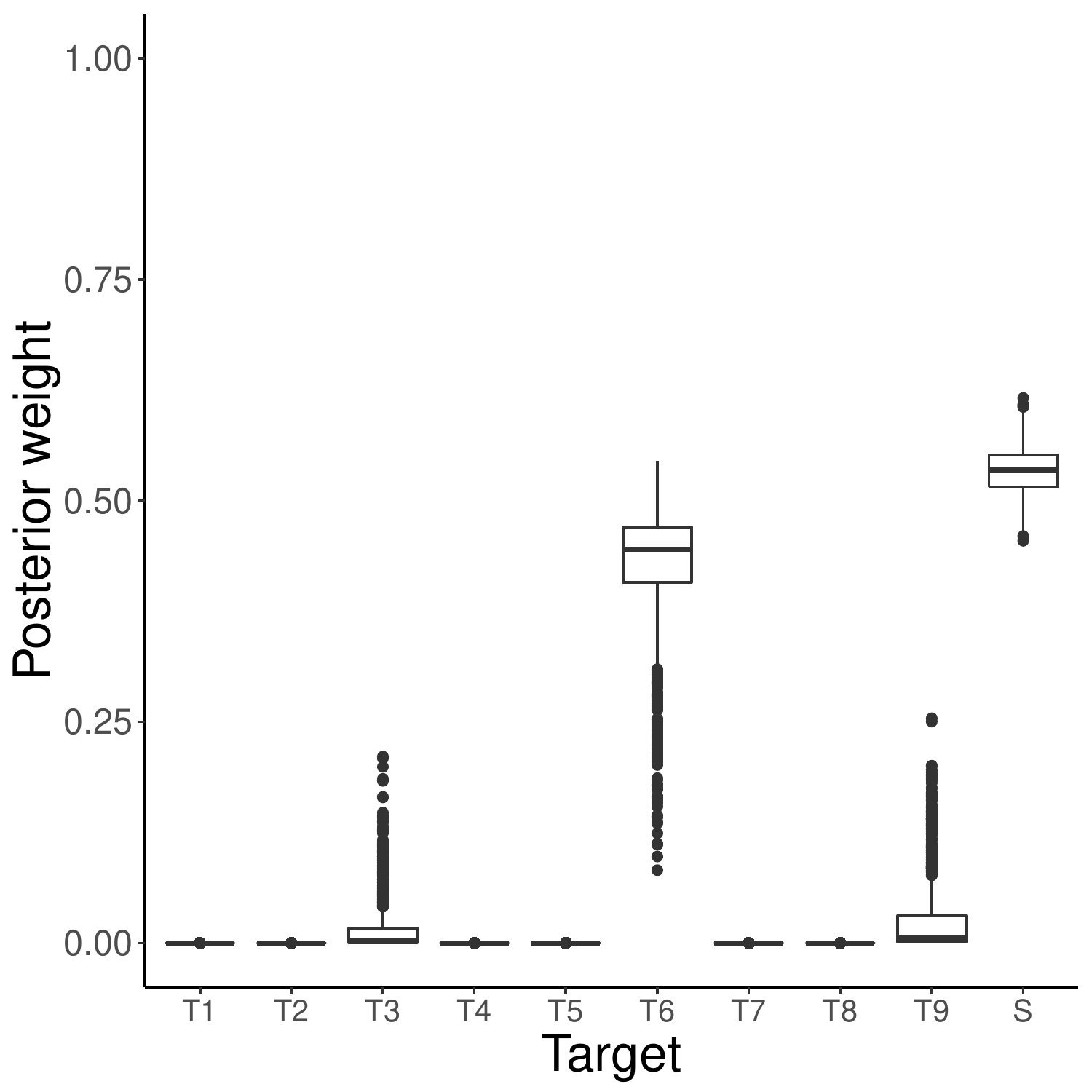}
  	\label{supp:Fig:dataSim2:sub3}
  	}
  	\end{minipage}\hfill
  	\begin{minipage}[c]{0.5\linewidth}
    \centering
    \subfigure[][$\boldsymbol{\hat{\Sigma}}_{\text{TAS-info}}$  ($n = 3p/4$)]{
  	\includegraphics[width=0.8\textwidth]{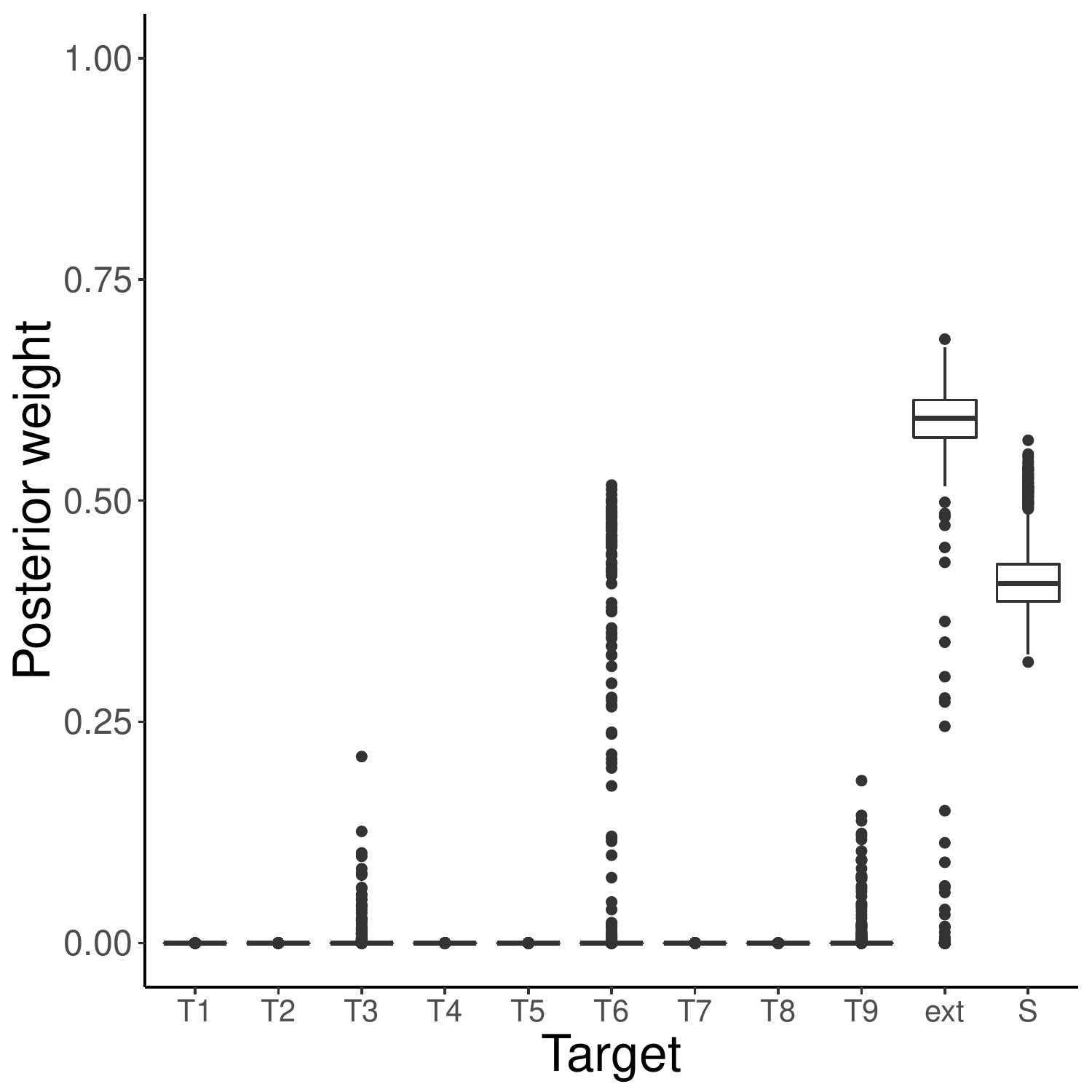}
  	\label{supp:Fig:dataSim2:sub4}
  	}
  	\end{minipage}\hfill
  	\caption{Target-specific posterior weights (see equation \eqref{Eq:TAS_weights}) obtained for estimators $\boldsymbol{\hat{\Sigma}}_{\text{TAS}}$ and $\boldsymbol{\hat{\Sigma}}_{\text{TAS-info}}$ across the $1,000$ random data partitions of the breast cancer data set when $n \in \{ p/4, 3p/4 \}$. The target ``ext'' in $\boldsymbol{\hat{\Sigma}}_{\text{TAS-info}}$ stands for the shrinkage target $\boldsymbol{\hat{\Sigma}}_{\text{ext}}$ estimated from external data.}
\label{supp:Fig:dataSim2}
\end{figure}
\vfill\null

\newpage
\null\vfill
\begin{figure}[ht]
  	\begin{minipage}[c]{0.5\linewidth}
    \centering
    \subfigure[][$\boldsymbol{\hat{\Sigma}}_{\text{TAS}}$ ($n = p/4$)]{
  	\includegraphics[width=0.8\textwidth]{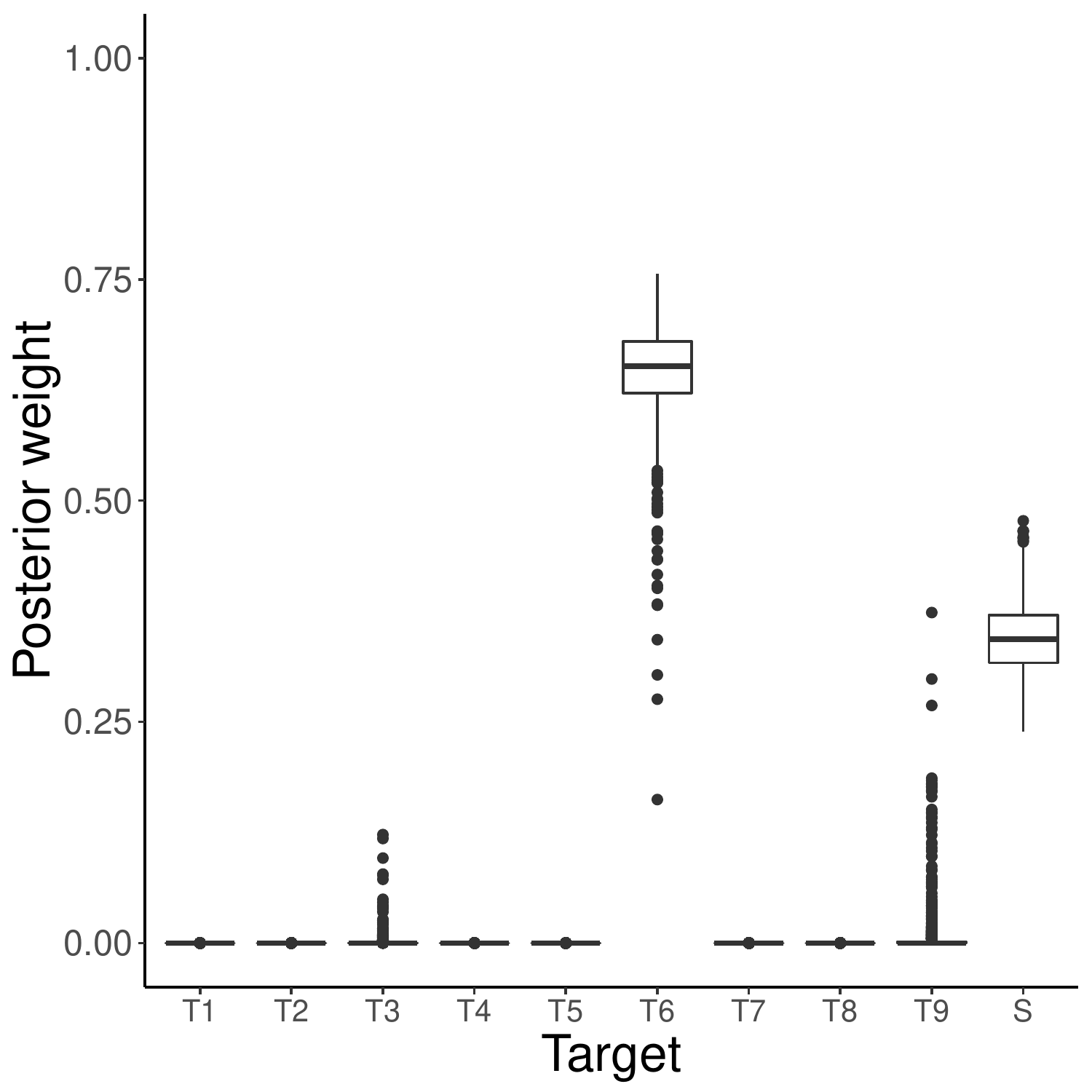}
  	\label{supp:Fig:dataSim3:sub1}
  	}
  	\end{minipage}\hfill
  	\begin{minipage}[c]{0.5\linewidth}
    \centering
    \subfigure[][$\boldsymbol{\hat{\Sigma}}_{\text{TAS-info}}$ ($n = p/4$)]{
  	\includegraphics[width=0.8\textwidth]{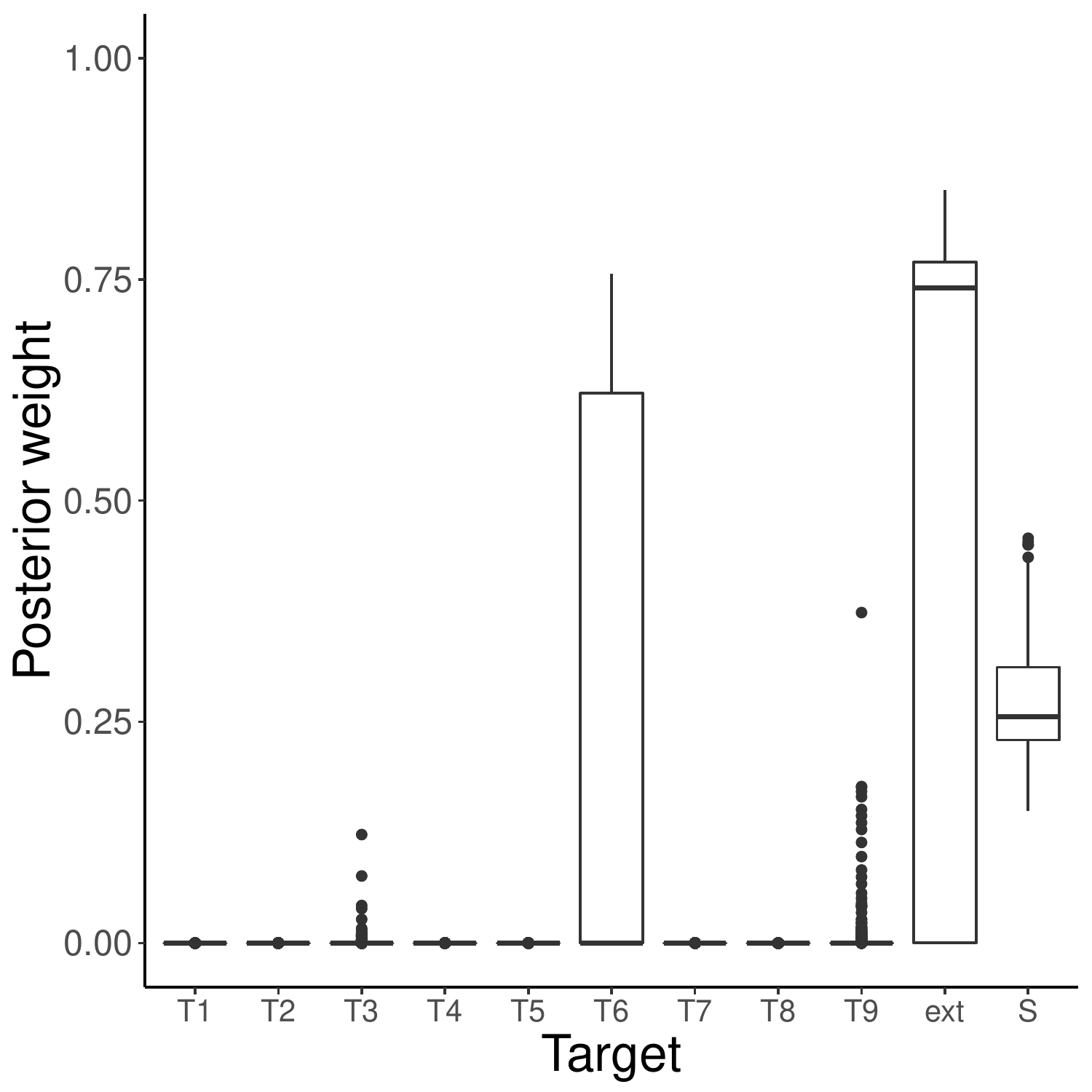}
  	\label{supp:Fig:dataSim3:sub2}
  	}
  	\end{minipage}\hfill
  	  	\begin{minipage}[c]{0.5\linewidth}
    \centering
    \subfigure[][$\boldsymbol{\hat{\Sigma}}_{\text{TAS}}$ ($n = 3p/4$)]{
  	\includegraphics[width=0.8\textwidth]{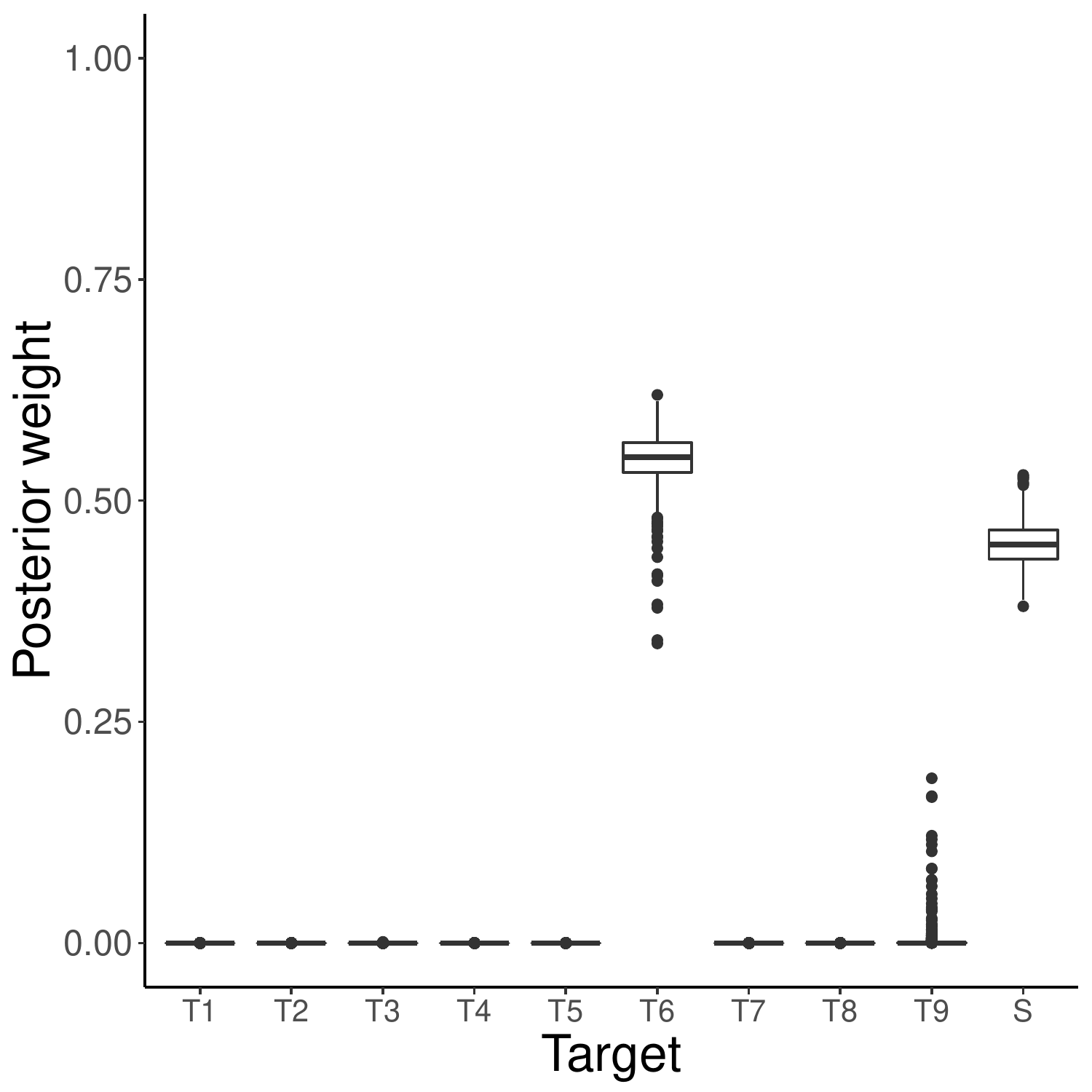}
  	\label{supp:Fig:dataSim3:sub3}
  	}
  	\end{minipage}\hfill
  	\begin{minipage}[c]{0.5\linewidth}
    \centering
    \subfigure[][$\boldsymbol{\hat{\Sigma}}_{\text{TAS-info}}$  ($n = 3p/4$)]{
  	\includegraphics[width=0.8\textwidth]{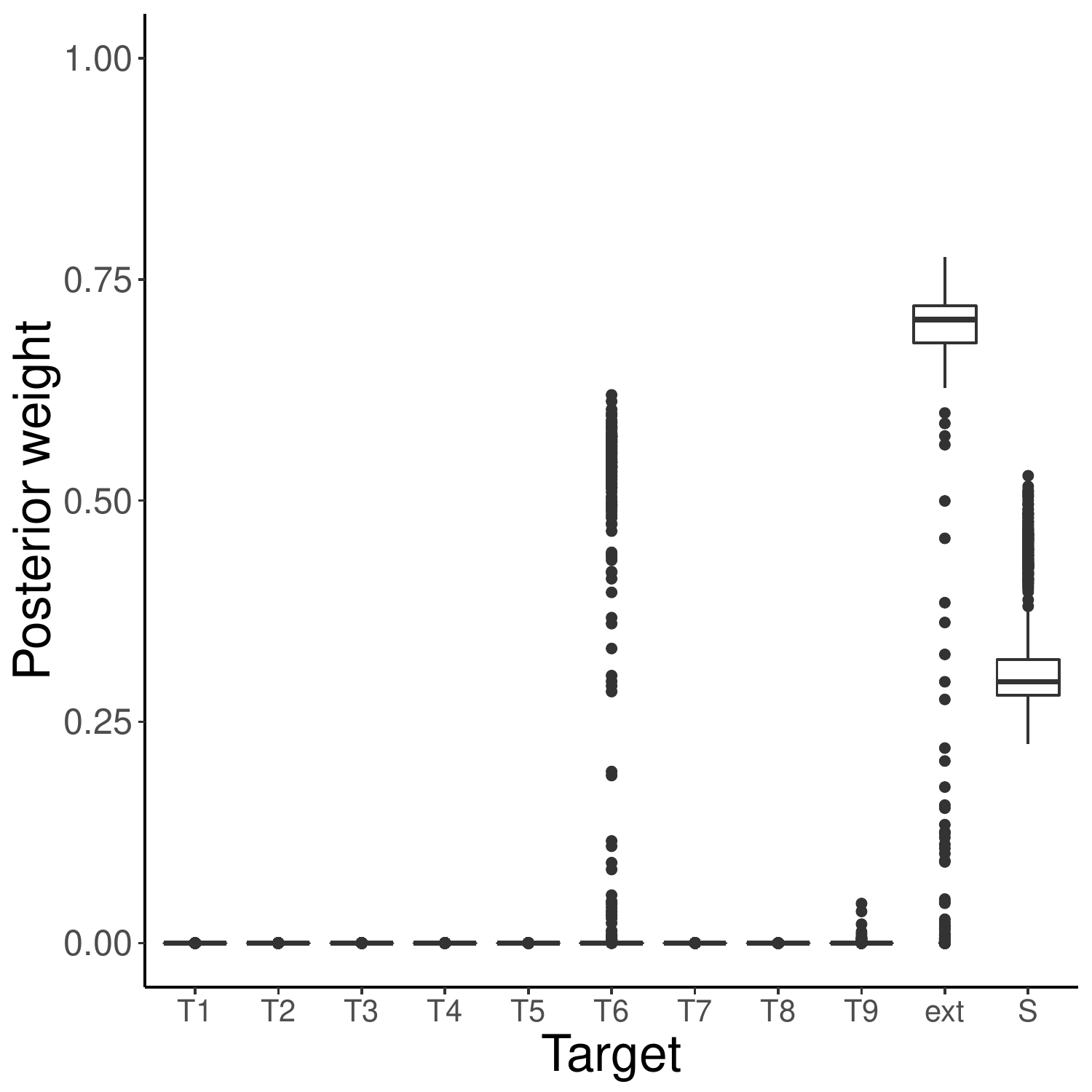}
  	\label{supp:Fig:dataSim3:sub4}
  	}
  	\end{minipage}\hfill
  	\caption{Target-specific posterior weights (see equation \eqref{Eq:TAS_weights}) obtained for estimators $\boldsymbol{\hat{\Sigma}}_{\text{TAS}}$ and $\boldsymbol{\hat{\Sigma}}_{\text{TAS-info}}$ across the $1,000$ random data partitions of the ovarian cancer data set when $n \in \{ p/4, 3p/4 \}$. The target ``ext'' in $\boldsymbol{\hat{\Sigma}}_{\text{TAS-info}}$ stands for the shrinkage target $\boldsymbol{\hat{\Sigma}}_{\text{ext}}$ estimated from external data. }
\label{supp:Fig:dataSim3}
\end{figure}
\vfill\null

\newpage
\section{Assumption of normality}
\label{supp:norm}

\begin{figure}[ht]
	\centering
	\includegraphics[width=0.45\textwidth]{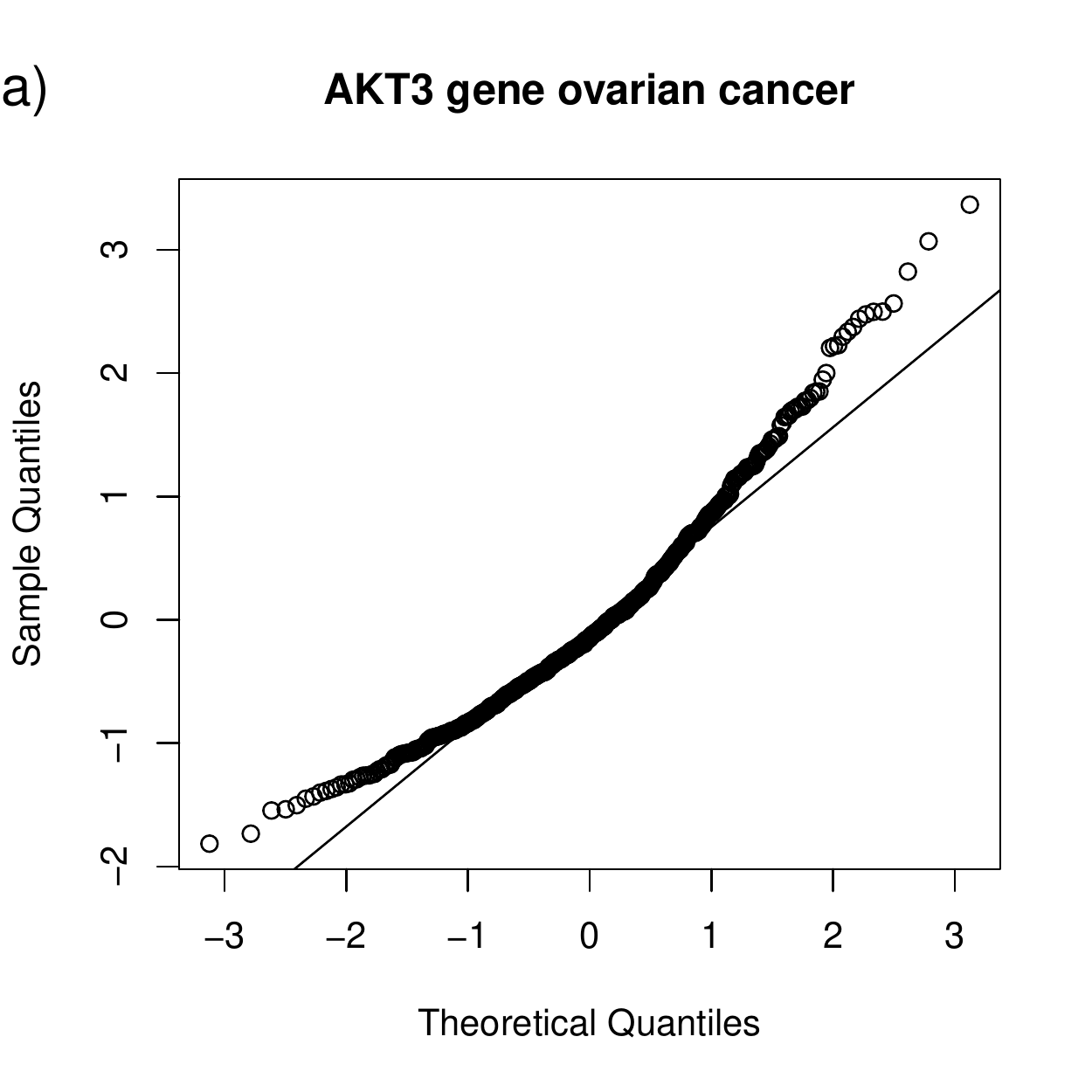}
	\includegraphics[width=0.45\textwidth]{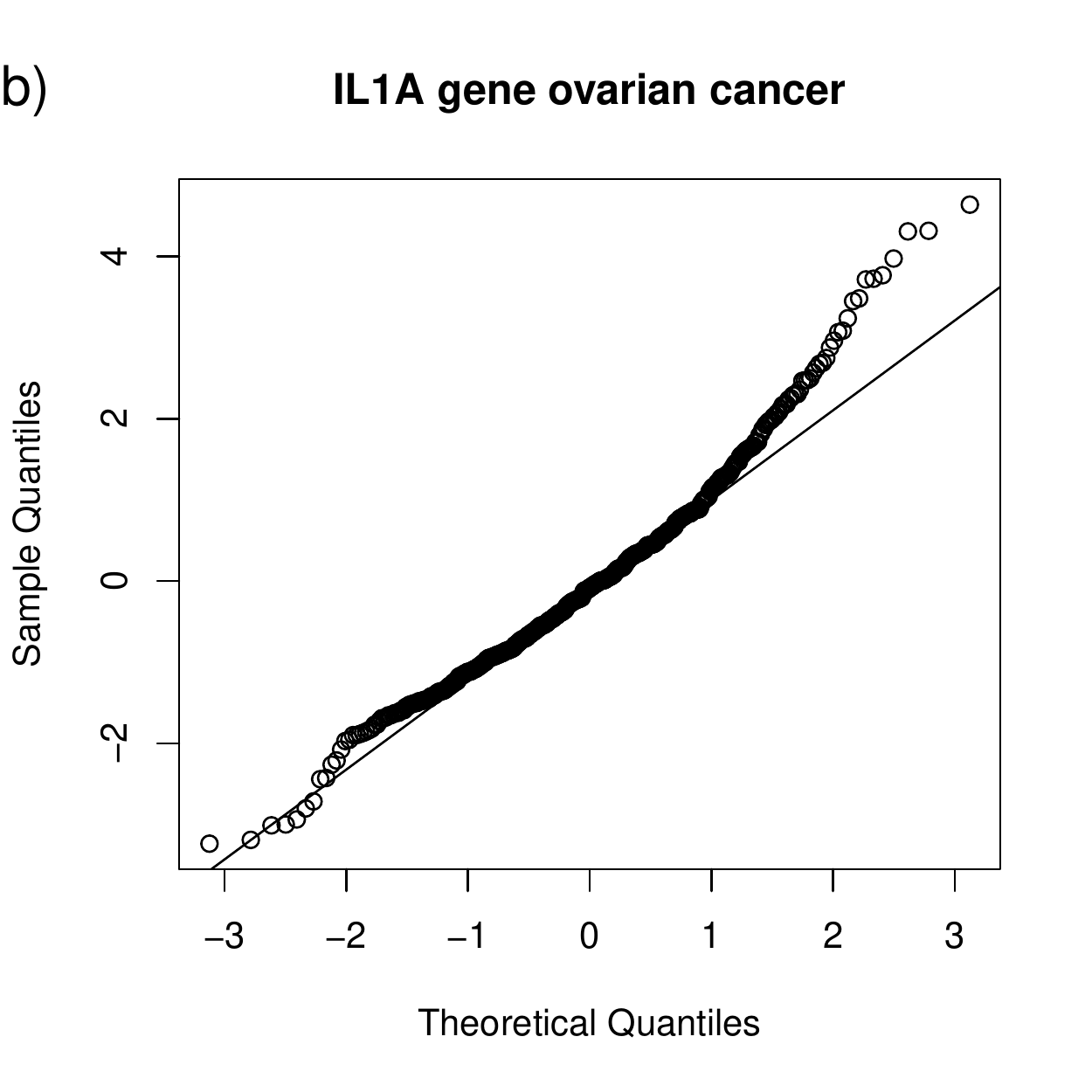}
	\includegraphics[width=0.45\textwidth]{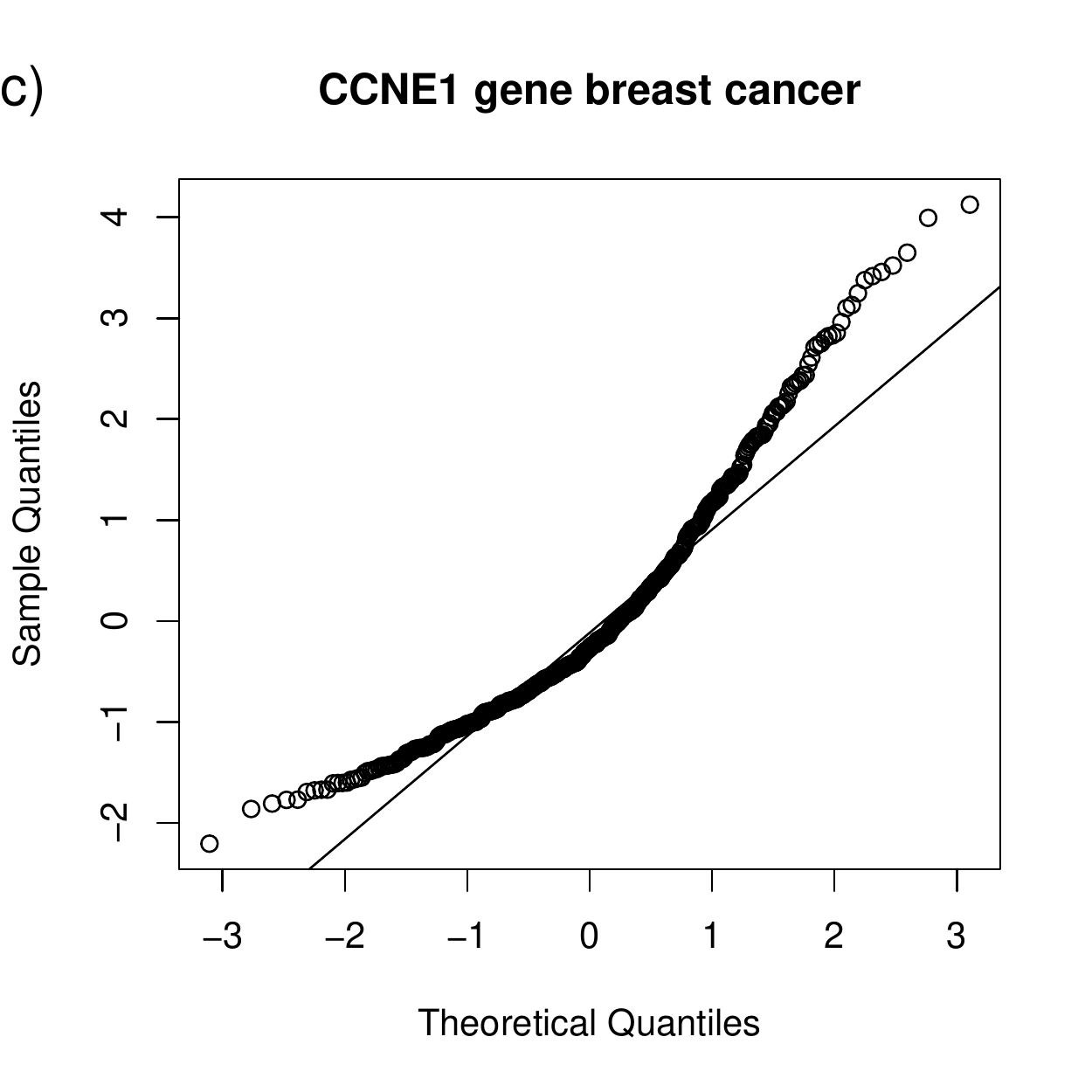}
	\includegraphics[width=0.45\textwidth]{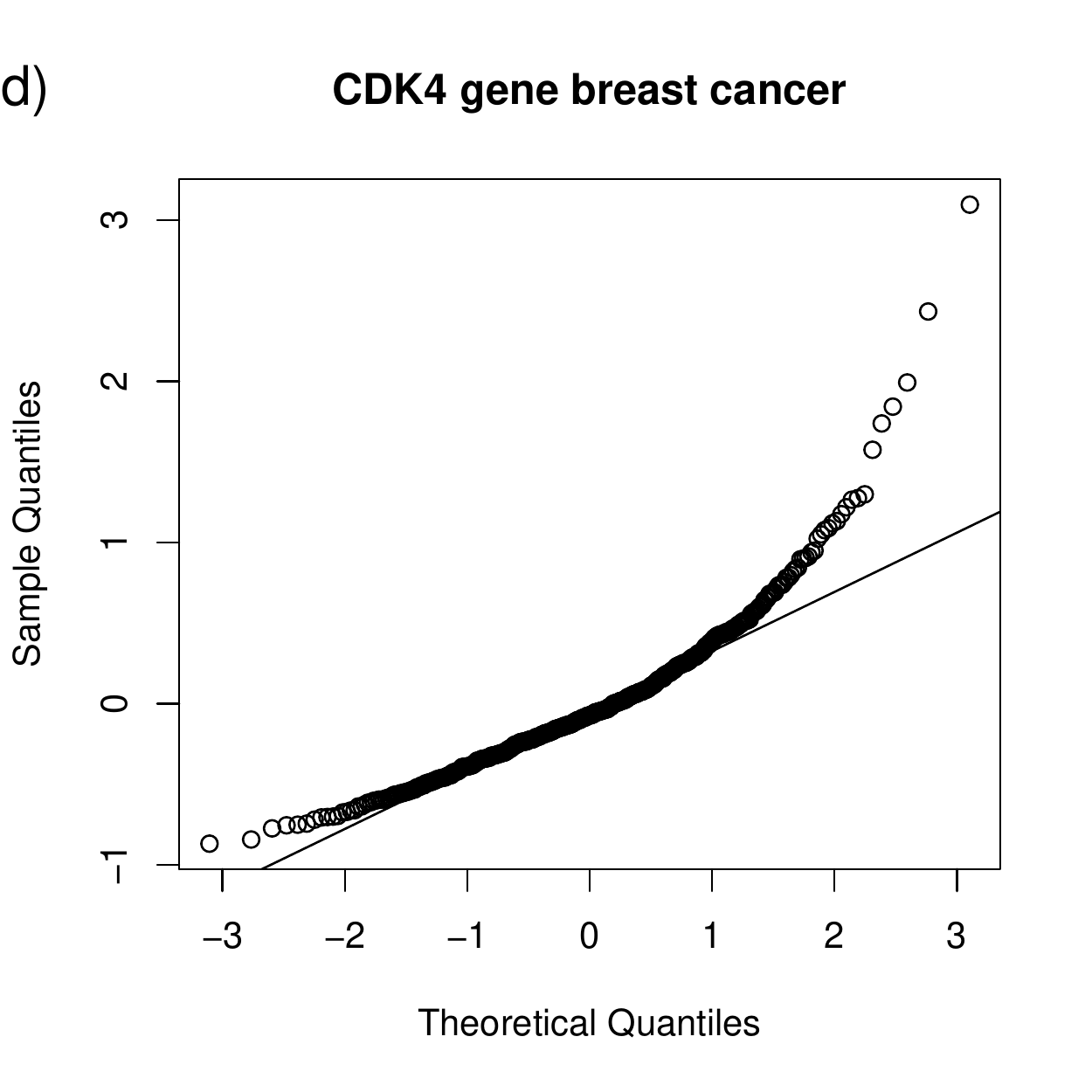}
\caption{Normal Quantile-Quantile plots for two genes from TCGA datasets. Sub-figures (a) and (b) show the departure from normality for genes AKT3 and IL1A in the ovarian cancer data whereas, sub-figures (c) and (d) show the departure from normality for genes CCNE1 and CDK4 in the breast cancer data.}
\label{supp:Fig:mvtnormality}
\end{figure}

Figure \ref{supp:Fig:mvtnormality} provides normal Quantile-Quantile plots for the expression levels of four different genes in two different cancer data sets from TCGA. This provides strong evidence to suggest that the Gaussian assumption does not hold (even for individual genes).\\

\bigskip

\newpage
\section{The PANCAN32 data set}
\label{supp:pancan}

\null
\vfill\null

\begin{table}[ht]\small
\centering
\begin{tabular}{cccc}
	\hline\hline\\[-7pt]
     & Cancer type & TCPA acronym & $n$\\[2pt] \hline\\[-7pt]
    1 & Adrenocortical carcinoma & ACC & 46 \\  
    2 & Bladder urothelial carcinoma & BLCA & 344 \\    
    3 & Breast invasive carcinoma & BRCA & 874\\
    4 & Cervical squamous cell carcinoma and endocervical adenocarcinoma & CESC & 171 \\
    5 & Cholangiocarcinoma & CHOL & 30\\
    6 & Colon adenocarcinoma & COAD & 357 \\
    7 & Lymphoid neoplasm niffuse large B-cell lymphoma & DLBC & 33 \\
    8 & Esophageal carcinoma & ESCA & 126 \\
    9 & Glioblastoma multiforme & GBM & 205 \\
    10 & Head and neck squamous cell carcinoma & HNSC & 346\\
    11 & Kidney chromophobe & KICH & 63 \\
    12 & Kidney renal clear cell carcinoma & KIRC & 445 \\
    13 & Kidney renal papillary cell carcinoma & KIRP & 208 \\
    14 & Brain lower grade glioma & LGG & 427 \\
    15 & Liver hepatocellular carcinoma & LIHC & 184 \\
    16 & Lung adenocarcinoma & LUAD & 362 \\
    17 & Lung squamous cell carcinoma & LUSC & 325 \\
    18 & Mesothelioma & MESO & 61 \\
    19 & Ovarian serous cystadenocarcinoma & OV & 411\\
    20 & Pancreatic adenocarcinoma & PAAD & 105\\
    21 & Pheochromocytoma and paraganglioma & PCPG & 80\\
    22 & Prostate adenocarcinoma & PRAD & 351 \\
    23 & Rectum adenocarcinoma & READ & 130 \\
    24 & Sarcoma & SARC & 221 \\
    25 & Skin cutaneous melanoma & SKCM & 353 \\
    26 & Stomach adenocarcinoma & STAD & 392 \\
    27 & Testicular germ cell tumors & TGCT & 118 \\
    28 & Thyroid carcinoma & THCA & 372 \\
    29 & Thymoma & THYM & 90 \\
    30 & Uterine corpus endometrial carcinoma & UCEC & 404\\
    31 & Uterine carcinosarcoma & UCS & 48 \\
    32 & Uveal melanoma & UVM & 12\\[2pt]
    \hline\hline
\end{tabular}
\caption{Cancer types and number of samples in the PANCAN32 protein expression data set from The Cancer Proteome Atlas.}
\label{supp:Tab:cancer}
\end{table}
\vfill\null

\end{document}